\documentclass[prd,preprint,superscriptaddress,tightenlines,nofootinbib,
  eqsecnum,showpacs]{revtex4}

\usepackage{amsmath}
\usepackage{amsfonts}
\usepackage{amssymb}
\usepackage{bm}
\usepackage{hyperref}
\usepackage{mathrsfs}
\usepackage{graphicx}

\usepackage{ulem}
\usepackage[usenames]{color}

\definecolor{darkgreen}{rgb}{0,0.5,0}

\hypersetup{
    bookmarks=true,         % show bookmarks bar?
    unicode=false,          % non-Latin characters in Acrobat???s bookmarks
    pdftoolbar=true,        % show Acrobat???s toolbar?
    pdfmenubar=true,        % show Acrobat???s menu?
    pdffitwindow=false,     % window fit to page when opened
    pdfstartview={FitH},    % fits the width of the page to the window
    pdftitle={My title},    % title
    pdfauthor={Author},     % author
    pdfsubject={Subject},   % subject of the document
    pdfcreator={Creator},   % creator of the document
    pdfproducer={Producer}, % producer of the document
    pdfkeywords={keyword1} {key2} {key3}, % list of keywords
    pdfnewwindow=true,      % links in new window
    colorlinks=true,       % false: boxed links; true: colored links
    linkcolor=red,          % color of internal links
    citecolor=cyan,        % color of links to bibliography
    filecolor=magenta,      % color of file links
    urlcolor=darkgreen,           % color of external links
    linktocpage=true
}

\allowdisplaybreaks
\DeclareSymbolFontAlphabet{\mathrsfs}{rsfs}
\DeclareMathAlphabet{\mathcal}{OMS}{cmsy}{m}{n}

\newcommand{\ud}{\mathrm{d}}
\newcommand{\ui}{\mathrm{i}} 
\newcommand{\ue}{\mathrm{e}} 
\newcommand{\beq}{\begin{equation}}
\newcommand{\eeq}{\end{equation}}

\newcounter{theorem} 

\begin{document}

\title{Center-of-Mass Equations of Motion and Conserved Integrals of Compact Binary Systems at the Fourth Post-Newtonian Order} 

\author{Laura Bernard}\email{laura.bernard@tecnico.ulisboa.pt}
\affiliation{CENTRA, Departamento de F\'{\i}sica, Instituto Superior T{\'e}cnico -- IST, Universidade de Lisboa -- UL, Avenida Rovisco Pais 1, 1049 Lisboa, Portugal}

\author{Luc Blanchet}\email{luc.blanchet@iap.fr}
\affiliation{$\mathcal{G}\mathbb{R}\varepsilon{\mathbb{C}}\mathcal{O}$, Institut d'Astrophysique de Paris,\\ UMR 7095, CNRS, Sorbonne Universit{\'e}s \& UPMC Univ Paris 6,\\ 98\textsuperscript{bis} boulevard Arago, 75014 Paris, France}

\author{Guillaume Faye}\email{guillaume.faye@iap.fr}
\affiliation{$\mathcal{G}\mathbb{R}\varepsilon{\mathbb{C}}\mathcal{O}$, Institut d'Astrophysique de Paris,\\ UMR 7095, CNRS, Sorbonne Universit{\'e}s \& UPMC Univ Paris 6,\\ 98\textsuperscript{bis} boulevard Arago, 75014 Paris, France}

\author{Tanguy Marchand}\email{tanguy.marchand@iap.fr}
\affiliation{$\mathcal{G}\mathbb{R}\varepsilon{\mathbb{C}}\mathcal{O}$, Institut d'Astrophysique de Paris,\\ UMR 7095, CNRS, Sorbonne Universit{\'e}s \& UPMC Univ Paris 6,\\ 98\textsuperscript{bis} boulevard Arago, 75014 Paris, France}
\affiliation{Laboratoire APC -- Astroparticule et Cosmologie, \\ Universit{\'e} Paris Diderot Paris 7, 75013 Paris, France}

\date{\today}

\begin{abstract}
  The dynamics of compact binary systems at the fourth post-Newtonian (4PN)
  approximation of general relativity has been recently completed in a
  self-consistent way. In this paper, we compute the ten Poincar\'e constants
  of the motion and present the equations of motion in the frame of the center
  of mass (CM), together with the corresponding CM Lagrangian, conserved
  energy and conserved angular momentum. Next, we investigate the reduction of
  the CM dynamics to the case of quasi-circular orbits. The non local (in
  time) tail effect at the 4PN order is consistently included, as well as the
  relevant radiation-reaction dissipative contributions to the energy and
  angular momentum.
\end{abstract}

\pacs{04.25.Nx, 04.30.-w, 97.60.Jd, 97.60.Lf}

\maketitle

\section{Introduction} 
\label{sec:intro}

\subsection{Context and motivation}
\label{sec:motiv}

The problem of the dynamics (\textit{i.e.}, the derivation of the equations of
motion) of two compact objects without spin --- neutron stars or black holes
--- has recently been fully resolved at the fourth post-Newtonian (4PN)
approximation of general relativity in a self-consistent way~\cite{BBBFMa, BBBFMb, BBBFMc, MBBF17}.\footnote{As usual we refer to
  post-Newtonian orders as $n$PN $\sim (v/c)^{2n}$ beyond the Newtonian
  acceleration. See Refs.~\cite{LD17, EIH, Fock, CE70, OO73b, OO74a, DD81b,
    Dhouches, S85, DS85, GKop86, Will, BFP98, IFA00} for ``historical'' works
  on the problem of the equations of motion and Refs.~\cite{JaraS98, JaraS99,
    BFeom, ABF01, BI03CM, DJSdim, BDE04, itoh1, itoh2, FS3PN} for key results at the
  previous 3PN order.} The first attacks of this problem were made by the ADM Hamiltonian formalism of general relativity~\cite{JaraS12, JaraS13,
  JaraS15} and, independently, by using the effective field theory
(EFT)~\cite{GR06, FS4PN, FMSS16}. Then, the ADM Hamiltonian was
completed~\cite{DJS14, DJS16} to include the crucial contribution of near-zone
(conservative) gravitational wave tails at the 4PN order (known from
Refs.~\cite{BD88, B93, B97}). The 4PN tail term was also previously computed
in the EFT approach~\cite{FStail, GLPR16}. Our own work~\cite{BBBFMa, BBBFMb,
  BBBFMc, MBBF17} is based on the construction of the Fokker action of point
particles in harmonic coordinates.\footnote{See, in particular,
  Ref.~\cite{MBBF17} for a synthetic overview of our method and main results.}
Our end result is in complete agreement, modulo an unphysical shift of the
dynamical variables, with that of the ADM Hamiltonian formalism~\cite{JaraS12,
  JaraS13, JaraS15, DJS14, DJS16}.

The ultimate goals of high post-Newtonian (3PN or 4PN) calculations are for
(i) improving the template bank used in the data analysis of compact binary
coalescence (CBC) events in gravitational-wave detectors, and (ii)
facilitating the comparisons or match with the results of numerical
relativity (NR). However, in order to reach these goals the equations of
motion have to be combined with a gravitational wave generation
formalism~\cite{Bliving14}.

A salient feature of the 4PN dynamics is the appearance of infra-red (IR)
divergences, in addition to the usual ultra-violet (UV) ones, due to the model
of point particles describing compact objects~\cite{BDE04}. The IR divergences
entangled the resolution of the problem, as they yielded for a while the
presence of ``ambiguity'' parameters, both in the ADM Hamiltonian~\cite{DJS14,
  DJS16} and the Fokker action~\cite{BBBFMa, BBBFMb}. Finally, the ambiguities
in the Fokker action could be fixed from first principles~\cite{BBBFMc,
  MBBF17}, by systematic use of dimensional regularization to cure both UV and
IR divergences (notably, a consistent computation of the 4PN tail effect in $d$
dimensions), combined with a matching between the near-zone and far-zone
fields. Similarly, one expects that the EFT derivation~\cite{FS4PN, FMSS16,
  FStail,GLPR16}, when it includes all the ``instantaneous'' (non-tail)
contributions, should be free of any ambiguity parameter as well~\cite{PR17}.
By contrast, the ADM Hamiltonian formalism~\cite{DJS14, DJS16} is still facing
one ambiguity parameter, probably because of the lack of consistent matching
between the near-zone and far-zone fields in this formalism, which
specifically deals with the near-zone dynamics but has not yet been developed
to study wave generation~\cite{S85}.

Anyway, the ambiguity parameters are uniquely fixed by comparison to
independent gravitational self-force (GSF) calculations, based on black hole perturbations valid in the small
mass ratio limit, of the conserved energy~\cite{BDLW10b, LBW12, LBB12,
  BiniD13} and the periastron advance~\cite{BDS10, D10sf, Letal11, DJS15eob,
  vdM16} for circular orbits. The final 4PN dynamics in either
Hamiltonian~\cite{JaraS12, JaraS13, JaraS15, DJS14, DJS16} or
Lagrangian~\cite{BBBFMa, BBBFMb, BBBFMc, MBBF17} guise fully reproduces the
relevant GSF results.

Another striking feature of the 4PN dynamics is the non-locality in time due
to the appearance of the tail effect at that order. As discussed
in~\cite{BBBFMb}, the non-locality entails crucial new contributions in the
conserved energy and angular momentum: For instance, the conserved energy for
a non-local Hamiltonian contains extra terms with respect to the on-shell
value of that Hamiltonian, because of the appearance of functional (instead of ordinary)
derivatives in the Hamilton's equations. Our final results carefully take into
account the non-local character of the 4PN dynamics brought by the tail term.
Note that the ``first law of compact binary mechanics'', which is
important in this context because it allows the connection between PN and GSF
results, remains valid for the non-local 4PN dynamics~\cite{BL17}.

The purpose of the present paper is to provide explicit consequences of the
4PN equations of motion that should be useful in various practical
applications, such as direct comparisons to NR calculations~\cite{Boyle08} or
GSF results (see, \textit{e.g.},~\cite{Det08}) during the inspiral phase of
CBCs, the improvement of phenomenological and effective-one-body
templates~\cite{Ajith11, DNorleans}, the study of accretion disks and jet
dynamics around spinning black-hole binaries~\cite{Campa12}, as well as the numerical
calculation of the orbits of non-precessing, spinning black holes via
asymptotic matching~\cite{Campa16}. Another important application is the
setup of accurate initial conditions for the grand-challenge binary black
hole problem in NR~\cite{Pret05, Camp06, Bak06}.

In this perspective, we shall derive the ten Noetherian conserved integrals of
the motion associated with the global Poincar\'e invariance of the Lagrangian
(which is \textit{manifest} in harmonic coordinates). This includes in
particular the integral of the center of mass (CM) associated with the
invariance under Lorentz boosts.\footnote{For lack of space, we shall only
  present the center-of-mass integral in a general frame but give the full
  conserved energy and angular momentum in the CM frame.} Moreover, we shall
obtain the harmonic-coordinates Lagrangian in the CM frame and provide
explicitly the CM relative acceleration. As is well known, this Lagrangian in
harmonic coordinates is a generalized one, depending on positions, velocities
and, from the 2PN order, on accelerations. At the 4PN order, after
adding suitable ``multi-zero'' terms and total time derivatives
(see~\cite{DS85}), the Lagrangian becomes linear in accelerations and does not
contain any derivative of accelerations~\cite{BBBFMa}. By performing shifts of
the dynamical variables, one can transform the generalized Lagrangian into an
ordinary one that corresponds to ADM, or ADM like coordinates, and construct
the corresponding ADM Hamiltonian which is then equivalent to the end result
of Refs.~\cite{DJS14, DJS16}. Finally, since the CBC orbits will
most likely be circular --- at least when the binaries are formed ``in the
field'' so that circularization will have enough time to proceed, we
present all relevant formulas for the reduction to the case of quasi-circular
orbits.

\subsection{Notation and conventions}
\label{sec:not}

The two ordinary coordinate trajectories in a harmonic coordinate system
$\{t, \mathbf{x}\}$ are denoted as $\bm{y}_A(t)$ (with $A=1,2$), the ordinary
velocities are $\bm{v}_A(t)=\ud \bm{y}_A/\ud t$, and the ordinary
accelerations $\bm{a}_A(t)=\ud \bm{v}_A/\ud t$. Here, boldface letters
indicate ordinary three-dimensional Euclidean vectors. In a general frame, the
orbital separation unit vector reads
$\bm{n}_{12} = (\bm{y}_1-\bm{y}_2)/r_{12}$ with
$r_{12}=\vert\bm{y}_1-\bm{y}_2\vert$. Ordinary Euclidean scalar products are
denoted, \textit{e.g.}, $(n_{12} v_1) = \bm{n}_{12}\cdot\bm{v}_1$. In the CM
frame, we rather introduce the alternative notations $\bm{n} = \bm{n}_{12}$
and $r=r_{12}$, together with the relative separation
$\bm{x}=\bm{y}_1-\bm{y}_2$ ($\bm{x}$ should not be confused with the spatial
harmonic coordinate $\mathbf{x}$), relative velocity
$\bm{v}=\bm{v}_1-\bm{v}_2$ and relative acceleration
$\bm{a}=\ud \bm{v}/\ud t$. We pose $v^2=(v v) = \bm{v}^2$ and
$\dot{r}=(n v) = \bm{n}\cdot\bm{v}$. The orbital frequency $\omega$ defined by
$v^2 = \dot{r}^2 + r^2\omega^2$ will be used mostly for quasi-circular orbits,
in which case $\dot{r}=\mathcal{O}(c^{-5})$ and
$v=r\omega[1+\mathcal{O}(c^{-10})]$. The two masses are denoted $m_A$, the
total mass $m$ and the reduced mass $\mu$; the symmetric mass ratio
\begin{equation}\label{nu}
\nu = \frac{\mu}{m} = \frac{m_1m_2}{(m_1+m_2)^2}\,,
\end{equation} 
is such that $0<\nu\leqslant\frac{1}{4}$. We also pose $X_A=\frac{m_A}{m}$, hence
$X_1X_2=\nu$ and $X_1+X_2=1$. In the case of quasi-circular orbits, we introduce
the relevant small PN parameters\footnote{All formulas will include the
  required powers of the gravitational constant $G$ and speed of light $c$.}
\begin{equation}\label{gamx}
\gamma=\frac{G m}{r c^2}\quad\text{and}\quad x=\left(\frac{G m \omega}{c^3}\right)^{2/3}\,.
\end{equation} 
While $\gamma$ is a gauge-dependent parameter, here defined in harmonic
coordinates, the frequency-dependent parameter $x$ is invariant in a large
class of coordinate systems: those that are asymptotically flat at infinity.
Upper indices $(s)$ denote $s$-th time derivatives; the symmetric-trace-free
(STF) projection is indicated by angular brackets surrounding indices,
\textit{e.g.},
$x^{\langle i}x^{j\rangle}=x^i x^j - \frac{1}{3}\delta^{ij}\bm{x}^2$;
$\epsilon_{ijk}$ is the totally anti-symmetric Levi-Civita symbol, whereas
$\gamma_\text{E}$ denotes Euler's constant.

Any physical quantity $Q$ investigated in this paper will be the sum of (i) many
\textit{instantaneous} (local-in-time) terms up to the 4PN order, (ii) the non-local
conservative tail term arising at the 4PN order, and, in most cases, (iii)
dissipative radiation-reaction terms present at the 2.5PN, 3.5PN, as well as the 4PN
orders. Accordingly, we write
\begin{equation}\label{Qfull}
Q = Q^\text{inst} + Q^\text{tail} + Q^\text{diss}\,.
\end{equation}
Since the long explicit results presented in the present article will concern the
instantaneous part of the dynamics, we adopt the convention that any
$Q^\text{inst}$ is directly given by the sequence of its PN coefficients
$Q_{n\text{PN}}$:
\begin{equation}\label{notationQ}
Q^\text{inst} = Q_\text{N} +
\frac{1}{c^2}Q_\text{1PN} +
\frac{1}{c^4}Q_\text{2PN} +
\frac{1}{c^6}Q_\text{3PN} +
\frac{1}{c^8}Q_\text{4PN} + \mathcal{O}\left(\frac{1}{c^{10}}\right)\,.
\end{equation}
Furthermore, as the 4PN coefficient is especially lengthy, we split it
into non-linear contributions corresponding to increasing powers of $G$:
\begin{equation}\label{notationQG}
Q_\text{4PN} = Q^{(0)}_\text{4PN} + G\,Q^{(1)}_\text{4PN} + G^2 Q^{(2)}_\text{4PN} 
+ G^3 Q^{(3)}_\text{4PN}+ G^4 Q^{(4)}_\text{4PN}+ G^5 Q^{(5)}_\text{4PN} \,.
\end{equation}
At 4PN order, there are no terms beyond the quintic non-linearities
$\propto G^5$. When relevant, the dissipative terms will be simply indicated
as $Q_\text{2.5PN}$ (for instance). We generally do not write a PN coefficient when it is zero.

Our results for the instantaneous part of the dynamics concern the CM
equations of motion and Lagrangian in Sec.~\ref{sec:Lag}, as well as the CM
conserved energy and angular momentum in Sec.~\ref{sec:cons}. The non-local
4PN tail part of the dynamics is investigated in Sec.~\ref{sec:tails},
basically following Ref.~\cite{BBBFMb}. The reduction to
quasi-circular orbits is achieved in Sec.~\ref{sec:EOMcirc} (including tail terms), while we add the
dissipative radiation-reaction 2.5PN, 3.5PN and 4PN terms in
Sec.~\ref{sec:diss}. Finally, we recap our general-frame 4PN Lagrangian in the
Appendix~\ref{sec:Lgen} and present the general-frame integral of the CM in
App.~\ref{sec:Gi}.

\section{Center-of-mass equations of motion and Lagrangian} 
\label{sec:Lag} 

Starting from the general-frame Lagrangian of Refs.~\cite{BBBFMa,BBBFMb} (see
App.~\ref{sec:Lgen} for a recap), we obtain the integral of the center of mass
$\bm{G}$. We then define the CM frame by solving iteratively the equation
$\bm{G} = 0$ (with order reduction of the accelerations). This gives explicit
expressions for the CM variables $\bm{y}_A$ and $\bm{v}_A$ as functions of the
relative position $\bm{x}=\bm{y}_1-\bm{y}_2$ and velocity
$\bm{v}=\bm{v}_1-\bm{v}_2$. The complete expressions of the CM integral
$\bm{G}$ and of the CM trajectories $\bm{y}_A$ are given in App.~\ref{sec:Gi}.
As we shall find in Sec.~\ref{sec:tails}, there are no tail contributions in
the CM trajectories $\bm{y}_A$ nor in the CM velocities
$\bm{v}_A$.

Next, we insert the CM variables [Eqs.~\eqref{y1y2}--\eqref{Q} in
App.~\ref{sec:Gi}] into the equations of motion derived from the variation of
the general-frame Lagrangian, from which we obtain the relative
acceleration $\bm{a}= \bm{a}_1-\bm{a}_2$ as a functional
of $\bm{n}$, $r$, $\dot{r}=(nv)$ and $\bm{v}$. The (instantaneous part of the)
acceleration takes the form
\begin{equation}\label{dvdt}
\bm{a} = -\frac{G m}{r^2}\Big[\big(1+A\big)\,\bm{n} +
    B\,\bm{v} \Big] \,,
\end{equation}
where the various PN coefficients are given by (see Sec.~\ref{sec:not} for our conventions)
\begin{subequations}\label{A}
\begin{align}
A_\text{1PN} &= -\frac{3\,\dot{r}^2\,\nu}{2} + v^2 +
3\,\nu\,v^2-\frac{G m}{r}\left(4 +2\,\nu \right) \,,\\
%%%%%%%%%%%%%%%%%%%%%%%%%%%%%%%%%%%%%%%%%%%%%%%%%%%%%%%%%%%%%%%%%
A_\text{2PN} &= \frac{15\,\dot{r}^4\,\nu}{8} -
\frac{45\,\dot{r}^4\,\nu^2}{8} - \frac{9\,\dot{r}^2\,\nu\,v^2}{2} +
6\,\dot{r}^2\,\nu^2\,v^2 + 3\,\nu\,v^4 -
4\,\nu^2\,v^4 \nonumber\\ &\quad + \frac{G m}{r}\left(
-2\,\dot{r}^2 - 25\,\dot{r}^2\,\nu - 2\,\dot{r}^2\,\nu^2 -
\frac{13\,\nu\,v^2}{2} + 2\,\nu^2\,v^2 \right) \nonumber\\
&\quad + \frac{G^2m^2}{r^2}\,\left( 9 + \frac{87\,\nu}{4}
\right)\,,\\
%%%%%%%%%%%%%%%%%%%%%%%%%%%%%%%%%%%%%%%%%%%%%%%%%%%%%%%%%%%%%%%%%
A_\text{3PN} &= -\frac{35\,\dot{r}^6\,\nu}{16} +
\frac{175\,\dot{r}^6\,\nu^2}{16} -
\frac{175\,\dot{r}^6\,\nu^3}{16}+\frac{15\,\dot{r}^4\,\nu\,v^2}{2}
\nonumber\\&\quad - \frac{135\,\dot{r}^4\,\nu^2\,v^2}{4} +
\frac{255\,\dot{r}^4\,\nu^3\,v^2}{8} -
\frac{15\,\dot{r}^2\,\nu\,v^4}{2} +
\frac{237\,\dot{r}^2\,\nu^2\,v^4}{8} \nonumber\\ &\quad
-\frac{45\,\dot{r}^2\,\nu^3\,v^4}{2} + \frac{11\,\nu\,v^6}{4} -
\frac{49\,\nu^2\,v^6}{4} + 13\,\nu^3\,v^6 \nonumber\\ &\quad +
\frac{G m}{r}\left( 79\,\dot{r}^4\,\nu -
\frac{69\,\dot{r}^4\,\nu^2}{2} - 30\,\dot{r}^4\,\nu^3 -
121\,\dot{r}^2\,\nu\,v^2 + 16\,\dot{r}^2\,\nu^2\,v^2
\right.\nonumber\\&\quad\quad\quad \left.+
20\,\dot{r}^2\,\nu^3\,v^2+\frac{75\,\nu\,v^4}{4} + 8\,\nu^2\,v^4 -
10\,\nu^3\,v^4 \right) \nonumber\\ &\quad +
\frac{G^2m^2}{r^2}\,\left( \dot{r}^2 +
\frac{32573\,\dot{r}^2\,\nu}{168} + \frac{11\,\dot{r}^2\,\nu^2}{8} -
7\,\dot{r}^2\,\nu^3 + \frac{615\,\dot{r}^2\,\nu\,\pi^2}{64} -
\frac{26987\,\nu\,v^2}{840} \right.\nonumber\\&\quad\quad\quad
+\left.  \nu^3\,v^2 - \frac{123\,\nu\,\pi^2\,v^2}{64} -
110\,\dot{r}^2\,\nu\,\ln \Big(\frac{r}{r'_0}\Big) + 22\,\nu\,v^2\,\ln
\Big(\frac{r}{r'_0}\Big) \right)\nonumber\\&\quad
+\frac{G^3m^3}{r^3}\left( -16 - \frac{437\,\nu}{4} -
\frac{71\,\nu^2}{2} + \frac{41\,\nu\,{\pi }^2}{16} \right)\,,
\end{align}
up to 3PN order, together with, for the 4PN terms,
\begin{align}
%%%%%%%%%%%%%%%%%%%%%%%%%%%%%%%%%%%%%%%%%%%%%%%%%%%%%%%%%%%%%%%%%
A^{(0)}_\text{4PN} &= \left(\frac{315}{128} \nu -  \frac{2205}{128} \nu^2 +
                     \frac{2205}{64} \nu^3 -  \frac{2205}{128} \nu^4\right)
                     \dot{r}^8 + \left(- \frac{175}{16} \nu + \frac{595}{8}
                     \nu^2 -  \frac{2415}{16} \nu^3 \right.\nonumber\\
&\left.\quad + \frac{735}{8} \nu^4\right) \dot{r}^6 v^{2} +
  \left(\frac{135}{8} \nu -  \frac{1875}{16} \nu^2 + \frac{4035}{16} \nu^3 -
  \frac{1335}{8} \nu^4\right) \dot{r}^4 v^{4} + \left(- \frac{21}{2} \nu +
  \frac{1191}{16} \nu^2\right.\nonumber\\
&\quad\left. -  \frac{327}{2} \nu^3 + 99 \nu^4\right) \dot{r}^2 v^{6} +
  \left(\frac{21}{8} \nu -  \frac{175}{8} \nu^2 + 61 \nu^3  - 54 \nu^4\right)
  v^{8}
 \,,\\
%%%%%%%%%%%%%%%%%%%%%%%%%%%%%%%%%%%%%%%%%%%%%%%%%%%%%%%%%%%%%%%%%
A^{(1)}_\text{4PN} &= \frac{m}{r} \left(\frac{2973}{40} \nu \dot{r}^6
+ 407 \nu^2 \dot{r}^6+ \frac{181}{2} \nu^3 \dot{r}^6 - 86 \nu^4 \dot{r}^6 
+ \frac{1497}{32} \nu \dot{r}^4 v^{2}
 -  \frac{1627}{2} \nu^2 \dot{r}^4 v^{2} \right.\nonumber\\
& \quad- 81 \nu^3 \dot{r}^4 v^{2} + 228 \nu^4 \dot{r}^4 v^{2} 
-  \frac{2583}{16} \nu \dot{r}^2 v^{4} + \frac{1009}{2} \nu^2 \dot{r}^2 v^{4} 
+ 47 \nu^3 \dot{r}^2 v^{4} - 104 \nu^4 \dot{r}^2 v^{4}\nonumber\\
&\quad\left. + \frac{1067}{32} \nu v^{6} - 58 \nu^2 v^{6} - 44 \nu^3 v^{6} 
+ 58 \nu^4 v^{6}\right)
 \,,\\
%%%%%%%%%%%%%%%%%%%%%%%%%%%%%%%%%%%%%%%%%%%%%%%%%%%%%%%%%%%%%%%%%
A^{(2)}_\text{4PN} &= \frac{m^2}{r^2} \left(\frac{2094751}{960} \nu \dot{r}^4
 + \frac{45255}{1024} \pi^2 \nu \dot{r}^4
 + \frac{326101}{96} \nu^2 \dot{r}^4
 -  \frac{4305}{128} \pi^2 \nu^2 \dot{r}^4
 -  \frac{1959}{32} \nu^3 \dot{r}^4\right.\nonumber\\
& \quad - 126 \nu^4 \dot{r}^4
 + 385 \nu^2 \ln\Big(\frac{r}{r'_{0}}\Big) \dot{r}^4
 + 385 \nu \ln\Big(\frac{r}{r''_{0}}\Big) \dot{r}^4
 - 1540 \nu^2 \ln\Big(\frac{r}{r''_{0}}\Big) \dot{r}^4
 -  \frac{1636681}{1120} \nu \dot{r}^2 v^{2}\nonumber\\
& \quad -  \frac{12585}{512} \pi^2 \nu \dot{r}^2 v^{2}
 -  \frac{255461}{112} \nu^2 \dot{r}^2 v^{2}
 + \frac{3075}{128} \pi^2 \nu^2 \dot{r}^2 v^{2}
 -  \frac{309}{4} \nu^3 \dot{r}^2 v^{2}
 + 63 \nu^4 \dot{r}^2 v^{2}\nonumber\\
&\quad  - 330 \nu \ln\Big(\frac{r}{r'_{0}}\Big) \dot{r}^2 v^{2}
 - 275 \nu^2 \ln\Big(\frac{r}{r'_{0}}\Big) \dot{r}^2 v^{2}
 - 275 \nu \ln\Big(\frac{r}{r''_{0}}\Big) \dot{r}^2 v^{2}
 + 1100 \nu^2 \ln\Big(\frac{r}{r''_{0}}\Big) \dot{r}^2 v^{2}\nonumber\\
&\quad  + \frac{1096941}{11200} \nu v^{4}
 + \frac{1155}{1024} \pi^2 \nu v^{4}
 + \frac{7263}{70} \nu^2 v^{4}
 -  \frac{123}{64} \pi^2 \nu^2 v^{4}
 + \frac{145}{2} \nu^3 v^{4}
 - 16 \nu^4 v^{4}\nonumber\\
&\quad \left. + 66 \nu \ln\Big(\frac{r}{r'_{0}}\Big) v^{4}
 + 22 \nu^2 \ln\Big(\frac{r}{r'_{0}}\Big) v^{4}
 + 22 \nu \ln\Big(\frac{r}{r''_{0}}\Big) v^{4}
 - 88 \nu^2 \ln\Big(\frac{r}{r''_{0}}\Big) v^{4}\right) \,,\\
%%%%%%%%%%%%%%%%%%%%%%%%%%%%%%%%%%%%%%%%%%%%%%%%%%%%%%%%%%%%%%%%%
A^{(3)}_\text{4PN} &= \frac{m^3}{r^3} \left(-2 \dot{r}^2
 + \frac{1297943}{8400} \nu \dot{r}^2
 -  \frac{2969}{16} \pi^2 \nu \dot{r}^2
 + \frac{1255151}{840} \nu^2 \dot{r}^2
 + \frac{7095}{32} \pi^2 \nu^2 \dot{r}^2
 - 17 \nu^3 \dot{r}^2 \right.\nonumber\\
& \quad - 24 \nu^4 \dot{r}^2
 + 384 \ln\Big(\frac{r}{r'_{0}}\Big) \dot{r}^2
 - 920 \nu \ln\Big(\frac{r}{r'_{0}}\Big) \dot{r}^2
 + 3100 \nu^2 \ln\Big(\frac{r}{r'_{0}}\Big) \dot{r}^2
 - 384 \ln\Big(\frac{r}{r''_{0}}\Big) \dot{r}^2\nonumber\\
& \quad + 3152 \nu \ln\Big(\frac{r}{r''_{0}}\Big) \dot{r}^2
 - 6464 \nu^2 \ln\Big(\frac{r}{r''_{0}}\Big) \dot{r}^2
 + \frac{1237279}{25200} \nu v^{2}
 + \frac{3835}{96} \pi^2 \nu v^{2}
 -  \frac{693947}{2520} \nu^2 v^{2}\nonumber\\
&\quad  -  \frac{229}{8} \pi^2 \nu^2 v^{2}
 + \frac{19}{2} \nu^3 v^{2}
 - 64 \ln\Big(\frac{r}{r'_{0}}\Big) v^{2}
 + 80 \nu \ln\Big(\frac{r}{r'_{0}}\Big) v^{2}
 -  \frac{1616}{3} \nu^2 \ln\Big(\frac{r}{r'_{0}}\Big) v^{2}\nonumber\\
&\quad\left. + 64 \ln\Big(\frac{r}{r''_{0}}\Big) v^{2}
 -  \frac{1576}{3} \nu \ln\Big(\frac{r}{r''_{0}}\Big) v^{2}
 + \frac{3232}{3} \nu^2 \ln\Big(\frac{r}{r''_{0}}\Big) v^{2}\right) \,,\\
%%%%%%%%%%%%%%%%%%%%%%%%%%%%%%%%%%%%%%%%%%%%%%%%%%%%%%%%%%%%%%%%%
A^{(4)}_\text{4PN} &= \frac{m^4}{r^4} \left(25
 + \frac{6625537}{12600} \nu
 -  \frac{4543}{96} \pi^2 \nu
 + \frac{477763}{720} \nu^2
 + \frac{3}{4} \pi^2 \nu^2
 + 16 \ln\Big(\frac{r}{r'_{0}}\Big)
 - 20 \nu \ln\Big(\frac{r}{r'_{0}}\Big) \right. \nonumber\\
& \quad \left. + 98 \nu^2 \ln\Big(\frac{r}{r'_{0}}\Big)
 - 16 \ln\Big(\frac{r}{r''_{0}}\Big)
 + \frac{394}{3} \nu \ln\Big(\frac{r}{r''_{0}}\Big)
 -  \frac{808}{3} \nu^2 \ln\Big(\frac{r}{r''_{0}}\Big)\right)\,.
%%%%%%%%%%%%%%%%%%%%%%%%%%%%%%%%%%%%%%%%%%%%%%%%%%%%%%%%%%%%%%%%%
\end{align}\end{subequations}
Similarly:
\begin{subequations}\label{B}
\begin{align}
B_\text{1PN} &= -4\,\dot{r} + 2\,\dot{r}\,\nu\,,\\
%%%%%%%%%%%%%%%%%%%%%%%%%%%%%%%%%%%%%%%%%%%%%%%%%%%%%%%%%%%%%%%%%
B_\text{2PN} &= \frac{9\,\dot{r}^3\,\nu}{2} +
3\,\dot{r}^3\,\nu^2 -\frac{15\,\dot{r}\,\nu\,v^2}{2} -
2\,\dot{r}\,\nu^2\,v^2\nonumber\\ &\quad +
\frac{G m}{r}\left( 2\,\dot{r} + \frac{41\,\dot{r}\,\nu}{2} +
4\,\dot{r}\,\nu^2 \right)\,,\\
%%%%%%%%%%%%%%%%%%%%%%%%%%%%%%%%%%%%%%%%%%%%%%%%%%%%%%%%%%%%%%%%%%
B_\text{3PN} &= -\frac{45\,\dot{r}^5\,\nu}{8} +
15\,\dot{r}^5\,\nu^2 + \frac{15\,\dot{r}^5\,\nu^3}{4} +
12\,\dot{r}^3\,\nu\,v^2 \nonumber\\&\quad
- \frac{111\,\dot{r}^3\,\nu^2\,v^2}{4}
-12\,\dot{r}^3\,\nu^3\,v^2 -\frac{65\,\dot{r}\,\nu\,v^4}{8} +
19\,\dot{r}\,\nu^2\,v^4 + 6\,\dot{r}\,\nu^3\,v^4
\nonumber\\&\quad + \frac{G m}{r}\left(
\frac{329\,\dot{r}^3\,\nu}{6} + \frac{59\,\dot{r}^3\,\nu^2}{2} +
18\,\dot{r}^3\,\nu^3 - 15\,\dot{r}\,\nu\,v^2 - 27\,\dot{r}\,\nu^2\,v^2
- 10\,\dot{r}\,\nu^3\,v^2 \right) \nonumber\\&\quad
+ \frac{G^2m^2}{r^2}\,\left( -4\,\dot{r} -
\frac{18169\,\dot{r}\,\nu}{840} + 25\,\dot{r}\,\nu^2 +
8\,\dot{r}\,\nu^3 - \frac{123\,\dot{r}\,\nu\,\pi^2}{32} 
+ 44\,\dot{r}\,\nu\,\ln \Big(\frac{r}{r'_0}\Big)
\right)\,,\\
%%%%%%%%%%%%%%%%%%%%%%%%%%%%%%%%%%%%%%%%%%%%%%%%%%%%%%%%%%%%%%%%%
B^{(0)}_\text{4PN} &= \left(\frac{105}{16} \nu
 -  \frac{245}{8} \nu^2
 + \frac{385}{16} \nu^3
 + \frac{35}{8} \nu^4\right) \dot{r}^7
 + \left(- \frac{165}{8} \nu
 + \frac{1665}{16} \nu^2
 -  \frac{1725}{16} \nu^3
 -  \frac{105}{4} \nu^4\right) \dot{r}^5 v^{2}\nonumber\\
&\quad + \left(\frac{45}{2} \nu
 -  \frac{1869}{16} \nu^2
 + 129 \nu^3
 + 54 \nu^4\right) \dot{r}^3 v^{4}
 + \left(- \frac{157}{16} \nu
 + 54 \nu^2
 - 69 \nu^3
 - 24 \nu^4\right) \dot{r} v^{6}\,,\\
%%%%%%%%%%%%%%%%%%%%%%%%%%%%%%%%%%%%%%%%%%%%%%%%%%%%%%%%%%%%%%%%%
B^{(1)}_\text{4PN} &= \frac{m}{r} \left(- \frac{54319}{160} \nu \dot{r}^5
 -  \frac{901}{8} \nu^2 \dot{r}^5
 + 60 \nu^3 \dot{r}^5
 + 30 \nu^4 \dot{r}^5
 + \frac{25943}{48} \nu \dot{r}^3 v^{2}
 + \frac{1199}{12} \nu^2 \dot{r}^3 v^{2} \right. \nonumber\\
&\quad \left. -  \frac{349}{2} \nu^3 \dot{r}^3 v^{2}
 - 98 \nu^4 \dot{r}^3 v^{2}
 -  \frac{5725}{32} \nu \dot{r} v^{4}
 -  \frac{389}{8} \nu^2 \dot{r} v^{4}
 + 118 \nu^3 \dot{r} v^{4}
 + 44 \nu^4 \dot{r} v^{4}\right)
 \,,\\
%%%%%%%%%%%%%%%%%%%%%%%%%%%%%%%%%%%%%%%%%%%%%%%%%%%%%%%%%%%%%%%%%
B^{(2)}_\text{4PN} &= \frac{m^2}{r^2} \left(- \frac{9130111}{3360} \nu \dot{r}^3
 -  \frac{4695}{256} \pi^2 \nu \dot{r}^3
 -  \frac{184613}{112} \nu^2 \dot{r}^3
 + \frac{1845}{64} \pi^2 \nu^2 \dot{r}^3
 + \frac{209}{2} \nu^3 \dot{r}^3 \right.\nonumber\\
&\quad + 74 \nu^4 \dot{r}^3
 + 660 \nu \ln\Big(\frac{r}{r'_{0}}\Big) \dot{r}^3
 - 330 \nu^2 \ln\Big(\frac{r}{r'_{0}}\Big) \dot{r}^3
 - 220 \nu \ln\Big(\frac{r}{r''_{0}}\Big) \dot{r}^3
 + 880 \nu^2 \ln\Big(\frac{r}{r''_{0}}\Big) \dot{r}^3\nonumber\\
&\quad + \frac{8692601}{5600} \nu \dot{r} v^{2}
 + \frac{1455}{256} \pi^2 \nu \dot{r} v^{2}
 + \frac{58557}{70} \nu^2 \dot{r} v^{2}
 -  \frac{123}{8} \pi^2 \nu^2 \dot{r} v^{2}
 - 70 \nu^3 \dot{r} v^{2}
 - 34 \nu^4 \dot{r} v^{2}\nonumber\\
&\quad \left. - 264 \nu \ln\Big(\frac{r}{r'_{0}}\Big) \dot{r} v^{2}
 + 176 \nu^2 \ln\Big(\frac{r}{r'_{0}}\Big) \dot{r} v^{2}
 + 110 \nu \ln\Big(\frac{r}{r''_{0}}\Big) \dot{r} v^{2}
 - 440 \nu^2 \ln\Big(\frac{r}{r''_{0}}\Big) \dot{r} v^{2}\right) \,,\\
%%%%%%%%%%%%%%%%%%%%%%%%%%%%%%%%%%%%%%%%%%%%%%%%%%%%%%%%%%%%%%%%%
B^{(3)}_\text{4PN} &= \frac{m^3}{r^3} \left(2
 -  \frac{619267}{525} \nu
 + \frac{791}{16} \pi^2 \nu
 -  \frac{28406}{45} \nu^2
 -  \frac{2201}{32} \pi^2 \nu^2
 + 66 \nu^3
 + 16 \nu^4
 - 128 \ln\Big(\frac{r}{r'_{0}}\Big) \right. \nonumber\\
&\quad\left. + 600 \nu \ln\Big(\frac{r}{r'_{0}}\Big)
 -  \frac{3188}{3} \nu^2 \ln\Big(\frac{r}{r'_{0}}\Big)
 + 128 \ln\Big(\frac{r}{r''_{0}}\Big)
 -  \frac{3284}{3} \nu \ln\Big(\frac{r}{r''_{0}}\Big)
 + \frac{6992}{3} \nu^2 \ln\Big(\frac{r}{r''_{0}}\Big)\right) \dot{r}\,.
\end{align}\end{subequations}
The tail part of the acceleration is presented in Sec.~\ref{sec:tails},
while the dissipative radiation-reaction terms are displayed in
Sec.~\ref{sec:diss} [see Eqs.~\eqref{ABdiss}].

Recall that the 4PN Lagrangian in harmonic coordinates~\cite{BBBFMa, BBBFMb}
is a generalized one, depending on positions, velocities and accelerations
(from the 2PN order). It also contains some logarithms and associated arbitrary gauge
constants entering their arguments. These constants, denoted $r'_A$ (one for
each particle), do not affect physical gauge invariant results. The CM
equations of motion~\eqref{dvdt}--\eqref{B} depend on these constants as well,
through the two combinations $r'_0$ and $r''_0$ defined by
\begin{subequations}\label{r'0r''0}
\begin{align}
\ln r'_0 &= X_1 \ln r'_1 + X_2 \ln r'_2\,,\\
\label{r''0} \ln r''_0 &= \frac{X_1^2 \ln r'_1 - X_2^2 \ln r'_2}{X_1-X_2}\,. 
\end{align} 
\end{subequations}
The combination $r'_0$ represents the one that appears at 3PN order; at the
4PN order, both combinations appear at once. Note that our definition of the constant $r''_0$ in Eq~\eqref{r''0} involves $X_{1}-X_{2}$ in the denominator. However, the equal mass limit $X_{1}=X_{2}$ is always well-defined since in all equations, like~\eqref{A}--\eqref{B}, the logarithm of $r''_0$ is always multiplied by a factor $X_{1}-X_{2}=\pm\sqrt{1-4\nu}$.

The previous CM equations of motion actually derive from a Lagrangian. This
Lagrangian, which is also a generalized one, can be constructed as follows. We
start from the general-frame Lagrangian, which is a functional of $\bm{y}_A$,
$\bm{v}_A$ and $\bm{a}_A$, and admits the CM integral
$\bm{G}[\bm{y}_A, \bm{v}_A]$ explicitly given by
Eqs.~\eqref{G3PN}--\eqref{G4PN} in App.~\ref{sec:Gi}. Then, we perform the
change of variables $(\bm{y}_1, \bm{y}_2)\longrightarrow (\bm{x}, \bm{G})$,
where we recall that $\bm{x} = \bm{y}_1 - \bm{y}_2$. Since to Newtonian order
we have $\bm{G} = m_1\,\bm{y}_1 + m_2\,\bm{y}_2 + \mathcal{O}(c^{-2})$, we
find for instance
$\bm{y}_1 = X_2\bm{x}+\frac{1}{m}\bm{G} + \mathcal{O}(c^{-2})$ and
$\bm{v}_1 = X_2\bm{v}+\frac{1}{m}\frac{\ud \bm{G}}{\ud t} +
\mathcal{O}(c^{-2})$.
Proceeding iteratively with the help of Eqs.~\eqref{G3PN}--\eqref{G4PN}, it is
easy to see that the old variables $\bm{y}_A$ are obtained as functionals of
the new variables $(\bm{x}, \bm{G})$ and their derivatives up to some high
differentiation order depending on the PN order. In the process, we do not
perform any order reduction of accelerations, so that we get
\begin{equation}
\bm{y}_A = \bm{y}_A\bigl[\bm{x}, \bm{v}, \bm{a}, \cdots; \bm{G},
\hbox{$\frac{\ud \bm{G}}{\ud t}$}, \hbox{$\frac{\ud^2 \bm{G}}{\ud t^2}$},
\cdots\bigr]\,.
\end{equation} 
Plugging those relations into the original general-frame Lagrangian and
performing time derivatives, but still without order reduction of the
accelerations, yields an equivalent Lagrangian, which is now ``doubly
generalized'' in terms of the two types of variables, \textit{i.e.}
\begin{equation}\label{Ldouble}
L = L\bigl[\bm{x}, \bm{v}, \bm{a}, \cdots; \bm{G}, \hbox{$\frac{\ud
    \bm{G}}{\ud t}$}, \hbox{$\frac{\ud^2 \bm{G}}{\ud t^2}$}, \cdots\bigr]\,.
\end{equation} 
The ensuing equations of motion read
\begin{equation}\label{eomx}
\frac{\delta L}{\delta \bm{x}} \equiv \frac{\partial L}{\partial \bm{x}}
-\frac{\ud}{\ud t}\left(\frac{\partial L}{\partial \bm{v}}\right) +
\frac{\ud^2}{\ud t^2}\left(\frac{\partial L}{\partial \bm{a}}\right) + \cdots
= \bm{0}\,,
\end{equation} 
together with the equation $\frac{\delta L}{\delta \bm{G}} = 0$,
which is necessarily equivalent to the conservation law for the CM integral, hence we have
\begin{equation}\label{d2G}
\frac{\delta L}{\delta \bm{G}} = \bm{0} \quad\Longleftrightarrow\quad \frac{\ud^2\bm{G}}{\ud t^2} = \bm{0}\,.
\end{equation} 
As a result, we can choose $\bm{G}=0$ as a solution of these equations. The CM
equations of motion are then given by Eq.~\eqref{eomx} in which we pose,
everywhere, $\bm{G}=\bm{0}$, $\frac{\ud\bm{G}}{\ud t} = \bm{0}$, $\cdots$; these equations are nothing but the CM equations of motion~\eqref{dvdt}--\eqref{B}.
Now it is clear, since Eqs.~\eqref{eomx} and~\eqref{d2G} are independent, that
those CM equations of motion derive precisely from the
Lagrangian~\eqref{Ldouble} in which we set, everywhere, $\bm{G}=\bm{0}$,
$\frac{\ud\bm{G}}{\ud t} = \bm{0}$, $\cdots$, hence the CM Lagrangian is
\begin{equation}\label{LdoubleCM}
L_\text{CM} = L\bigl[\bm{x}, \bm{v}, \bm{a}, \cdots; \bm{0}, \bm{0}, \cdots\bigr]\,.
\end{equation} 
At this stage, we follow the standard procedure of subtracting
``multi-zero'' terms and total time derivatives (without performing any shift), which reduces the Lagrangian to one that is linear in accelerations and deprived of
time derivatives of accelerations. Defining as usual the reduced CM Lagrangian as
$\mathcal{L}=L_\text{CM}/\mu$, we explicitly get
\begin{subequations}\label{L}\begin{align}
\mathcal{L}_\text{N} &= \frac{v^2}{2} + \frac{G m}{r}\,,\\
%%%%%%%%%%%%%%%%%%%%%%%%%%%%%%%%%%%%%%%%%%%%%%%%%%%%%%%%%%%%%%%%%
\mathcal{L}_\text{1PN} &= \frac{v^4}{8} - \frac{3\,\nu\,v^4}{8} +
\frac{G m}{r}\,\left( \frac{\dot{r}^2\,\nu}{2} + \frac{3\,v^2}{2} +
\frac{\nu\,v^2}{2} \right)-\frac{G^2m^2}{2\,r^2}\,,\\ 
%%%%%%%%%%%%%%%%%%%%%%%%%%%%%%%%%%%%%%%%%%%%%%%%%%%%%%%%%%%%%%%%%
\mathcal{L}_\text{2PN} &= \frac{v^6}{16} - \frac{7\,\nu\,v^6}{16} +
\frac{13\,\nu^2\,v^6}{16} \nonumber\\ &\quad~ + \frac{G m}{r}\,\left(
\frac{3\,\dot{r}^4\,\nu^2}{8} - \frac{\dot{r}^2\,a_n\,\nu\,r}{8} +
\frac{\dot{r}^2\,\nu\,v^2}{4} - \frac{5\,\dot{r}^2\,\nu^2\,v^2}{4} +
\frac{7\,a_n\,\nu\,r\,v^2}{8} \right.\nonumber\\ &\quad\quad\quad~
+ \left.\frac{7\,v^4}{8} - \frac{5\,\nu\,v^4}{4} -
\frac{9\,\nu^2\,v^4}{8} - \frac{7\,\dot{r}\,\nu\,r\,a_v}{4} \right)
\nonumber\\ &\quad~ +\frac{G^2m^2}{r^2}\,\left( \frac{\dot{r}^2}{2} +
\frac{41\,\dot{r}^2\,\nu}{8} + \frac{3\,\dot{r}^2\,\nu^2}{2} +
\frac{7\,v^2}{4} - \frac{27\,\nu\,v^2}{8} + \frac{\nu^2\,v^2}{2}
\right) \nonumber\\ &\quad~ +\frac{G^3m^3}{r^3}\,\left( \frac{1}{2} +
\frac{15\,\nu}{4} \right)\,,\\ 
%%%%%%%%%%%%%%%%%%%%%%%%%%%%%%%%%%%%%%%%%%%%%%%%%%%%%%%%%%%%%%%%%
\mathcal{L}_\text{3PN} &= \frac{5\,v^8}{128} -
\frac{59\,\nu\,v^8}{128} + \frac{119\,\nu^2\,v^8}{64} -
\frac{323\,\nu^3\,v^8}{128} \nonumber\\ &\quad~ + \frac{G
  m}{r}\,\left( \frac{5\,\dot{r}^6\,\nu^3}{16} +
\frac{\dot{r}^4\,a_n\,\nu\,r}{16} -
\frac{5\,\dot{r}^4\,a_n\,\nu^2\,r}{16} -
\frac{3\,\dot{r}^4\,\nu\,v^2}{16}
\right.\nonumber\\ &\quad\quad\quad~\left.+
\frac{7\,\dot{r}^4\,\nu^2\,v^2}{4} -
\frac{33\,\dot{r}^4\,\nu^3\,v^2}{16} -
\frac{3\,\dot{r}^2\,a_n\,\nu\,r\,v^2}{16} -
\frac{\dot{r}^2\,a_n\,\nu^2\,r\,v^2}{16}
\right.\nonumber\\ &\quad\quad\quad~\left.+
\frac{5\,\dot{r}^2\,\nu\,v^4}{8} - 3\,\dot{r}^2\,\nu^2\,v^4
+\frac{75\,\dot{r}^2\,\nu^3\,v^4}{16} + \frac{7\,a_n\,\nu\,r\,v^4}{8}
\right.\nonumber\\ &\quad\quad\quad~\left.-
\frac{7\,a_n\,\nu^2\,r\,v^4}{2} + \frac{11\,v^6}{16} -
\frac{55\,\nu\,v^6}{16} + \frac{5\,\nu^2\,v^6}{2}
\right.\nonumber\\ &\quad\quad\quad~ +\left.
\frac{65\,\nu^3\,v^6}{16} + \frac{5\,\dot{r}^3\,\nu\,r\,a_v}{12} -
\frac{13\,\dot{r}^3\,\nu^2\,r\,a_v}{8}
\right.\nonumber\\ &\quad\quad\quad~\left.-
\frac{37\,\dot{r}\,\nu\,r\,v^2\,a_v}{8} +
\frac{35\,\dot{r}\,\nu^2\,r\,v^2\,a_v}{4} \right) \nonumber\\ &\quad~
+ \frac{G^2m^2}{r^2}\,\left( -\frac{109\,\dot{r}^4\,\nu}{144} -
\frac{259\,\dot{r}^4\,\nu^2}{36} + 2\,\dot{r}^4\,\nu^3 -
\frac{17\,\dot{r}^2\,a_n\,\nu\,r}{6}
\right.\nonumber\\ &\quad\quad\quad~ +\left.
\frac{97\,\dot{r}^2\,a_n\,\nu^2\,r}{12} +\frac{\dot{r}^2\,v^2}{4} -
\frac{41\,\dot{r}^2\,\nu\,v^2}{6} -
\frac{2287\,\dot{r}^2\,\nu^2\,v^2}{48}
\right.\nonumber\\ &\quad\quad\quad~ -\left.
\frac{27\,\dot{r}^2\,\nu^3\,v^2}{4} + \frac{203\,a_n\,\nu\,r\,v^2}{12}
+ \frac{149\,a_n\,\nu^2\,r\,v^2}{6}
\right.\nonumber\\ &\quad\quad\quad~ +\left. \frac{45\,v^4}{16} +
\frac{53\,\nu\,v^4}{24} + \frac{617\,\nu^2\,v^4}{24} -
\frac{9\,\nu^3\,v^4}{4} \right.\nonumber\\ &\quad\quad\quad~
-\left. \frac{235\,\dot{r}\,\nu\,r\,a_v}{24} +
\frac{235\,\dot{r}\,\nu^2\,r\,a_v}{6} \right) \nonumber\\ &\quad~ +
\frac{G^3m^3}{r^3}\,\left( \frac{3\,\dot{r}^2}{2} -
\frac{12041\,\dot{r}^2\,\nu}{420} + \frac{37\,\dot{r}^2\,\nu^2}{4} +
\frac{7\,\dot{r}^2\,\nu^3}{2} - \frac{123\,\dot{r}^2\,\nu\,{\pi
  }^2}{64} \right.\nonumber\\ &\quad\quad\quad~
+\left. \frac{5\,v^2}{4} + \frac{387\,\nu\,v^2}{70} -
\frac{7\,\nu^2\,v^2}{4} + \frac{\nu^3\,v^2}{2} + \frac{41\,\nu\,{\pi
  }^2\,v^2}{64} \right.\nonumber\\ &\quad\quad\quad~ \left. +
22\,\dot{r}^2\,\nu\,\ln \Big(\frac{r}{r'_0}\Big) -
\frac{22\,\nu\,v^2}{3}\ln \Big(\frac{r}{r'_0}\Big)
\right)\nonumber\\ &\quad~ + \frac{G^4m^4}{r^4}\,\left( -\frac{3}{8}-
\frac{18469\,\nu}{840} + \frac{22\,\nu}{3}\ln \Big(\frac{r}{r'_0}\Big)
\right) \,,\\
%%%%%%%%%%%%%%%%%%%%%%%%%%%%%%%%%%%%%%%%%%%%%%%%%%%%%%%%%%%%%%%%%
\mathcal{L}^{(0)}_\text{4PN} &=\frac{7}{256} v^{10}
 -  \frac{121}{256} \nu v^{10}
 + \frac{785}{256} \nu^2 v^{10}
 -  \frac{1127}{128} \nu^3 v^{10}
 + \frac{2415}{256} \nu^4 v^{10} \,,\\
%%%%%%%%%%%%%%%%%%%%%%%%%%%%%%%%%%%%%%%%%%%%%%%%%%%%%%%%%%%%%%%%%
\mathcal{L}^{(1)}_\text{4PN} &= \frac{m}{r} 
\left(\frac{23}{20} \nu^2 a_{v} r \dot{r}^5
 -  \frac{5}{128} \nu \dot{r}^8
 + \frac{35}{128} \nu^2 \dot{r}^8
 -  \frac{35}{64} \nu^3 \dot{r}^8
 + \frac{35}{128} \nu^4 \dot{r}^8
 + \frac{7}{4} \nu a_{v} r \dot{r}^3 v^{2}\right.\nonumber\\
& \quad + \frac{361}{24} \nu^3 a_{v} r \dot{r}^3 v^{2}
 + \frac{19}{32} \nu a_{n} r \dot{r}^4 v^{2}
 + \frac{85}{32} \nu^3 a_{n} r \dot{r}^4 v^{2}
 -  \frac{5}{16} \nu \dot{r}^6 v^{2}
 -  \frac{31}{16} \nu^2 \dot{r}^6 v^{2}\nonumber\\
&\quad + \frac{45}{32} \nu^3 \dot{r}^6 v^{2}
 -  \frac{85}{32} \nu^4 \dot{r}^6 v^{2}
 + \frac{341}{4} \nu^2 a_{v} r \dot{r} v^{4}
 + \frac{245}{32} \nu^2 a_{n} r \dot{r}^2 v^{4}
 -  \frac{17}{64} \nu \dot{r}^4 v^{4}
 -  \frac{11}{8} \nu^2 \dot{r}^4 v^{4}\nonumber\\
&\quad -  \frac{193}{16} \nu^3 \dot{r}^4 v^{4}
 + \frac{693}{64} \nu^4 \dot{r}^4 v^{4}
 + \frac{217}{96} \nu a_{n} r v^{6}
 + \frac{2261}{96} \nu^3 a_{n} r v^{6}
 -  \frac{11}{48} \nu \dot{r}^2 v^{6}
 -  \frac{29}{4} \nu^2 \dot{r}^2 v^{6}\nonumber\\
&\quad + \frac{1021}{96} \nu^3 \dot{r}^2 v^{6}
 -  \frac{665}{32} \nu^4 \dot{r}^2 v^{6}
 + \frac{75}{128} v^{8}
 -  \frac{1595}{384} \nu v^{8}
 + \frac{2917}{128} \nu^2 v^{8}
 + \frac{493}{192} \nu^3 v^{8}\nonumber\\
&\quad\left. -  \frac{2261}{128} \nu^4 v^{8}\right) \,,\\
%%%%%%%%%%%%%%%%%%%%%%%%%%%%%%%%%%%%%%%%%%%%%%%%%%%%%%%%%%%%%%%%%
\mathcal{L}^{(2)}_\text{4PN} &= \frac{m^2}{r^2} 
\left(\frac{5407}{288} \nu a_{v} r \dot{r}^3
 -  \frac{5531}{3200} \nu \dot{r}^6
 -  \frac{487}{40} \nu^2 \dot{r}^6
 + \frac{13}{5} \nu^3 \dot{r}^6
 + \frac{3}{2} \nu^4 \dot{r}^6
 + \frac{11497}{48} \nu^2 a_{v} r \dot{r} v^{2}\right.\nonumber\\
&\quad + \frac{469}{4} \nu^3 a_{v} r \dot{r} v^{2}
 + \frac{517}{48} \nu^2 a_{n} r \dot{r}^2 v^{2}
 -  \frac{61183}{5760} \nu \dot{r}^4 v^{2}
 + \frac{3079}{96} \nu^2 \dot{r}^4 v^{2}
 -  \frac{161}{8} \nu^3 \dot{r}^4 v^{2}\nonumber\\
&\quad -  \frac{27}{2} \nu^4 \dot{r}^4 v^{2}
 + \frac{14627}{384} \nu a_{n} r v^{4}
 + \frac{3}{16} \dot{r}^2 v^{4}
 -  \frac{12091}{640} \nu \dot{r}^2 v^{4}
 -  \frac{22045}{192} \nu^2 \dot{r}^2 v^{4}
 + \frac{255}{64} \nu^3 \dot{r}^2 v^{4}\nonumber\\
&\quad\left. + \frac{569}{16} \nu^4 \dot{r}^2 v^{4}
 + \frac{115}{32} v^{6}
 + \frac{593}{120} \nu v^{6}
 + \frac{9467}{192} \nu^2 v^{6}
 + \frac{599}{64} \nu^3 v^{6}
 + \frac{195}{16} \nu^4 v^{6}\right)
 \,,\\
%%%%%%%%%%%%%%%%%%%%%%%%%%%%%%%%%%%%%%%%%%%%%%%%%%%%%%%%%%%%%%%%%
\mathcal{L}^{(3)}_\text{4PN} &= \frac{m^3}{r^3} \left(\frac{4937}{1260} 
\nu^2 a_{v} r \dot{r}
 -  \frac{41}{32} \pi^2 \nu^2 a_{v} r \dot{r}
 + \frac{44}{3} \nu^2 a_{v} r \ln\Big(\frac{r}{r'_{0}}\Big) \dot{r}
 + \frac{22}{3} \nu a_{v} r \ln\Big(\frac{r}{r''_{0}}\Big) \dot{r}
 -  \frac{246373}{2240} \nu \dot{r}^4\right.\nonumber\\
&\quad -  \frac{2155}{1024} \pi^2 \nu \dot{r}^4
 -  \frac{210733}{2016} \nu^2 \dot{r}^4
 + \frac{205}{128} \pi^2 \nu^2 \dot{r}^4
 + \frac{367}{32} \nu^3 \dot{r}^4
 + \frac{29}{4} \nu^4 \dot{r}^4
 -  \frac{55}{3} \nu^2 \ln\Big(\frac{r}{r'_{0}}\Big) \dot{r}^4\nonumber\\
&\quad -  \frac{55}{3} \nu \ln\Big(\frac{r}{r''_{0}}\Big) \dot{r}^4
 + \frac{220}{3} \nu^2 \ln\Big(\frac{r}{r''_{0}}\Big) \dot{r}^4
 + \frac{229319}{6300} \nu a_{n} r v^{2}
 -  \frac{21}{32} \pi^2 \nu a_{n} r v^{2}
 + \frac{49}{4} \nu^3 a_{n} r v^{2}\nonumber\\
&\quad + 44 \nu a_{n} r \ln\Big(\frac{r}{r'_{0}}\Big) v^{2}
 + \frac{44}{3} \nu^2 a_{n} r \ln\Big(\frac{r}{r''_{0}}\Big) v^{2}
 + \frac{7}{4} \dot{r}^2 v^{2}
 + \frac{516319}{4200} \nu \dot{r}^2 v^{2}
 + \frac{447}{512} \pi^2 \nu \dot{r}^2 v^{2}\nonumber\\
&\quad + \frac{53099}{560} \nu^2 \dot{r}^2 v^{2}
 + \frac{123}{64} \pi^2 \nu^2 \dot{r}^2 v^{2}
 -  \frac{1003}{16} \nu^3 \dot{r}^2 v^{2}
 -  \frac{47}{2} \nu^4 \dot{r}^2 v^{2}
 - 55 \nu \ln\Big(\frac{r}{r'_{0}}\Big) \dot{r}^2 v^{2}\nonumber\\
&\quad - 22 \nu^2 \ln\Big(\frac{r}{r'_{0}}\Big) \dot{r}^2 v^{2}
 + 11 \nu \ln\Big(\frac{r}{r''_{0}}\Big) \dot{r}^2 v^{2}
 - 88 \nu^2 \ln\Big(\frac{r}{r''_{0}}\Big) \dot{r}^2 v^{2}
 + \frac{91}{16} v^{4}
 -  \frac{166703}{20160} \nu v^{4}\nonumber\\
&\quad + \frac{133}{1024} \pi^2 \nu v^{4}
 + \frac{10601}{3360} \nu^2 v^{4}
 -  \frac{123}{128} \pi^2 \nu^2 v^{4}
 + \frac{567}{32} \nu^3 v^{4}
 -  \frac{15}{4} \nu^4 v^{4}
 + \frac{55}{3} \nu \ln\Big(\frac{r}{r'_{0}}\Big) v^{4}\nonumber\\
&\quad\left. + 11 \nu^2 \ln\Big(\frac{r}{r'_{0}}\Big) v^{4}
 + \frac{44}{3} \nu^2 \ln\Big(\frac{r}{r''_{0}}\Big) v^{4}\right) \,,\\
%%%%%%%%%%%%%%%%%%%%%%%%%%%%%%%%%%%%%%%%%%%%%%%%%%%%%%%%%%%%%%%%%
\mathcal{L}^{(4)}_\text{4PN} &=\frac{m^4}{r^4} \left(\frac{9}{4} \dot{r}^2
 -  \frac{245971}{4200} \nu \dot{r}^2
 + \frac{2771}{96} \pi^2 \nu \dot{r}^2
 -  \frac{8089}{140} \nu^2 \dot{r}^2
 - 44 \pi^2 \nu^2 \dot{r}^2
 + \frac{185}{8} \nu^3 \dot{r}^2
 + \frac{15}{2} \nu^4 \dot{r}^2\right.\nonumber\\
&\quad - 64 \ln\Big(\frac{r}{r'_{0}}\Big) \dot{r}^2
 + \frac{482}{3} \nu \ln\Big(\frac{r}{r'_{0}}\Big) \dot{r}^2
 - 436 \nu^2 \ln\Big(\frac{r}{r'_{0}}\Big) \dot{r}^2
 + 64 \ln\Big(\frac{r}{r''_{0}}\Big) \dot{r}^2
 -  \frac{1499}{3} \nu \ln\Big(\frac{r}{r''_{0}}\Big) \dot{r}^2\nonumber\\
&\quad + \frac{2924}{3} \nu^2 \ln\Big(\frac{r}{r''_{0}}\Big) \dot{r}^2
 + \frac{15}{16} v^{2}
 + \frac{2039993}{50400} \nu v^{2}
 -  \frac{191}{32} \pi^2 \nu v^{2}
 -  \frac{52907}{1008} \nu^2 v^{2}
 + 11 \pi^2 \nu^2 v^{2}\nonumber\\
&\quad -  \frac{1}{8} \nu^3 v^{2}
 + \frac{1}{2} \nu^4 v^{2}
 + 16 \ln\Big(\frac{r}{r'_{0}}\Big) v^{2}
 -  \frac{71}{3} \nu \ln\Big(\frac{r}{r'_{0}}\Big) v^{2}
 + \frac{349}{3} \nu^2 \ln\Big(\frac{r}{r'_{0}}\Big) v^{2}
 - 16 \ln\Big(\frac{r}{r''_{0}}\Big) v^{2}\nonumber\\
&\quad\left. + 124 \nu \ln\Big(\frac{r}{r''_{0}}\Big) v^{2}
 - 240 \nu^2 \ln\Big(\frac{r}{r''_{0}}\Big) v^{2}\right)
 \,,\\
%%%%%%%%%%%%%%%%%%%%%%%%%%%%%%%%%%%%%%%%%%%%%%%%%%%%%%%%%%%%%%%%%
\mathcal{L}^{(5)}_\text{4PN} &=\frac{m^5}{r^5} \left(\frac{3}{8}
 + \frac{1697177}{25200} \nu
 + \frac{105}{32} \pi^2 \nu
 + \frac{55111}{720} \nu^2
 - 11 \pi^2 \nu^2
 - 16 \ln\Big(\frac{r}{r'_{0}}\Big)
 + \frac{82}{3} \nu \ln\Big(\frac{r}{r'_{0}}\Big)\right.\nonumber\\
&\quad\left. - 120 \nu^2 \ln\Big(\frac{r}{r'_{0}}\Big)
 + 16 \ln\Big(\frac{r}{r''_{0}}\Big)
 - 124 \nu \ln\Big(\frac{r}{r''_{0}}\Big)
 + 240 \nu^2 \ln\Big(\frac{r}{r''_{0}}\Big)\right)
 \,.
\end{align}\end{subequations}
Up to 3PN order, we recover the known result~\cite{BI03CM}. The CM Lagrangian in harmonic coordinates still depends on accelerations
starting at 2PN order, through $a_n=(an)=\bm{a}\cdot\bm{n}$ and
$a_v=(av)=\bm{a}\cdot\bm{v}$, as well as logarithms of $r/r'_0$ and $r/r''_0$.
After applying the contact transformation or shift of Ref.~\cite{BBBFMa} (when reduced
to the CM frame) the previous Lagrangian will be transformed into an
\textit{ordinary}, non harmonic Lagrangian, depending only on positions and velocities, and,
furthermore, the logarithmic terms $\ln(r/r'_0)$ and $\ln(r/r''_0)$ therein
will disappear (see Sec.~V B of~\cite{BBBFMa} for more details).

\section{Center-of-mass energy and angular momentum} 
\label{sec:cons} 

The conserved energy and angular momentum may be split into instantaneous,
tail and dissipative radiation reaction parts; following our conventions, we
first give the instantaneous contributions, before discussing the interesting 4PN
tails in Sec.~\ref{sec:tails} and the dissipative terms in Sec.\ref{sec:diss}.
The reduced CM energy is defined by $\mathcal{E}=E/\mu$ and the reduced CM angular
momentum by $\mathcal{J}=J/J_\text{N}$, which is the Euclidean norm
$J=\vert\bm{J}\vert$ rescaled by that of the Newtonian angular momentum
$\bm{J}_\text{N}=\mu\,\bm{x}\times\bm{v}$. We have
\begin{subequations}\label{E}\begin{align}
\mathcal{E}_\text{N} &= \frac{v^2}{2}-\frac{G m}{r}\,,\\ 
%%%%%%%%%%%%%%%%%%%%%%%%%%%%%%%%%%%%%%%%%%%%%%%%%%%%%%%%%%%%%%%%%
\mathcal{E}_\text{1PN} &= \frac{3\,v^4}{8} - \frac{9\,\nu\,v^4}{8} +
\frac{G m}{r}\,\left( \frac{\dot{r}^2\,\nu}{2} + \frac{3\,v^2}{2} +
\frac{\nu\,v^2}{2} \right)+\frac{G^2m^2}{2r^2}\,,\\ 
%%%%%%%%%%%%%%%%%%%%%%%%%%%%%%%%%%%%%%%%%%%%%%%%%%%%%%%%%%%%%%%%%
\mathcal{E}_\text{2PN} &= \frac{5\,v^6}{16} - \frac{35\,\nu\,v^6}{16} +
\frac{65\,\nu^2\,v^6}{16} \nonumber\\ &\quad~ + \frac{G m}{r}\,\left(
-\frac{3\,\dot{r}^4\,\nu}{8} + \frac{9\,\dot{r}^4\,\nu^2}{8} +
\frac{\dot{r}^2\,\nu\,v^2}{4} - \frac{15\,\dot{r}^2\,\nu^2\,v^2}{4} +
\frac{21\,v^4}{8} - \frac{23\,\nu\,v^4}{8} - \frac{27\,\nu^2\,v^4}{8}
\right)\nonumber\\ &\quad~ +\frac{G^2m^2}{r^2}\,\left(
\frac{\dot{r}^2}{2} + \frac{69\,\dot{r}^2\,\nu}{8} +
\frac{3\,\dot{r}^2\,\nu^2}{2} + \frac{7\,v^2}{4} -
\frac{55\,\nu\,v^2}{8} + \frac{\nu^2\,v^2}{2} \right) \nonumber\\
&\quad~+\frac{G^3m^3}{r^3}\,\left( -\frac{1}{2}- \frac{15\,\nu}{4}
\right)\,,\\ 
%%%%%%%%%%%%%%%%%%%%%%%%%%%%%%%%%%%%%%%%%%%%%%%%%%%%%%%%%%%%%%%%%
\mathcal{E}_\text{3PN} &= \frac{35\,v^8}{128} -
\frac{413\,\nu\,v^8}{128} + \frac{833\,\nu^2\,v^8}{64} -
\frac{2261\,\nu^3\,v^8}{128} \nonumber\\ &\quad~ + \frac{G
  m}{r}\,\left( \frac{5\,\dot{r}^6\,\nu}{16} -
\frac{25\,\dot{r}^6\,\nu^2}{16} + \frac{25\,\dot{r}^6\,\nu^3}{16} -
\frac{9\,\dot{r}^4\,\nu\,v^2}{16} +
\frac{21\,\dot{r}^4\,\nu^2\,v^2}{4}
\right.\nonumber\\ &\quad\quad\quad~
\left. -\frac{165\,\dot{r}^4\,\nu^3\,v^2}{16} -
\frac{21\,\dot{r}^2\,\nu\,v^4}{16} -
\frac{75\,\dot{r}^2\,\nu^2\,v^4}{16} +
\frac{375\,\dot{r}^2\,\nu^3\,v^4}{16}
\right.\nonumber\\ &\quad\quad\quad~ \left. + \frac{55\,v^6}{16} -
\frac{215\,\nu\,v^6}{16} + \frac{29\,\nu^2\,v^6}{4} +
\frac{325\,\nu^3\,v^6}{16} \right) \nonumber\\ &\quad~ +
\frac{G^2m^2}{r^2}\,\left( -\frac{731\,\dot{r}^4\,\nu}{48} +
\frac{41\,\dot{r}^4\,\nu^2}{4} + 6\,\dot{r}^4\,\nu^3 +
\frac{3\,\dot{r}^2\,v^2}{4} + \frac{31\,\dot{r}^2\,\nu\,v^2}{2}
\right.\nonumber\\ &\quad\quad\quad~
\left. -\frac{815\,\dot{r}^2\,\nu^2\,v^2}{16} -
\frac{81\,\dot{r}^2\,\nu^3\,v^2}{4} + \frac{135\,v^4}{16} -
\frac{97\,\nu\,v^4}{8} + \frac{203\,\nu^2\,v^4}{8} -
\frac{27\,\nu^3\,v^4}{4} \right) \nonumber\\ &\quad~ +
\frac{G^3m^3}{r^3}\,\left( \frac{3\,\dot{r}^2}{2} +
\frac{803\,\dot{r}^2\,\nu}{840} + \frac{51\,\dot{r}^2\,\nu^2}{4} +
\frac{7\,\dot{r}^2\,\nu^3}{2} - \frac{123\,\dot{r}^2\,\nu\,{\pi
  }^2}{64} + \frac{5\,v^2}{4} \right.\nonumber\\ &\quad\quad\quad~
\left. -\frac{6747\,\nu\,v^2}{280} - \frac{21\,\nu^2\,v^2}{4} +
\frac{\nu^3\,v^2}{2} + \frac{41\,\nu\,\pi^2\,v^2}{64}
\right.\nonumber\\ &\quad\quad\quad~ \left. +
22\,\dot{r}^2\,\nu\,\ln \Big(\frac{r}{r'_0}\Big) -
\frac{22\,\nu\,v^2\,}{3}\ln \Big(\frac{r}{r'_0}\Big)
\right)\nonumber\\ &\qquad~ + \frac{G^4m^4}{r^4}\,\left( \frac{3}{8} +
\frac{18469\,\nu}{840} - \frac{22\,\nu}{3}\,\ln
\Big(\frac{r}{r'_0}\Big) \right) \,,\\
%%%%%%%%%%%%%%%%%%%%%%%%%%%%%%%%%%%%%%%%%%%%%%%%%%%%%%%%%%%%%%%%%
\mathcal{E}^{(0)}_\text{4PN} &= \left(\frac{63}{256}
 -  \frac{1089}{256} \nu
 + \frac{7065}{256} \nu^2
 -  \frac{10143}{128} \nu^3
 + \frac{21735}{256} \nu^4\right) v^{10}
 \,,\\
%%%%%%%%%%%%%%%%%%%%%%%%%%%%%%%%%%%%%%%%%%%%%%%%%%%%%%%%%%%%%%%%%
\mathcal{E}^{(1)}_\text{4PN} &= \frac{m}{r} \left(- \frac{35}{128} \nu \dot{r}^8
 + \frac{245}{128} \nu^2 \dot{r}^8
 -  \frac{245}{64} \nu^3 \dot{r}^8
 + \frac{245}{128} \nu^4 \dot{r}^8
 + \frac{25}{32} \nu \dot{r}^6 v^{2}
 -  \frac{125}{16} \nu^2 \dot{r}^6 v^{2}\right.\nonumber\\
&\quad + \frac{185}{8} \nu^3 \dot{r}^6 v^{2}
 -  \frac{595}{32} \nu^4 \dot{r}^6 v^{2}
 + \frac{27}{64} \nu \dot{r}^4 v^{4}
 + \frac{243}{32} \nu^2 \dot{r}^4 v^{4}
 -  \frac{1683}{32} \nu^3 \dot{r}^4 v^{4}
 + \frac{4851}{64} \nu^4 \dot{r}^4 v^{4}\nonumber\\
&\quad -  \frac{147}{32} \nu \dot{r}^2 v^{6}
 + \frac{369}{32} \nu^2 \dot{r}^2 v^{6}
 + \frac{423}{8} \nu^3 \dot{r}^2 v^{6}
 -  \frac{4655}{32} \nu^4 \dot{r}^2 v^{6}
 + \frac{525}{128} v^{8}
 -  \frac{4011}{128} \nu v^{8}\nonumber\\
&\quad\left. + \frac{9507}{128} \nu^2 v^{8}
 -  \frac{357}{64} \nu^3 v^{8}
 -  \frac{15827}{128} \nu^4 v^{8}\right) \,,\\
%%%%%%%%%%%%%%%%%%%%%%%%%%%%%%%%%%%%%%%%%%%%%%%%%%%%%%%%%%%%%%%%%
\mathcal{E}^{(2)}_\text{4PN} &= \frac{m^2}{r^2} \left(- \frac{4771}{640} 
\nu \dot{r}^6
 -  \frac{461}{8} \nu^2 \dot{r}^6
 -  \frac{17}{2} \nu^3 \dot{r}^6
 + \frac{15}{2} \nu^4 \dot{r}^6
 + \frac{5347}{384} \nu \dot{r}^4 v^{2}
 + \frac{19465}{96} \nu^2 \dot{r}^4 v^{2}\right.\nonumber\\
&\quad -  \frac{439}{8} \nu^3 \dot{r}^4 v^{2}
 -  \frac{135}{2} \nu^4 \dot{r}^4 v^{2}
 + \frac{15}{16} \dot{r}^2 v^{4}
 -  \frac{5893}{128} \nu \dot{r}^2 v^{4}
 -  \frac{12995}{64} \nu^2 \dot{r}^2 v^{4}
 + \frac{18511}{64} \nu^3 \dot{r}^2 v^{4}\nonumber\\
&\left.\quad + \frac{2845}{16} \nu^4 \dot{r}^2 v^{4}
 + \frac{575}{32} v^{6}
 -  \frac{4489}{128} \nu v^{6}
 + \frac{5129}{64} \nu^2 v^{6}
 -  \frac{8289}{64} \nu^3 v^{6}
 + \frac{975}{16} \nu^4 v^{6}\right)
 \,,\\
%%%%%%%%%%%%%%%%%%%%%%%%%%%%%%%%%%%%%%%%%%%%%%%%%%%%%%%%%%%%%%%%%
\mathcal{E}^{(3)}_\text{4PN} &= \frac{m^3}{r^3} 
\left(- \frac{2599207}{6720} \nu \dot{r}^4
 -  \frac{6465}{1024} \pi^2 \nu \dot{r}^4
 -  \frac{103205}{224} \nu^2 \dot{r}^4
 + \frac{615}{128} \pi^2 \nu^2 \dot{r}^4
 + \frac{69}{32} \nu^3 \dot{r}^4
 + \frac{87}{4} \nu^4 \dot{r}^4\right.\nonumber\\
&\quad - 55 \nu^2 \ln\Big(\frac{r}{r'_{0}}\Big) \dot{r}^4
 - 55 \nu \ln\Big(\frac{r}{r''_{0}}\Big) \dot{r}^4
 + 220 \nu^2 \ln\Big(\frac{r}{r''_{0}}\Big) \dot{r}^4
 + \frac{21}{4} \dot{r}^2 v^{2}
 + \frac{1086923}{1680} \nu \dot{r}^2 v^{2}\nonumber\\
&\quad + \frac{333}{512} \pi^2 \nu \dot{r}^2 v^{2}
 + \frac{206013}{560} \nu^2 \dot{r}^2 v^{2}
 + \frac{123}{64} \pi^2 \nu^2 \dot{r}^2 v^{2}
 -  \frac{2437}{16} \nu^3 \dot{r}^2 v^{2}
 -  \frac{141}{2} \nu^4 \dot{r}^2 v^{2}\nonumber\\
&\quad - 33 \nu \ln\Big(\frac{r}{r'_{0}}\Big) \dot{r}^2 v^{2}
 - 22 \nu^2 \ln\Big(\frac{r}{r'_{0}}\Big) \dot{r}^2 v^{2}
 + 55 \nu \ln\Big(\frac{r}{r''_{0}}\Big) \dot{r}^2 v^{2}
 - 220 \nu^2 \ln\Big(\frac{r}{r''_{0}}\Big) \dot{r}^2 v^{2}
 + \frac{273}{16} v^{4}\nonumber\\
&\quad -  \frac{22649399}{100800} \nu v^{4}
 + \frac{1071}{1024} \pi^2 \nu v^{4}
 + \frac{521063}{10080} \nu^2 v^{4}
 -  \frac{205}{128} \pi^2 \nu^2 v^{4}
 + \frac{2373}{32} \nu^3 v^{4}
 -  \frac{45}{4} \nu^4 v^{4}\nonumber\\
&\quad \left. + 11 \nu \ln\Big(\frac{r}{r'_{0}}\Big) v^{4}
 + \frac{55}{3} \nu^2 \ln\Big(\frac{r}{r'_{0}}\Big) v^{4}
 -  \frac{22}{3} \nu \ln\Big(\frac{r}{r''_{0}}\Big) v^{4}
 + \frac{88}{3} \nu^2 \ln\Big(\frac{r}{r''_{0}}\Big) v^{4}\right) \,,\\
%%%%%%%%%%%%%%%%%%%%%%%%%%%%%%%%%%%%%%%%%%%%%%%%%%%%%%%%%%%%%%%%%
\mathcal{E}^{(4)}_\text{4PN} &= \frac{m^4}{r^4} \left(\frac{9}{4} \dot{r}^2
 -  \frac{1622437}{12600} \nu \dot{r}^2
 + \frac{2645}{96} \pi^2 \nu \dot{r}^2
 -  \frac{289351}{2520} \nu^2 \dot{r}^2
 -  \frac{1367}{32} \pi^2 \nu^2 \dot{r}^2
 + \frac{213}{8} \nu^3 \dot{r}^2\right.\nonumber\\
&\quad + \frac{15}{2} \nu^4 \dot{r}^2
 - 64 \ln\Big(\frac{r}{r'_{0}}\Big) \dot{r}^2
 + \frac{746}{3} \nu \ln\Big(\frac{r}{r'_{0}}\Big) \dot{r}^2
 -  \frac{1352}{3} \nu^2 \ln\Big(\frac{r}{r'_{0}}\Big) \dot{r}^2
 + 64 \ln\Big(\frac{r}{r''_{0}}\Big) \dot{r}^2\nonumber\\
&\quad - 507 \nu \ln\Big(\frac{r}{r''_{0}}\Big) \dot{r}^2
 + 1004 \nu^2 \ln\Big(\frac{r}{r''_{0}}\Big) \dot{r}^2
 + \frac{15}{16} v^{2}
 + \frac{1859363}{16800} \nu v^{2}
 -  \frac{149}{32} \pi^2 \nu v^{2}\nonumber\\
&\quad + \frac{22963}{5040} \nu^2 v^{2}
 + \frac{311}{32} \pi^2 \nu^2 v^{2}
 -  \frac{29}{8} \nu^3 v^{2}
 + \frac{1}{2} \nu^4 v^{2}
 + 16 \ln\Big(\frac{r}{r'_{0}}\Big) v^{2}
 -  \frac{335}{3} \nu \ln\Big(\frac{r}{r'_{0}}\Big) v^{2}\nonumber\\
&\quad\left. + 131 \nu^2 \ln\Big(\frac{r}{r'_{0}}\Big) v^{2}
 - 16 \ln\Big(\frac{r}{r''_{0}}\Big) v^{2}
 + \frac{394}{3} \nu \ln\Big(\frac{r}{r''_{0}}\Big) v^{2}
 -  \frac{808}{3} \nu^2 \ln\Big(\frac{r}{r''_{0}}\Big) v^{2}\right) \,,\\
%%%%%%%%%%%%%%%%%%%%%%%%%%%%%%%%%%%%%%%%%%%%%%%%%%%%%%%%%%%%%%%%%
\mathcal{E}^{(5)}_\text{4PN} &= \frac{m^5}{r^5} \left(- \frac{3}{8}
 -  \frac{1697177}{25200} \nu
 -  \frac{105}{32} \pi^2 \nu
 -  \frac{55111}{720} \nu^2
 + 11 \pi^2 \nu^2
 + 16 \ln\Big(\frac{r}{r'_{0}}\Big)
 -  \frac{82}{3} \nu \ln\Big(\frac{r}{r'_{0}}\Big)\right.\nonumber\\
&\left.\quad + 120 \nu^2 \ln\Big(\frac{r}{r'_{0}}\Big)
 - 16 \ln\Big(\frac{r}{r''_{0}}\Big)
 + 124 \nu \ln\Big(\frac{r}{r''_{0}}\Big)
 - 240 \nu^2 \ln\Big(\frac{r}{r''_{0}}\Big)\right)
 \,,
\end{align}\end{subequations}
and
\begin{subequations}\label{J}\begin{align}
\mathcal{J}_\text{N} &= 1\,,\\
%%%%%%%%%%%%%%%%%%%%%%%%%%%%%%%%%%%%%%%%%%%%%%%%%%%%%%%%%%%%%%%%%
\mathcal{J}_\text{1PN} &= \left(1 - 3\, \nu \right)\frac{v^2}{2} +
\frac{G m}{r}\,\left( 3 + \nu \right)\,,\\ 
%%%%%%%%%%%%%%%%%%%%%%%%%%%%%%%%%%%%%%%%%%%%%%%%%%%%%%%%%%%%%%%%%
\mathcal{J}_\text{2PN} &= \frac{3\,v^4}{8} - \frac{21\,\nu\,v^4}{8} +
\frac{39\,\nu^2\,v^4}{8} \nonumber\\ &\quad~ + \frac{G m}{r}\,\left(
-\dot{r}^2\,\nu - \frac{5\,\dot{r}^2\,\nu^2}{2} + \frac{7\,v^2}{2} -
5\,\nu\,v^2 - \frac{9\,\nu^2\,v^2}{2} \right)\nonumber\\ &\quad~
+\frac{G^2m^2}{r^2}\,\left( \frac{7}{2} - \frac{41\,\nu}{4} + \nu^2
\right)\,,\\ 
%%%%%%%%%%%%%%%%%%%%%%%%%%%%%%%%%%%%%%%%%%%%%%%%%%%%%%%%%%%%%%%%%
%\mathcal{J}_\text{2.5PN} &=
%-\frac{8\,G^3m^3\,\dot{r}\,\nu^2}{5\,r^2}\,,\\
%%%%%%%%%%%%%%%%%%%%%%%%%%%%%%%%%%%%%%%%%%%%%%%%%%%%%%%%%%%%%%%%%
\mathcal{J}_\text{3PN} &= \frac{5\,v^6}{16} -
\frac{59\,\nu\,v^6}{16} + \frac{119\,\nu^2\,v^6}{8} -
\frac{323\,\nu^3\,v^6}{16} \nonumber\\ &\quad~ + \frac{G m}{r}\,\left(
\frac{3\,\dot{r}^4\,\nu}{4} - \frac{3\,\dot{r}^4\,\nu^2}{4}
-\frac{33\,\dot{r}^4\,\nu^3}{8} - 3\,\dot{r}^2\,\nu\,v^2 +
\frac{7\,\dot{r}^2\,\nu^2\,v^2}{4} \right.\nonumber\\
&\quad\quad\quad~ +\left.  \frac{75\,\dot{r}^2\,\nu^3\,v^2}{4} +
\frac{33\,v^4}{8} - \frac{71\,\nu\,v^4}{4} + \frac{53\,\nu^2\,v^4}{4}
+ \frac{195\,\nu^3\,v^4}{8} \right) \nonumber\\ &\quad~ +
\frac{G^2m^2}{r^2}\,\left( \frac{\dot{r}^2}{2} -
\frac{287\,\dot{r}^2\,\nu}{24} - \frac{317\,\dot{r}^2\,\nu^2}{8} -
\frac{27\,\dot{r}^2\,\nu^3}{2} + \frac{45\,v^2}{4} \right.\nonumber\\
&\quad\quad\quad~ -\left.  \frac{161\,\nu\,v^2}{6} +
\frac{105\,\nu^2\,v^2}{4} - 9\,\nu^3\,v^2 \right) \nonumber\\
&\quad~ + \frac{G^3m^3}{r^3}\,\left( \frac{5}{2} -
\frac{5199\,\nu}{280} - 7\,\nu^2 + \nu^3 + \frac{41\,\nu\,{\pi
}^2}{32} - \frac{44\,\nu}{3}\ln \Big(\frac{r}{r'_0}\Big)
\right)\,,\\ 
%%%%%%%%%%%%%%%%%%%%%%%%%%%%%%%%%%%%%%%%%%%%%%%%%%%%%%%%%%%%%%%%%
\mathcal{J}^{(0)}_\text{4PN} &= \left(\frac{35}{128}
 -  \frac{605}{128} \nu
 + \frac{3925}{128} \nu^2
 -  \frac{5635}{64} \nu^3
 + \frac{12075}{128} \nu^4\right) v^{8} \,,\\
%%%%%%%%%%%%%%%%%%%%%%%%%%%%%%%%%%%%%%%%%%%%%%%%%%%%%%%%%%%%%%%%%
\mathcal{J}^{(1)}_\text{4PN} &= \frac{m}{r} \left(- \frac{5}{8} \nu \dot{r}^6
 + \frac{15}{8} \nu^2 \dot{r}^6
 + \frac{45}{16} \nu^3 \dot{r}^6
 -  \frac{85}{16} \nu^4 \dot{r}^6
 + 3 \nu \dot{r}^4 v^{2}
 -  \frac{45}{4} \nu^2 \dot{r}^4 v^{2}
 -  \frac{135}{16} \nu^3 \dot{r}^4 v^{2}\right.\nonumber\\
&\quad + \frac{693}{16} \nu^4 \dot{r}^4 v^{2}
 -  \frac{53}{8} \nu \dot{r}^2 v^{4}
 + \frac{423}{16} \nu^2 \dot{r}^2 v^{4}
 + \frac{299}{16} \nu^3 \dot{r}^2 v^{4}
 -  \frac{1995}{16} \nu^4 \dot{r}^2 v^{4}
 + \frac{75}{16} v^{6}\nonumber\\
&\quad\left. -  \frac{151}{4} \nu v^{6}
 + \frac{1553}{16} \nu^2 v^{6}
 -  \frac{425}{16} \nu^3 v^{6}
 -  \frac{2261}{16} \nu^4 v^{6}\right)
 \,,\\
%%%%%%%%%%%%%%%%%%%%%%%%%%%%%%%%%%%%%%%%%%%%%%%%%%%%%%%%%%%%%%%%%
\mathcal{J}^{(2)}_\text{4PN} &= \frac{m^2}{r^2} \left(\frac{14773}{320} \nu \dot{r}^4
 + \frac{3235}{48} \nu^2 \dot{r}^4
 -  \frac{155}{4} \nu^3 \dot{r}^4
 - 27 \nu^4 \dot{r}^4
 + \frac{3}{4} \dot{r}^2 v^{2}
 -  \frac{5551}{60} \nu \dot{r}^2 v^{2}\right.\nonumber\\
&\quad -  \frac{256}{3} \nu^2 \dot{r}^2 v^{2}
 + \frac{4459}{16} \nu^3 \dot{r}^2 v^{2}
 + \frac{569}{4} \nu^4 \dot{r}^2 v^{2}
 + \frac{345}{16} v^{4}
 -  \frac{65491}{960} \nu v^{4}
 + \frac{12427}{96} \nu^2 v^{4}\nonumber\\
&\left.\quad -  \frac{3845}{32} \nu^3 v^{4}
 + \frac{585}{8} \nu^4 v^{4}\right)
 \,,\\
%%%%%%%%%%%%%%%%%%%%%%%%%%%%%%%%%%%%%%%%%%%%%%%%%%%%%%%%%%%%%%%%%
\mathcal{J}^{(3)}_\text{4PN} &= \frac{m^3}{r^3} \left(\frac{7}{2} \dot{r}^2
 + \frac{7775977}{16800} \nu \dot{r}^2
 + \frac{447}{256} \pi^2 \nu \dot{r}^2
 + \frac{121449}{560} \nu^2 \dot{r}^2
 -  \frac{1025}{8} \nu^3 \dot{r}^2
 - 47 \nu^4 \dot{r}^2 \right.\nonumber\\
&\quad - 110 \nu \ln\Big(\frac{r}{r'_{0}}\Big) \dot{r}^2
 + 44 \nu \ln\Big(\frac{r}{r''_{0}}\Big) \dot{r}^2
 - 176 \nu^2 \ln\Big(\frac{r}{r''_{0}}\Big) \dot{r}^2
 + \frac{91}{4} v^{2}
 -  \frac{13576009}{50400} \nu v^{2}\nonumber\\
&\quad + \frac{469}{256} \pi^2 \nu v^{2}
 + \frac{276433}{5040} \nu^2 v^{2}
 -  \frac{41}{16} \pi^2 \nu^2 v^{2}
 + \frac{637}{8} \nu^3 v^{2}
 - 15 \nu^4 v^{2}
 -  \frac{44}{3} \nu \ln\Big(\frac{r}{r'_{0}}\Big) v^{2}\nonumber\\
&\left. \quad + \frac{88}{3} \nu^2 \ln\Big(\frac{r}{r'_{0}}\Big) v^{2}
 -  \frac{22}{3} \nu \ln\Big(\frac{r}{r''_{0}}\Big) v^{2}
 + \frac{88}{3} \nu^2 \ln\Big(\frac{r}{r''_{0}}\Big) v^{2}\right) \,,\\
%%%%%%%%%%%%%%%%%%%%%%%%%%%%%%%%%%%%%%%%%%%%%%%%%%%%%%%%%%%%%%%%%
\mathcal{J}^{(4)}_\text{4PN} &= \frac{m^4}{r^4} \left(\frac{15}{8}
 + \frac{3809041}{25200} \nu
 -  \frac{85}{8} \pi^2 \nu
 -  \frac{20131}{420} \nu^2
 + \frac{663}{32} \pi^2 \nu^2
 -  \frac{15}{4} \nu^3
 + \nu^4
 + 32 \ln\Big(\frac{r}{r'_{0}}\Big) \right.\nonumber\\
&\quad\left. -  \frac{406}{3} \nu \ln\Big(\frac{r}{r'_{0}}\Big)
 + \frac{742}{3} \nu^2 \ln\Big(\frac{r}{r'_{0}}\Big)
 - 32 \ln\Big(\frac{r}{r''_{0}}\Big)
 + \frac{766}{3} \nu \ln\Big(\frac{r}{r''_{0}}\Big)
 -  \frac{1528}{3} \nu^2 \ln\Big(\frac{r}{r''_{0}}\Big)\right) \,.
\end{align}\end{subequations}

\section{Gravitational wave tails at 4PN order} 
\label{sec:tails} 

We start with the end result of Refs.~\cite{BBBFMa, BBBFMb}, where the
Fokker Lagrangian at the 4PN order in harmonic coordinates was obtained as the sum of
instantaneous and tail contributions: $L = L^\text{inst} + L^\text{tail}$.
This Lagrangian,  in
our approach, provides only the
conservative part of the dynamics and does not account for dissipative effects.\footnote{See~\cite{GLPR16} and references therein for an
  alternative approach in the EFT context where the Lagrangian describes at
  once conservative and dissipative effects.} See
App.~\ref{sec:Lgen} for a recap of its 4PN instantaneous terms. Now, the 4PN
tail piece is given as the non-local-in-time integral
\begin{align}\label{Ltail}
L^\text{tail} &= \frac{G^2M}{5c^8}
\,I_{ij}^{(3)}(t)\!\mathop{\text{Pf}}_{2r_{12}/c}
\int_{-\infty}^{+\infty} \frac{\ud t'}{\vert t-t'\vert}
I_{ij}^{(3)}(t')\nonumber\\&= \frac{G^2M}{5c^8}
\,I_{ij}^{(3)}(t) \int_0^{+\infty} \! \ud\tau \,
\ln{\left( \frac{c\tau}{2 r_{12}} \right)} \left[ I_{ij}^{(4)}(t-\tau)
  - I_{ij}^{(4)}(t+\tau) \right]\,.
\end{align}
In the above expression, $M$ is the constant ADM mass of the system, such that
$M=m+\mathcal{O}(c^{-2})$, $I_{ij}(t)$ are the components of the STF
quadrupole moment of the two particles at Newtonian order,
$I_{ij} = \sum_A m_A \,y_A^{\langle i} y_A^{j\rangle}$, and $I_{ij}^{(s)}(t)$
is the $s$-th time derivative of $I_{ij}(t)$ performed ``off-shell'',
\textit{i.e.}, without order reduction by means of the equations of motion. The tail
integral involves the Hadamard \textit{partie finie} (Pf) prescription, which
depends on some arbitrary scale, here chosen to be the relative separation
between the two particles $r_{12}=r_{12}(t)$ at time $t$ in harmonic
coordinates. For convenience, we introduce a special notation for the tail
factor:
\begin{equation}\label{tailfactor}
\mathcal{T}_{ij}^{(s)}(t) = \mathop{\text{Pf}}_{2r_{12}/c}
\int_{-\infty}^{+\infty} \frac{\ud t'}{\vert t-t'\vert}
I_{ij}^{(s)}(t')\,.
\end{equation}
Beware that, because of the presence of the time-dependent Hadamard scale
$r_{12}$ we have adopted, $\mathcal{T}_{ij}^{(s)}$ is not the time derivative of
$\mathcal{T}_{ij}^{(s-1)}$.

Varying the Lagrangian with respect to $\bm{y}_A$ leads to the conservative
part of the acceleration $\bm{a}_A=\ud \bm{v}_A/\ud t$, which similarly decomposes
into instantaneous and tail parts as
$\bm{a}_A^\text{inst} + \bm{a}_A^\text{tail}$. The instantaneous part contains
many terms up to the 4PN order; in the present section, we analyze the effect of
including the 4PN tail piece~\eqref{Ltail} into the Lagrangian. Since this piece is non
local, its variation involves functional derivatives rather than ordinary ones.
We find for the acceleration of particle 1:
\begin{align}\label{acctailcons}
a_1^{i\,\text{tail}} =& -\frac{4G^2M}{5c^8} \,y_1^j
  \,\mathcal{T}_{ij}^{(6)}\nonumber\\ & +\frac{8G^2M}{5c^8}
  y_1^j\biggl[\left(I_{ij}^{(3)}\ln r_{12}\right)^{(3)}-I_{ij}^{(6)}\ln
    r_{12}\biggr] -\frac{2G^2M}{5m_1c^8}
  \frac{n_{12}^i}{r_{12}}\Bigl(I_{jk}^{(3)}\Bigr)^2\,.
\end{align}
The first term coincides with the conservative part of the known 4PN tail
effect~\cite{BD88, B93, B97}, while the other terms, which are in fact instantaneous,
come from the variation of the ``constant'' $r_{12}(t)$ in
Eq.~\eqref{Ltail}. In the CM frame, the tail part of the acceleration is
directly obtained by reducing the previous expression~\eqref{acctailcons} to
the CM. Indeed, there are no tail contributions in the CM
relations~\eqref{y1y2}, \textit{i.e.}, $\bm{y}_A^\text{tail}=0$ when
considered as a function of the relative variables $\bm{x}$ and $\bm{v}$.
Therefore, the tail acceleration
$a^{i\,\text{tail}} = a_1^{i\,\text{tail}} - a_2^{i\,\text{tail}}$ reads simply
(with $r=r_{12}$)\footnote{For completeness, let us give two useful results
  concerning the third time-derivative of the quadrupole moment in
  the CM variables:
\begin{align*}
I_{ij}^{(3)} &= \frac{2G m^2\nu}{r^2}\biggl[-4 n^{\langle
    i}v^{j\rangle} + 3\dot{r}\,n^{\langle
    i}n^{j\rangle}\biggr]\,,\\ \Bigl(I_{ij}^{(3)}\Bigr)^2 &=
\frac{8G^2 m^4\nu^2}{r^4}\biggl(4 v^2 -
\frac{11}{3}\dot{r}^2\biggr)\,.
\end{align*}}
\begin{align}\label{acctailCM}
a^{i\,\text{tail}} =& -\frac{4G^2M}{5c^8}
  x^j\,\mathcal{T}_{ij}^{(6)}\nonumber\\ & +\frac{8G^2M}{5c^8}
  x^j\left[\left(I_{ij}^{(3)}\ln r\right)^{(3)}-I_{ij}^{(6)}\ln
    r\right] -\frac{2G^2}{5c^8\nu}
  \frac{n^i}{r}\Bigl(I_{jk}^{(3)}\Bigr)^2\,.
\end{align}
The instantaneous part of the relative acceleration in the CM frame has
already been
displayed in Eqs.~\eqref{dvdt}--\eqref{B}. The non local tail integral
in~\eqref{acctailCM} cannot be expressed in analytic closed form for general
orbits, but it can usefully be written using Fourier series
[see~\eqref{tailfactorFourier} below]. For circular orbits, the tail factor does
reduce to a simple closed-form expression as in Eq.~\eqref{T6}. The
dissipative part of the equations of motion (including notably that associated
with the tail term) will be investigated in Sec.~\ref{sec:diss}.

Given the accelerations $\bm{a}_A$ as functionals of the positions and
velocities of the particles for general orbits, we compute the associated
conserved (Noetherian) energy by forming the combination
$\sum m_A \bm{v}_A\cdot\bm{a}_A$, where $\bm{a}_A$ is the explicit expression
of the acceleration, including the tail
term~\eqref{acctailcons}, and writing it in the form of a total derivative,
say $-\ud E_0/\ud t$. The looked-for energy will then be the sum of kinetic
and ``potential'' contributions, $E = \sum \frac{1}{2} m_A v_A^2 + E_0$,
with $E_0$ starting  at leading order with the usual Newtonian potential energy
$-G m_1 m_2/r_{12}$.

Applying this method to the instantaneous (local) part of the acceleration,
$\bm{a}_A^\text{inst}$, we readily find that there is, evidently, a
well defined notion of ``instantaneous'' conserved energy, say
$E^\text{inst}=\sum \frac{1}{2} m_A v_A^2 + E_0^\text{inst}$. We are
neglecting the radiation reaction piece, which will be added in
Sec.~\ref{sec:diss}. The instantaneous part of the energy, $E^\text{inst}$,
comprises many terms up to the 4PN order. It was provided in the CM frame in
Sec.~\ref{sec:cons}.

Looking next for the same combination as before but for the tail part of the
acceleration, explicitly given by~\eqref{acctailcons}, we get, in a first
stage, after a series of operations by parts,
\begin{equation}\label{combtail}
\sum_A m_A v_A^i (a_A^i)^\text{tail} = - \frac{\ud E_0^\text{tail}}{\ud t} -
H_1^\text{tail} \,.
\end{equation}
The first term, indeed, takes the form of the total time derivative of a
certain quantity:
\begin{align}\label{E0tail}
E_0^\text{tail} &= \frac{2G^2M}{5c^8}\Bigl[
  I_{ij}^{(1)}\,\mathcal{T}_{ij}^{(5)} -
  I_{ij}^{(2)}\,\mathcal{T}_{ij}^{(4)} +
  \frac{1}{2}I_{ij}^{(3)}\,\mathcal{T}_{ij}^{(3)}\Bigr]\nonumber\\& -
\frac{4G^2M}{5c^8}\left( 2I_{ij}^{(1)}\,I_{ij}^{(4)} (\ln
r_{12})^{(1)} - I_{ij}^{(2)}\,I_{ij}^{(3)} (\ln r_{12})^{(1)} +
I_{ij}^{(1)}\,I_{ij}^{(3)} (\ln r_{12})^{(2)}
\right)\,,
\end{align}
where we have used the notation~\eqref{tailfactor} for tails, the
non-tail terms coming from the derivation of the Hadamard partie finie scale
$r_{12}$. However, performing simple operations by parts does not allow to
recast the second term in~\eqref{combtail} into the requested form. It
remains like a ``flux'' at this stage, given by
\begin{equation}\label{Ftail}
H_1^\text{tail} = \frac{G^2M}{5c^8}\,\left[
  I_{ij}^{(3)}\,\mathcal{T}_{ij}^{(4)} -
  I_{ij}^{(4)}\,\mathcal{T}_{ij}^{(3)}\right]\,.
\end{equation}
Note that $r_{12}$ cancels out from the two
terms in the right-hand side of~\eqref{Ftail}. Because of the
``flux''~\eqref{Ftail}, it appears that the problem of finding the complete total
time derivative defining the energy in the non-local case is more complicated.

The problem has been solved in Ref.~\cite{BBBFMb} by resorting to (discrete)
Fourier series. Let us decompose the components of the quadrupole moment as
\begin{equation}\label{Fourier}
I_{ij}(t) =
\sum_{p=-\infty}^{+\infty}\,\mathop{{\mathcal{I}}}_{p}{}_{\!\!ij}\,\ue^{\ui
  \,p\,\ell} \quad\Longleftrightarrow\quad \mathop{{\mathcal{I}}}_{p}{}_{\!\!ij} =
\int_0^{2\pi}\frac{\ud\ell}{2\pi}\,I_{ij}\,\ue^{-\ui\, p\,\ell}\,,
\end{equation}
where $\ell=n(t-t_0)$ is the mean anomaly, $n=2\pi/P$ the radial frequency,
$P$ the orbital period and $t_0$ some instant of passage at
periastron.\footnote{The ``azimuthal'' frequency $\omega$, averaged over one
  orbit, agrees with the radial frequency, as there is no precession at
  Newtonian order: $\omega=n$.} The Fourier coefficients
${}_p\mathcal{I}_{ij}$ depend on $n$ as well as the orbit's eccentricity $e$, and
satisfy ${}_{p}{\mathcal{I}}_{ij}={}_{-p}{\overline{\mathcal{I}}}_{ij}$. They
are available as linear combinations of Bessel functions (see two
possible forms presented in App.~B of~\cite{BBBFMb} and App.~A
of~\cite{ABIQ08tail}). With those notations, we have
\begin{equation}\label{tailfactorFourier}
\mathcal{T}_{ij}^{(s)} = -2 \sum_{p=-\infty}^{+\infty} (\ui
p\,n)^s\,\mathop{{\mathcal{I}}}_{p}{}_{\!\!ij} \Bigl(\ln\left(2\vert
p\vert n r\right) + \gamma_\text{E} \Bigr)\ue^{\ui p \ell}\,.
\end{equation}
Now, it was shown in Sec.~III A of~\cite{BBBFMb} that the
``flux''~\eqref{Ftail} does admit a first integral in the sense of Fourier
series, \textit{i.e.}, there exists some $E_1^\text{tail}$ such that
\begin{equation}
\frac{\ud E_1^\text{tail}}{\ud t} = H_1^\text{tail}\,.
\end{equation}
Moreover, we found that $E_1^\text{tail}$ contains a crucial ``DC'' contribution,
\textit{i.e.}, \textit{constant} in time, which is furthermore directly
related, quite remarkably, to the total averaged gravitational-wave energy flux
$\mathcal{F}_\text{GW} =
\frac{G}{5c^5}\,\langle\bigl(\hat{I}_{ij}^{(3)}\bigr)^2\rangle$.
More precisely,
\begin{equation}\label{EDC}
E_\text{DC}^\text{tail} = - \frac{2G^2M n^6}{5c^8}
\sum_{p=-\infty}^{+\infty}\,
\vert\mathop{{\mathcal{I}}}_{p}{}_{\!\!ij}\vert^2 p^6 = -
\frac{2G M}{c^3}\,\mathcal{F}_\text{GW}\,.
\end{equation}
In addition, there is an ``AC'' contribution, \textit{i.e.},
oscillating with \textit{zero time average}, which is given by a double Fourier
series (over $p$, $q$ such that $p+q \not= 0$):
\begin{equation}\label{EAC}
E_\text{AC}^\text{tail} = \frac{G^2M n^6}{5c^8}
\sum_{p+q \not=
    0}\,\mathop{{\mathcal{I}}}_{p}{}_{\!\!ij}\mathop{{\mathcal{I}}}_{q}{}_{\!\!ij}
  \,\frac{p^3q^3(p-q)}{p+q}
  \ln\left|\frac{p}{q}\right|\,\ue^{\ui(p+q)\ell}\,.
\end{equation}
In conclusion, we have
$E_1^\text{tail} = E_\text{DC}^\text{tail} + E_\text{AC}^\text{tail}$. To
emphasize this result, let us observe that the presence of the latter DC and
AC contributions in the energy implies that, in a Hamiltonian formalism,
the conserved energy is not equal to the value of the Hamiltonian
``on-shell'', \textit{i.e.}, computed along trajectories satisfying the
corresponding Hamilton's equations. This is due to the fact that, for a
non-local dynamics, the Hamiltonian equations involve functional derivatives
instead of ordinary ones. Finally, the complete contributions of tails to the
conserved energy, which are to be added to the instantaneous contribution
$E^\text{inst}$ given by Eqs.~\eqref{E} in Sec.~\ref{sec:cons}, consist of the
three terms computed above:
\begin{equation}\label{Etailthree}
E^\text{tail} = E_0^\text{tail} + E_\text{DC}^\text{tail} + E_\text{AC}^\text{tail}\,.
\end{equation}

Next, we look for the tail contributions to the conserved integral of the
angular momentum. Proceeding in a similar way as for the energy, we form the combination
$\sum m_A \,\epsilon_{ijk}\,y_A^j\,a_A^{k\,\text{tail}}$ and perform a series
of operations by parts yielding
\begin{equation}\label{combJ}
\sum_A m_A \,\epsilon_{ijk}\,y_A^j\,a_A^{k\,\text{tail}} 
= - \frac{\ud J_0^{i\,\text{tail}}}{\ud t} - K_1^{i\,\text{tail}} \,,
\end{equation}
where 
\begin{subequations}\label{JKtail}
\begin{align}
J_0^{i\,\text{tail}} &= \frac{4G^2M}{5c^8}\,\epsilon_{ijk}\left[
    I_{jl}\,\mathcal{T}_{kl}^{(5)} -
    I_{jl}^{(1)}\,\mathcal{T}_{kl}^{(4)} +
    I_{jl}^{(2)}\,\mathcal{T}_{kl}^{(3)}\right]\nonumber\\& -
  \frac{8G^2M}{5c^8}\,\epsilon_{ijk}\left( 2I_{jl}\,I_{kl}^{(4)}
  (\ln r_{12})^{(1)} - I_{jl}^{(1)}\,I_{kl}^{(3)} (\ln r_{12})^{(1)} +
  I_{jl}\,I_{kl}^{(3)} (\ln r_{12})^{(2)}
  \right)\,,\\ 
%%%%%%%%%%%%%%%%%%%%%%%%%%%%%%%%%%%%%%%%%%%%%
K_1^{i\,\text{tail}} &=
  -\frac{4G^2M}{5c^8}\,\epsilon_{ijk}\,I_{jl}^{(3)}
  \,\mathcal{T}_{kl}^{(3)}\,.
\end{align}
\end{subequations}
We then rely on Ref.~\cite{BBBFMb} to transform the second term in
Eq.~\eqref{combJ} into a total time derivative after decomposing it as a
Fourier series. The computation parallels that for the energy and permits constructing
some $J_1^{i\,\text{tail}}$ that satisfies
\begin{equation}\label{balanceJ}
\frac{\ud J_1^{i\,\text{tail}}}{\ud t} = K_1^{i\,\text{tail}}\,.
\end{equation}
Like for the energy, $J_1^{i\,\text{tail}}$ is made of DC and AC contributions
\textit{i.e.},
$J_1^{i\,\text{tail}} = J_\text{DC}^{i\,\text{tail}} +
J_\text{AC}^{i\,\text{tail}}$.
The DC contribution, remarkably, is proportional to the gravitational-wave
flux of angular momentum. Explicitly, we have\footnote{The norms of the vectors~\eqref{JDCAC} are obtained by projection perpendicular to the
  orbital plane. For instance,
\begin{equation*}
\ell^i\epsilon_{ijk}\mathop{{\mathcal{I}}}_{p}{}_{\!\!jl}
\mathop{{\mathcal{I}}}_{q}{}_{\!\!kl} =
\bigl(\mathop{{\mathcal{I}}}_{p}{}_{\!\!xx}-
\mathop{{\mathcal{I}}}_{p}{}_{\!\!yy}\bigr)\mathop{{\mathcal{I}}}_{q}{}_{\!\!xy}
- \mathop{{\mathcal{I}}}_{p}{}_{\!\!xy}
\bigl(\mathop{{\mathcal{I}}}_{q}{}_{\!\!xx}-
\mathop{{\mathcal{I}}}_{q}{}_{\!\!yy}\bigr)\,,
\end{equation*}
where $\bm{\ell}=\bm{n}\times\bm{\lambda} = (0,0,1)$ denotes the unit vector
orthogonal to the orbital plane, which is spanned by the two moving unit vectors
$\bm{n} = (\cos\varphi, \sin\varphi, 0)$ and
$\bm{\lambda} = (-\sin\varphi, \cos\varphi, 0)$, with $\varphi$ being the
orbital phase angle. The spatial coordinates $(x,y,z)$ are such that $(x,y)$
lies in this plane, with $z$ in the direction along $\bm{\ell}$
(\textit{i.e.}, in the sense of the orbital motion). Notice also that
\begin{equation*}
\frac{1}{2}\frac{\partial
  I_{ij}^{(s)}}{\partial \varphi} =
\ell^k\epsilon_{kl \langle i} I_{j \rangle l}^{(s)}\,.
\end{equation*}
}
\begin{subequations}\label{JDCAC}
\begin{align}
J_\text{DC}^{i\,\text{tail}} &= \frac{4 G^2M n^5}{5c^8} \sum_{p=-\infty}^{+\infty}
  \ui\epsilon_{ijk} \mathop{{\mathcal{I}}}_{p}{}_{\!\!jl}
  \mathop{{\mathcal{I}}}_{-p}{}_{\!\!kl}\,p^5 = -
\frac{2G M}{c^3}\,\mathcal{G}^i_\text{GW}\,,\\
%%%%%%%%%%%%%%%%%%%%%%%%%%%%%%%%%%%%%%%%%%%%%%%%%%%%%%%%
J_\text{AC}^{i\,\text{tail}} &= - \frac{4 G^2M n^5}{5c^8} \sum_{p+q \not=
    0}\ui\epsilon_{ijk}\mathop{{\mathcal{I}}}_{p}{}_{\!\!jl}
  \mathop{{\mathcal{I}}}_{q}{}_{\!\!kl} \,\frac{p^3q^3}{p+q}
  \ln\left|\frac{p}{q}\right|\,\ue^{\ui(p+q)\ell}\,.
\end{align}
\end{subequations}
The complete tail contribution to the angular momentum is finally
\begin{equation}\label{Jtailthree}
J^\text{tail} = J_0^\text{tail} + J_\text{DC}^\text{tail} + J_\text{AC}^\text{tail}\,,
\end{equation}
which is thus to be added to the instantaneous contributions presented in
Eqs.~\eqref{J}.

Finally, we obtain the tail parts corresponding to the linear momentum $\bm{P}$
and the CM position $\bm{G}$. The instantaneous part of the CM position,
$\bm{G}^\text{inst}$, is provided in App.~\ref{sec:Gi}. In these two cases the
analysis is simpler than for $E$ and $\bm{J}$ because the usual operations by
parts applied to the relevant combination $\sum m_A\, \bm{a}_A^\text{tail}$
directly lead to the requested total time derivatives and conservation laws. We
find
\begin{subequations}\label{PGtaildef}
\begin{align}
P^i_\text{tail} &= \frac{4G^2M}{5c^8}\,\left[
    I_{j}\,\mathcal{T}_{ij}^{(5)} -
    I_{j}^{(1)}\,\mathcal{T}_{ij}^{(4)}\right]\nonumber\\& -
  \frac{8G^2M}{5c^8}\left( 2I_{j}\,I_{ij}^{(4)} (\ln r_{12})^{(1)} -
  I_{j}^{(1)}\,I_{ij}^{(3)} (\ln r_{12})^{(1)} + I_{j}\,I_{ij}^{(3)}
  (\ln r_{12})^{(2)} \right)\,,\\
%%%%%%%%%%%%%%%%%%%%%%%%%%%%%%%%%%%%%%%%%%%%%%%%%%%%%%%%%%%%%%%%%%%%
G^i_\text{tail} &=
  \frac{4G^2M}{5c^8}\!\left[ I_{j}\,\mathcal{T}_{ij}^{(4)} - 2
    I_{j}^{(1)}\,\mathcal{T}_{ij}^{(3)}\right] -
  \frac{8G^2M}{5c^8}\,I_{j}\,I_{ij}^{(3)} (\ln
  r_{12})^{(1)}\,.
\end{align}
\end{subequations}
Note that the quadrupolar factors above are coupled to the Newtonian mass
dipole moment $I_i\equiv\sum_A m_A\,y_A^i$. At Newtonian order, the CM position reduces to the mass dipole moment,
$G^i=I_i + \mathcal{O}(c^{-2})$, hence we see from Eqs.~\eqref{PGtaildef} that
the tails will not affect the definition of the CM frame since we can pose
$I_i = 0$ in Eqs.~\eqref{PGtaildef} at the current PN order.

%To conclude this section, let us remind that the latter conservation laws
%concern only the conservative dynamics, as we have neglected the radiation reaction
%contributions in what precedes. By adding the dissipative radiation reaction terms in
%Sec.~\ref{sec:diss}, we shall recover the usual balance equations between the
%losses of the matter system and the corresponding fluxes.

\section{Equations of motion for circular orbits} 
\label{sec:EOMcirc} 

With the dynamics in the CM frame in hand, we are in the position to reduce
it to the case of circular orbits. The conservative part of the relative
acceleration is then given by the purely radial acceleration
$\bm{a} = -\omega^2 \bm{x}$, the physical content of which being entirely encoded into the relation between the orbital frequency $\omega$ and the
orbital separation $r$. All the results in this section consistently include the tail effect.

The tail term in the relative CM acceleration has been already shown in
Eq.~\eqref{acctailCM}, and the instantaneous terms in
Eqs.~\eqref{dvdt}--\eqref{B}. For circular orbits, the tail integral can be
evaluated in closed-form. Using $M=m$, we have at that order,
\begin{equation}\label{T6}
\mathcal{T}_{ij}^{(6)} = -64\frac{G^2m^3\nu}{r^6}\left(v^iv^j-\frac{G
  m}{r^3}\,x^ix^j\right)\Bigl[\ln\left(4\sqrt{\gamma}\right) +
  \gamma_\text{E}\Bigr]\,,
\end{equation}
where the PN parameter $\gamma$ has been defined in Eq.~\eqref{gamx}; we
recall that $\gamma_\text{E}$ is Euler's constant. Notice
that~\eqref{T6} is automatically STF as we have $v^2=\frac{G m}{r}$ for
circular orbits at Newtonian order. From Eq.~\eqref{acctailCM}, we readily
obtain the tail contribution as
$\bm{a}^\text{tail} = - \omega^2_\text{tail} \,\bm{x}$, where
\begin{equation}\label{omtail}
\omega^2_\text{tail} = \frac{128}{5}\frac{G
    m}{r^3}\gamma^4\nu\biggl[\ln\left(16 \gamma\right) +
    2\gamma_\text{E} + \frac{1}{2}\biggr]\,.
\end{equation}
On the other hand, the instantaneous contributions are computed by a
straightforward reduction of the equations of motion~\eqref{A}--\eqref{B} to
circular orbits, with $\dot{r}=0$ and $v^2=r^2\omega^2$ (since we neglect the
dissipative terms). Adding the tail contribution~\eqref{omtail}, we
get
\begin{align}\label{keplerlaw}
 \omega^2 &= \frac{G m}{r^3} \bigg\{ 1+(-3+\nu) \gamma + \left( 6 +
  \frac{41}{4}\nu + \nu^2 \right) \gamma^2 \nonumber\\ & \quad\quad +
  \left( -10 + \left[- \frac{75707}{840} + \frac{41}{64} \pi^2 + 22
    \ln \left( \frac{r}{r'_0}\right) \right]\nu + \frac{19}{2}\nu^2 +
  \nu^3 \right) \gamma^3 +
  \left(15+ 48 \ln\Big(\frac{r'_0}{r''_{0}}\Big) \right. \nonumber\\
  & \quad\quad\quad
\left. + \nu \left[\frac{19644217}{33600} + \frac{163}{1024} \pi^2
+ \frac{256}{5} \gamma_E + \frac{128}{5} \ln (16 \gamma ) 
+ 82 \ln\Big(\frac{r}{r'_{0}}\Big) - 
372 \ln\Big(\frac{r}{r''_{0}}\Big)\right] \right.\nonumber\\
&\quad\quad\quad \left.  + \nu^2 \left[\frac{44329}{336} 
-  \frac{1907}{64} \pi^2 -  \frac{992}{3} \ln\Big(\frac{r}{r'_{0}}\Big) 
+ 720 \ln\Big(\frac{r}{r''_{0}}\Big)\right] + \frac{51}{4} \nu^3 
+ \nu^4 \right)\gamma^4\biggr\}\,.
\end{align}
This is a gauge dependent result, as the separation $r$ refers to harmonic
coordinates. It depends on the gauge constants $r'_0$ and
$r''_0$ defined in~\eqref{r'0r''0}. Inverting~\eqref{keplerlaw}, we express
$\gamma=\frac{G m}{r c^2}$ as a function of the orbital frequency $\omega$ or,
rather, of the PN parameter $x\equiv (\frac{G m \omega}{c^3})^{2/3}$:
\begin{align}\label{gammax}
 \gamma &= x \bigg\{1  + x \left(1 -  \frac{1}{3} \nu\right) 
+ x^2 \left(1 -  \frac{65}{12} \nu\right)  \nonumber\\
 &\qquad  + x^3 \left(1 + \nu \left[- \frac{2203}{2520} 
-  \frac{41}{192} \pi^2  -  \frac{22}{3} \ln\Big(\frac{G m}{c^2 r'_{0}}\Big) 
+ \frac{22}{3} \ln(x)\right] + \frac{229}{36} \nu^2 
+ \frac{\nu^3}{81} \right) \nonumber\\
&\qquad + x^4 \left(1 + 16 \ln\Big(\frac{G m}{c^2 r'_{0}}\Big)  
- 16 \ln\Big(\frac{G m}{c^2 r''_{0}}\Big) -  \frac{1261}{324} \nu^3 
+ \frac{\nu^4}{243} \right.\nonumber\\
&\qquad\quad+ \nu \left[- \frac{2067859}{33600} - \frac{256}{15} \gamma_E 
-  \frac{5411}{3072} \pi^2 - 86 \ln\Big(\frac{G m}{c^2 r'_{0}}\Big) 
\right.\nonumber\\
&\qquad\qquad\quad \left. + 124 \ln\Big(\frac{G m}{c^2 r''_{0}}\Big) - \frac{256}{15} \ln 4 
- \frac{698}{15} \ln(x)\right]\nonumber\\
&\qquad\quad\left. + \nu^2 \left[\frac{153613}{15120} 
+ \frac{6049}{576} \pi^2 + \frac{1168}{9} \ln\Big(\frac{G m}{c^2 r'_{0}}\Big) 
- 240 \ln\Big(\frac{G m}{c^2 r''_{0}}\Big) 
+ \frac{992}{9} \ln(x)\right] \right)  \bigg\}.
 \end{align}

 Let us deal next with the conserved energy as a function of the separation $r$. All
 the instantaneous terms in the CM frame are presented in Eqs.~\eqref{E},
 while the tail
 part in the energy has been obtained as the sum of three terms in
 Eq.~\eqref{Etailthree}. For circular orbits, the term $E_0^\text{tail}$ does
 contribute, as well as the crucial constant DC term in~\eqref{Etailthree}; by
 contrast,
 the AC contribution vanishes in this case. Furthermore there are extra tail terms coming from the reduction to circular orbits. We are led to the 4PN energy for
 circular orbits as a function of $\gamma$:
\begin{align}\label{Ecircgam}
E &= -\frac{\mu c^2 \gamma}{2} \biggl\{ 1 + \left( - \frac{7}{4} +
  \frac{1}{4} \nu \right) \gamma + \left( - \frac{7}{8} + \frac{49}{8}
  \nu + \frac{1}{8} \nu^2 \right) \gamma^2  + \left(-\frac{235}{64} \right.
\nonumber\\ & \quad \left.
 + \left[\frac{46031}{2240} - \frac{123}{64}
    \pi^2 + \frac{22}{3} \ln \left( \frac{r}{r_0'} \right) \right] \nu
  + \frac{27}{32} \nu^2 + \frac{5}{64} \nu^3 \right) \gamma^3
 + \left(- \frac{649}{128} 
+ 16 \ln\Big(\frac{r'_0}{r''_0}\Big) \right. \nonumber\\ & \quad \quad \left. 
+ \nu \left[- \frac{3357833}{28800} 
+ \frac{384}{5}\gamma_E + \frac{192}{5}\ln (16 \gamma) 
+ \frac{14935}{1024}\pi^2+ 31 \ln\Big(\frac{r}{r'_{0}}\Big) 
- 124 \ln\Big(\frac{r}{r''_{0}}\Big)\right] \right. \nonumber\\
&\left.  \quad \quad+ \nu^2 \left[\frac{83959}{8064} 
-  \frac{957}{128} \pi^2 -  \frac{349}{3} \ln\Big(\frac{r}{r'_{0}}\Big) 
+ 240 \ln\Big(\frac{r}{r''_{0}}\Big)\right] + \frac{69}{64} \nu^3 
+ \frac{7}{128} \nu^4 \right)\gamma^4 \biggr\} \,.
\end{align}
This result is not yet the invariant we are looking for, as it still depends
on the constant $r'_0$ and $r''_0$. However, these constants are canceled when
we replace $\gamma$ by the frequency-related parameter $x$, using
Eq.~\eqref{gammax}. Finally, we arrive at~\cite{DJS14, DJS16, BBBFMa, BBBFMb, FSrevue}
\begin{align}\label{Ecirc}
	E &= -\frac{\mu c^2 x}{2} \biggl\{ 1 + \left( - \frac{3}{4} -
        \frac{\nu}{12} \right) x + \left( - \frac{27}{8} +
        \frac{19}{8} \nu - \frac{\nu^2}{24} \right) x^2 \nonumber
        \\ &\quad\quad + \left( - \frac{675}{64} + \biggl[
          \frac{34445}{576} - \frac{205}{96} \pi^2 \biggr] \nu -
        \frac{155}{96} \nu^2 - \frac{35}{5184} \nu^3 \right) x^3
        \nonumber \\ &\quad\quad + \left( - \frac{3969}{128} +
        \left[-\frac{123671}{5760}+\frac{9037}{1536}\pi^2 +
          \frac{896}{15}\gamma_\text{E}+ \frac{448}{15} \ln(16
          x)\right]\nu\right.\nonumber\\ & \quad\quad\quad \left.+
        \left[-\frac{498449}{3456}+\frac{3157}{576}\pi^2\right]\nu^2
        +\frac{301}{1728}\nu^3 + \frac{77}{31104}\nu^4\right) x^4
        \biggr\} \,.
\end{align}
Note the presence of the logarithmic term at the 4PN order, due to the
tail contribution defined in~\eqref{Etailthree}.\footnote{See~\cite{BDLW10b,
    LBW12} for the logarithm at the next 5PN order, which is associated
  with higher-order tail effects.} As for the tail contribution in
Eq.~\eqref{Ecirc}, it reads
\begin{equation}\label{Etildetail}
\tilde{E}^\text{tail} = - \frac{224}{15} \mu c^2 \nu x^5\biggl[\ln\left(16
  x\right) + 2\gamma_\text{E} + \frac{2}{7}\biggr] \,.
\end{equation}
This contribution is made of the ``direct'' tail piece $E^\text{tail}$ given
by~\eqref{Etailthree}, plus some extra tail terms due to circular-orbit reduction. In the small mass ratio limit $\nu\to 0$, the
result~\eqref{Ecirc} agrees with GSF calculations~\cite{BDLW10b,LBW12,LBB12,BiniD13}. 

The 4PN angular momentum for
circular orbits can be found either by a direct calculation, or from the
well-known ``thermodynamic'' relation
$\frac{\ud E}{\ud \omega} = \omega \, \frac{\ud J}{\ud \omega}$, which is a
particular case of the first law of compact binary mechanics; this law has been
derived up to the 4PN order, taking into account the non locality associated
with the tail effect~\cite{BL17}. We get
\begin{align}\label{Jcirc}
    J &= \frac{G \,\mu \,m}{c\,x^{1/2}} \biggl\{ 1 + \left(
    \frac{3}{2} + \frac{\nu}{6} \right) x + \left( \frac{27}{8} -
    \frac{19}{8} \nu + \frac{\nu^2}{24} \right) x^2 \nonumber
    \\ &\quad\quad + \left( \frac{135}{16} + \biggl[ -
      \frac{6889}{144} + \frac{41}{24} \pi^2 \biggr] \nu +
    \frac{31}{24} \nu^2 + \frac{7}{1296} \nu^3 \right) x^3 \nonumber
    \\ &\quad\quad + \left( \frac{2835}{128} +
    \left[\frac{98869}{5760}-\frac{6455}{1536}\pi^2 -
      \frac{128}{3}\gamma_\text{E} -
      \frac{64}{3}\ln(16x)\right]\nu\right.\nonumber\\ &
    \quad\quad\quad \left.+
    \left[\frac{356035}{3456}-\frac{2255}{576}\pi^2\right]\nu^2
    -\frac{215}{1728}\nu^3 - \frac{55}{31104}\nu^4\right) x^4 \biggr\}
    \,,
\end{align}
with the tail contribution therein\footnote{Notice that the tail
  contributions~\eqref{Etildetail} and~\eqref{Jtildetail} satisfy separately
  the first law:
  $\frac{\ud \tilde{E}^\text{tail}}{\ud \omega} = \omega \, \frac{\ud
    \tilde{J}^\text{tail}}{\ud \omega}$.}
\begin{equation}\label{Jtildetail}
\tilde{J}^\text{tail} = - \frac{64}{3} \frac{G\,\mu^2\,x^{7/2}}{c}\biggl[\ln\left(16
  x\right) + 2\gamma_\text{E} + \frac{1}{5}\biggr] \,.
\end{equation}

Finally, let us also recall the expression of the 4PN periastron advance in
the limiting case of circular orbits. It was computed in Refs.~\cite{DJS15eob,
  DJS16, BBBFMb} within the Hamiltonian formalism, dealing in particular
with the non-locality of the 4PN Hamiltonian. For the present paper, we have recomputed it using the Lagrangian formalism in harmonic coordinates:
\begin{align}\label{Kcirc}
K &= 1 + 3 x + \left(\frac{27}{2} - 7\nu\right) x^2 +
\left(\frac{135}{2}
+\left[-\frac{649}{4}+\frac{123}{32}\pi^2\right]\nu+ 7\nu^2\right) x^3
\nonumber\\& + \left(\frac{2835}{8}
+\left[-\frac{275941}{360}+\frac{48007}{3072}\pi^2 -
  \frac{1256}{15}\ln x - \frac{592}{15}\ln 2 - \frac{1458}{5}\ln 3 -
  \frac{2512}{15}\gamma_\text{E}\right]\nu
\right.\nonumber\\&\left.\quad+
\left[\frac{5861}{12}-\frac{451}{32}\pi^2\right]\nu^2 -
\frac{98}{27}\nu^3\right) x^4\,,
\end{align}
including the tail contribution 
\begin{equation}\label{Ktailcirc}
K^\text{tail} = \left(\frac{352}{5} - \frac{1256}{15}\ln x -
\frac{592}{15}\ln 2 - \frac{1458}{5}\ln 3 -
\frac{2512}{15}\gamma_\text{E}\right)\nu \,x^4\,.
\end{equation}
Again, the small mass-ratio limit of the periastron advance \eqref{Kcirc}
perfectly agrees with GSF results known from numerical~\cite{BDS10,
  Letal11, vdM16} and analytical~\cite{D10sf, DJS15eob, DJS16, BL17} works.
Nonetheless, let us remind that, thanks to our recent ambiguity-free
completion of the 4PN equations of motion in Refs.~\cite{BBBFMc, MBBF17}, the
formulas~\eqref{Ecirc},~\eqref{Jcirc} and~\eqref{Kcirc} have now been derived
from first principles without any reference to GSF calculations.

\section{Dissipative radiation reaction terms} 
\label{sec:diss} 

In this section, we add the dissipative, radiation-reaction driven part of the
dynamics. This includes the usual odd-parity 2.5PN and 3.5PN terms, but the
most interesting dissipative effect is the one associated with the tails and
occuring at the (formally even) 4PN order. The tail acceleration was obtained
in Eq.~\eqref{acctailcons}. It contains the tail integral
\begin{equation}\label{tailfactor6}
\mathcal{T}_{ij}^{(6)} = \mathop{\text{Pf}}_{2r_{12}/c}
\int_{-\infty}^{+\infty} \frac{\ud t'}{\vert t-t'\vert}
I_{ij}^{(6)}(t') = \int_0^{+\infty} \ud\tau
\ln\left(\frac{c\tau}{2r_{12}}\right)\left[I_{ij}^{(7)}(t-\tau) -
  I_{ij}^{(7)}(t+\tau)\right]\,,
\end{equation}
which only corresponds to the conservative part of the tail effect. Indeed, we
recognize a ``time-symmetric'' decomposition, which is
conservative in the sense that the corresponding acceleration is purely radial
in the case of circular orbits, as we have seen in Eq.~\eqref{omtail}. The
dissipative part is given by the corresponding ``time-antisymmetric''
combination, hence the dissipative tail acceleration is
\begin{equation}\label{acctaildiss}
a_1^i{}_\text{diss}^\text{tail} = -\frac{4G^2M}{5c^8} \,y_1^j
  \,\int_0^{+\infty} \ud\tau
  \ln\left(\frac{\tau}{2P}\right)\left[I_{ij}^{(7)}(t-\tau) +
    I_{ij}^{(7)}(t+\tau)\right]\,.
\end{equation}
In this expression, $P$ denotes an arbitrary scale, but it is easy to check
that this scale cancels out from the two terms of~\eqref{acctaildiss} so that we
can choose $P=r_{12}/c$. Thus, the complete tail part of the acceleration is
the sum of~\eqref{acctailcons} and~\eqref{acctaildiss}. It reads
\begin{align}\label{acctail}
a_1^i{}^\text{tail} =& -\frac{8G^2M}{5c^8} \,y_1^j \int_0^{+\infty}
  \ud\tau \ln\left(\frac{c\tau}{2r_{12}}\right) I_{ij}^{(7)}(t-\tau)
  \nonumber\\ & +\frac{8G^2M}{5c^8} y_1^j\left[\left(I_{ij}^{(3)}\ln
    r_{12}\right)^{(3)}-I_{ij}^{(6)}\ln r_{12}\right]
  -\frac{2G^2M}{5m_1c^8}
  \frac{n_{12}^i}{r_{12}}\Bigl(I_{jk}^{(3)}\Bigr)^2\,.
\end{align}
The non-local tail term agrees with the result found from first-principle
derivations of the near zone metric in Refs.~\cite{BD88, B93, B97}.

Besides the 4PN dissipative tail effect determined in Eq.~\eqref{acctaildiss},
we also need to include the well-known radiation reaction odd terms at the 2.5PN
and 3.5PN orders~\cite{IW93, IW95, PW02, KFS03, NB05, itoh3}, given here in
the CM frame, with the notation of~\eqref{dvdt}--\eqref{B}:
\begin{subequations}\label{ABdiss}
\begin{align}
A_\text{2.5PN} &= \frac{8
  G\,m\,\nu}{5r}\dot{r}\left[-\frac{17}{3} \frac{G m}{r} - 3
  v^2\right]\,,\\
%%%%%%%%%%%%%%%%%%%%%%%%%%%%%%%%%%%%%%%%%%%%%%%%%%%%%%%%%%%%%%%%%
B_\text{2.5PN} &= \frac{8 G\,m\,\nu}{5r}\left[ 3\frac{G
    m}{r} + v^2\right]\,,\\
%%%%%%%%%%%%%%%%%%%%%%%%%%%%%%%%%%%%%%%%%%%%%%%%%%%%%%%%%%%%%%%%%
A_\text{3.5PN} &= \frac{G m \nu}{r} \dot{r}\left[\frac{G^2
    m^2}{r^2} \, \left( \frac{3956}{35} + \frac{184}{5} \nu \right) +
  \frac{G m \,v^2}{r} \, \left( \frac{692}{35} - \frac{724}{15} \nu
  \right)\right. \nonumber\\ & \quad\quad + v^4 \, \left(
  \frac{366}{35} + 12 \nu \right) + \frac{G m \,\dot{r}^2}{r} \, \left(
  \frac{294}{5} + \frac{376}{5} \nu \right) \nonumber\\ &
  \quad\quad\left. - v^2 \dot{r}^2 \, \left( 114 + 12 \nu \right) + 112
  \dot{r}^4 \right]\,,\\
%%%%%%%%%%%%%%%%%%%%%%%%%%%%%%%%%%%%%%%%%%%%%%%%%%%%%%%%%%%%%%%%%%
B_\text{3.5PN} &= \frac{G m \nu}{r}\left[\frac{G^2
    m^2}{r^2} \, \left( - \frac{1060}{21} - \frac{104}{5}\nu \right) +
  \frac{G m v^2}{r} \, \left( \frac{164}{21} + \frac{148}{5} \nu
  \right) \right. \nonumber\\ & \quad\quad + v^4\, \left(
  - \frac{626}{35} - \frac{12}{5} \nu \right) + \frac{G m \dot{r}^2}{r} \,
  \left( - \frac{82}{3} - \frac{848}{15} \nu \right) \nonumber\\ &
  \quad\quad\left. + v^2 \dot{r}^2 \left( \frac{678}{5} +
  \frac{12}{5} \nu \right) - 120 \dot{r}^4 \right]\,.
\end{align}\end{subequations}
In the case of (quasi-)circular orbits, the 4PN equations of motion, including
the 2.5PN, 3.5PN and 4PN radiation reaction effects, become
\begin{equation}\label{eomcirc}
\bm{a} = -\omega^2 \bm{x} - \frac{32}{5}\frac{G^3m^3\nu}{c^5r^4}\left[1 +
  \left(-\frac{743}{336} - \frac{11}{4}\nu\right)\gamma + 4\pi
  \gamma^{3/2} \right] \bm{v} \,.
\end{equation}
The orbital frequency as a function of the separation $r$, with all
conservative terms, has been obtained in Eq.~\eqref{keplerlaw}. In the above
equation, witness the contribution of the radiation reaction 4PN tail effect
with coefficient $4\pi$.

With the radiation reaction terms added to the conservative acceleration, the
energy and angular momentum are no longer conserved. Their time derivatives
are now equal to (minus) the corresponding fluxes in gravitational
waves. As usual, in order to recover the familiar expressions for those
fluxes,\footnote{\textit{I.e.}, the familiar Einstein quadrupole formula at
  leading order, and its extension, at next-to-leading orders, built from an
  irreducible STF decomposition of the mass and current (radiative type)
  multipole moments (see \textit{e.g.} Eqs.~(68) in~\cite{Bliving14}).} we have
to transfer certain terms in the form of total time derivatives from the
right-hand side of the balance equations to the left-hand side. This implies
that the energy and angular momentum will also acquire certain
radiation-reaction contributions. The balance equations read
\begin{equation}\label{balance}
\frac{\ud E}{\ud t} = - F \,,\quad \frac{\ud \bm{J}}{\ud t}= - \bm{M} \,,
\end{equation} 
with purely dissipative energy and angular momentum fluxes $F$ and $\bm{M}$ in
the right-hand side. The conservative parts of the CM energy
$\mathcal{E}=E/\mu$ and angular momentum $\mathcal{J}=J/J_N$ have already been
provided in Eqs.~\eqref{E} and~\eqref{J}. We now present the 2.5PN and 3.5PN dissipative
contributions to the balance equation for energy~\cite{IW93,
  IW95, PW02, KFS03, NB05, itoh3}\footnote{As we are working in harmonic
  coordinates, a particular set of Iyer-Will parameters~\cite{IW93, IW95} has
  been selected, namely the one displayed in Eqs.~(5.10) of Ref.~\cite{NB05}.}
\begin{subequations}\label{Ediss}
\begin{align}
\mathcal{E}_\text{2.5PN} &= \frac{8 G^2 m^2\,\nu}{5r^2}\dot{r}v^2\,,\\
%%%%%%%%%%%%%%%%%%%%%%%%%%%%%%%%%%%%%%%%%%%%%%%%%%%%%%%%%%%%%%%%%%%%%%
\mathcal{E}_\text{3.5PN} &= - \frac{8 G^2 m^2\,\nu}{5r^2}
\dot{r}\biggl[ \left(\frac{271}{28}+6\nu\right)v^4 
+ \left(-\frac{77}{4}-\frac{3}{2}\nu\right)v^2\dot{r}^2 
+ \left(\frac{79}{14}-\frac{92}{7}\nu\right)v^2\frac{G m}{r} \nonumber\\ 
&\quad  + 10 \dot{r}^4 + \left(\frac{5}{42}+\frac{242}{21}\nu\right) 
\dot{r}^2\frac{G m}{r} + \left(-\frac{4}{21}+\frac{16}{21}\nu\right) 
\left(\frac{G m}{r}\right)^2\,\biggr]\,,
\end{align}\end{subequations}
together with the corresponding terms in the flux (with $\mathcal{F}=F/\mu$)
\begin{subequations}\label{Fdiss}
\begin{align}
\mathcal{F}_\text{2.5PN} &= \frac{8 G^3 m^3\,\nu}{5r^4}\left( 4 v^2 -
\frac{11}{3}\dot{r}^2\right)\,,\\
%%%%%%%%%%%%%%%%%%%%%%%%%%%%%%%%%%%%%%%%%%%%%%%%%%%%%%%%%%%%%%%%%%%%%%
\mathcal{F}_\text{3.5PN} &= \frac{8 G^3 m^3\,\nu}{5r^4}
\biggl[ \left(\frac{785}{84}-\frac{71}{7}\nu\right)v^4 
+ \left(-\frac{680}{21}+\frac{40}{21}\nu\right)v^2\frac{G m}{r} 
+ \left(-\frac{1487}{42}+\frac{232}{7}\nu\right)v^2\dot{r}^2 \nonumber\\
                         &\quad 
+ \left(\frac{734}{21}-\frac{10}{7}\nu\right)\dot{r}^2\frac{G m}{r} 
+ \left(\frac{687}{28}-\frac{155}{7}\nu\right)\dot{r}^4 
+ \left(\frac{4}{21}-\frac{16}{21}\nu\right)
\left(\frac{G m}{r}\right)^2 \,\biggr]\,.
\end{align}\end{subequations}
As was said, this expression is nothing but the standard ``irreducible'' expression
for the flux reduced to the case of binary motion in the CM frame. For the
angular momentum, we have
\begin{subequations}\label{Jdiss}
\begin{align}
\mathcal{J}_\text{2.5PN} &= - \frac{8 G^2 m^2\,\nu}{5r^2}\dot{r}\,,\\
%%%%%%%%%%%%%%%%%%%%%%%%%%%%%%%%%%%%%%%%%%%%%%%%%%%%%%%%%%%%%%%%%%%%%%
\mathcal{J}_\text{3.5PN} &= 
- \frac{8 G^2 \,m^2\,\nu}{5r^2}\dot{r}
\biggl[ \left(\frac{40}{3}-\frac{11}{21}\nu\right)v^2 
\nonumber\\ &\quad + \left(-\frac{439}{28}
+\frac{18}{7}\nu\right) \dot{r}^2
+ \left(-\frac{17}{21}-\frac{169}{21}\nu\right) 
\left(\frac{G m}{r}\right)\,\biggr]\,,
\end{align}\end{subequations}
while the flux contributions are (with $\mathcal{M}=M/J_\text{N}$ and $J_\text{N}=\mu\vert\bm{x}\times\bm{v}\vert$)
\begin{subequations}\label{Gdiss}
\begin{align}
\mathcal{M}_\text{2.5PN} &= \frac{8 G^2 m^2\,\nu}{5r^3}\left( 2 v^2 
+ 2\frac{G m}{r} - 3\dot{r}^2\right)\,,\\
%%%%%%%%%%%%%%%%%%%%%%%%%%%%%%%%%%%%%%%%%%%%%%%%%%%%%%%%%%%%%%%%%%%%%%
\mathcal{M}_\text{3.5PN} &= \frac{8 G^2 m^2\,\nu}{5r^3}
\biggl[ \left(\frac{307}{84}-\frac{137}{21}\nu\right)v^4 
+ \left(-\frac{58}{21}-\frac{95}{21}\nu\right)v^2\frac{G m}{r} 
+ \left(-\frac{37}{7}+\frac{277}{14}\nu\right)v^2\dot{r}^2 \nonumber\\ 
&\quad + \left(\frac{62}{7}+\frac{197}{42}\nu\right)\dot{r}^2\frac{G m}{r} 
+ \left(\frac{95}{28}-\frac{90}{7}\nu\right)\dot{r}^4 
+ \left(-\frac{745}{42}+\frac{\nu}{21}\right)
\left(\frac{G m}{r}\right)^2 \,\biggr]\,.
\end{align}\end{subequations}

We must still include the dissipative 4PN tail contributions to both fluxes.
From the expression of the corresponding acceleration in
Eq.~\eqref{acctaildiss}, we readily compute the terms in the right-hand sides of
the balance equations~\eqref{balance} as
\begin{subequations}\label{FHtaildiss}
\begin{align}
F^\text{tail}_\text{diss} &= \frac{2G^2M}{5c^8} \,I_{ij}^{(1)}(t)
\,\int_0^{+\infty} \ud\tau
\ln\left(\frac{c\tau}{2r_{12}}\right)\left[I_{ij}^{(7)}(t-\tau) +
  I_{ij}^{(7)}(t+\tau)\right]\,,\\
M^{i\,\text{tail}}_\text{diss} &= \frac{4G^2M}{5c^8} \,\epsilon_{ijk}\,I_{jl}(t)
\,\int_0^{+\infty} \ud\tau
\ln\left(\frac{c\tau}{2r_{12}}\right)\left[I_{kl}^{(7)}(t-\tau) +
  I_{kl}^{(7)}(t+\tau)\right]\,.
\end{align}
\end{subequations}
By performing some operations by parts, one could produce some total time
derivatives which could be transferred to the left-hand side of the balance equations,
where they would contribute as some dissipative 4PN terms to the energy and
angular momentum $E$ and $J$. We find, however, that such an equivalent way of
presenting our results is not so fruitful, so that we keep the
expressions~\eqref{FHtaildiss} as they are.

% There are similarly some 2.5PN, 3.5PN and 4PN dissipative effects in the
% linear momentum $\bm{P}$ and CM integral $\bm{G}$. Notably the flux of
% $\bm{P}$ starts at order 3.5PN only and describes the total recoil of the
% system.

\acknowledgments

L.Be. acknowledges financial support provided under the European Union's H2020
ERC Consolidator Grant ``Matter and strong-field gravity: New frontiers in
Einstein's theory'' grant agreement no. MaGRaTh646597.

\appendix

\section{Recap of the 4PN harmonic coordinates Lagrangian} 
\label{sec:Lgen} 

In this Appendix, we recapitulate our final result for the 4PN Lagrangian in an
arbitrary frame. The instantaneous terms up to 3PN order are well-known (see
\textit{e.g.} Eqs.~(5.2) in Ref.~\cite{BBBFMa}), but we redisplay them again
for completeness:
{\allowdisplaybreaks
\begin{subequations}\label{L3PN}
\begin{align}
L_\text{N} &= \frac{G m_1 m_2}{2 r_{12}} + \frac{m_1 v_1^2}{2} + 1
\leftrightarrow 2\,,\\
%%%%%%%%%%%%%%%%%%%%%%%%%%%%%%%%%%%%%%%%%%%%%%%%%%%%%%%%%%%%%%%%%
L_\text{1PN} &= - \frac{G^2 m_1^2 m_2}{2 r_{12}^2} + \frac{m_1
  v_1^4}{8} \nonumber \\ & + \frac{G m_1 m_2}{r_{12}} \left( -
\frac{1}{4} (n_{12}v_1) (n_{12}v_2) + \frac{3}{2} v_1^2 - \frac{7}{4}
(v_1v_2) \right) + 1 \leftrightarrow 2\,,\\
%%%%%%%%%%%%%%%%%%%%%%%%%%%%%%%%%%%%%%%%%%%%%%%%%%%%%%%%%%%%%%%%%
L_\text{2PN} &= \frac{G^3 m_1^3 m_2}{2 r_{12}^3} + \frac{19 G^3 m_1^2
  m_2^2}{8 r_{12}^3} \nonumber \\ &  + \frac{G^2 m_1^2
  m_2}{r_{12}^2} \left( \frac{7}{2} (n_{12}v_1)^2 - \frac{7}{2}
(n_{12}v_1) (n_{12}v_2) + \frac{1}{2}(n_{12}v_2)^2 + \frac{1}{4} v_1^2
- \frac{7}{4} (v_1v_2) + \frac{7}{4} v_2^2 \right) \nonumber \\ &
+ \frac{G m_1 m_2}{r_{12}} \bigg( \frac{3}{16}
(n_{12}v_1)^2 (n_{12}v_2)^2 - \frac{7}{8} (n_{12}v_2)^2 v_1^2 +
\frac{7}{8} v_1^4 + \frac{3}{4} (n_{12}v_1) (n_{12}v_2) (v_1v_2)
\nonumber \\ & \quad    - 2 v_1^2 (v_1v_2) +
\frac{1}{8} (v_1v_2)^2 + \frac{15}{16} v_1^2 v_2^2 \bigg) + \frac{m_1
  v_1^6}{16} \nonumber \\ &  + G m_1 m_2 \left( -
\frac{7}{4} (a_1 v_2) (n_{12}v_2) - \frac{1}{8} (n_{12} a_1)
(n_{12}v_2)^2 + \frac{7}{8} (n_{12} a_1) v_2^2 \right) + 1
\leftrightarrow 2\,,\\
%%%%%%%%%%%%%%%%%%%%%%%%%%%%%%%%%%%%%%%%%%%%%%%%%%%%%%%%%%%%%%%%%
L_\text{3PN} &= \frac{G^2 m_1^2 m_2}{r_{12}^2} \bigg( \frac{13}{18}
(n_{12}v_1)^4 + \frac{83}{18} (n_{12}v_1)^3 (n_{12}v_2) - \frac{35}{6}
(n_{12}v_1)^2 (n_{12}v_2)^2 - \frac{245}{24} (n_{12}v_1)^2 v_1^2
\nonumber \\ & \quad + \frac{179}{12} (n_{12}v_1) (n_{12}v_2) v_1^2 -
\frac{235}{24} (n_{12}v_2)^2 v_1^2 + \frac{373}{48} v_1^4 +
\frac{529}{24} (n_{12}v_1)^2 (v_1v_2) \nonumber \\ & \quad - \frac{97}{6}
(n_{12}v_1) (n_{12}v_2) (v_1v_2) - \frac{719}{24} v_1^2 (v_1v_2) +
\frac{463}{24} (v_1v_2)^2 - \frac{7}{24} (n_{12}v_1)^2 v_2^2 \nonumber
\\ & \quad - \frac{1}{2} (n_{12}v_1) (n_{12}v_2) v_2^2 + \frac{1}{4}
(n_{12}v_2)^2 v_2^2 + \frac{463}{48} v_1^2 v_2^2 - \frac{19}{2}
(v_1v_2) v_2^2 + \frac{45}{16} v_2^4 \bigg) \nonumber \\ & + G m_1 m_2
\bigg(\frac{3}{8} (a_1 v_2) (n_{12}v_1) (n_{12}v_2)^2 + \frac{5}{12}
(a_1 v_2) (n_{12}v_2)^3 + \frac{1}{8} (n_{12} a_1) (n_{12}v_1)
(n_{12}v_2)^3 \nonumber \\ & \quad + \frac{1}{16} (n_{12} a_1) (n_{12}v_2)^4
+ \frac{11}{4} (a_1 v_1) (n_{12}v_2) v_1^2 - (a_1 v_2) (n_{12}v_2)
v_1^2 \nonumber \\ & \quad - 2 (a_1 v_1) (n_{12}v_2) (v_1v_2) + \frac{1}{4}
(a_1 v_2) (n_{12}v_2) (v_1v_2) \nonumber \\ & \quad + \frac{3}{8} (n_{12}
a_1) (n_{12}v_2)^2 (v_1v_2) - \frac{5}{8} (n_{12} a_1) (n_{12}v_1)^2
v_2^2 + \frac{15}{8} (a_1 v_1) (n_{12}v_2) v_2^2 \nonumber \\ & \quad -
\frac{15}{8} (a_1 v_2) (n_{12}v_2) v_2^2 - \frac{1}{2} (n_{12} a_1)
(n_{12}v_1) (n_{12}v_2) v_2^2 \nonumber \\ & \quad - \frac{5}{16} (n_{12}
a_1) (n_{12}v_2)^2 v_2^2 \bigg) + \frac{5m_1 v_1^8}{128} \nonumber
\\ & + \frac{G^2 m_1^2 m_2}{r_{12}} \bigg( - \frac{235}{24} (a_2 v_1)
(n_{12}v_1) - \frac{29}{24} (n_{12} a_2) (n_{12}v_1)^2 -
\frac{235}{24} (a_1 v_2) (n_{12}v_2) \nonumber \\ & \quad - \frac{17}{6}
(n_{12} a_1) (n_{12}v_2)^2 + \frac{185}{16} (n_{12} a_1) v_1^2 -
\frac{235}{48} (n_{12} a_2) v_1^2 \nonumber \\ & \quad - \frac{185}{8}
(n_{12} a_1) (v_1v_2) + \frac{20}{3} (n_{12} a_1) v_2^2 \bigg)
\nonumber \\ & + \frac{G m_1 m_2}{r_{12}} \bigg( - \frac{5}{32}
(n_{12}v_1)^3 (n_{12}v_2)^3 + \frac{1}{8} (n_{12}v_1) (n_{12}v_2)^3
v_1^2 + \frac{5}{8} (n_{12}v_2)^4 v_1^2 \nonumber \\ & \quad - \frac{11}{16}
(n_{12}v_1) (n_{12}v_2) v_1^4 + \frac{1}{4} (n_{12}v_2)^2 v_1^4 +
\frac{11}{16} v_1^6 \nonumber \\ & \quad - \frac{15}{32} (n_{12}v_1)^2
(n_{12}v_2)^2 (v_1v_2) + (n_{12}v_1) (n_{12}v_2) v_1^2 (v_1v_2)
\nonumber \\ & \quad + \frac{3}{8} (n_{12}v_2)^2 v_1^2 (v_1v_2) -
\frac{13}{16} v_1^4 (v_1v_2) + \frac{5}{16} (n_{12}v_1) (n_{12}v_2)
(v_1v_2)^2 \nonumber \\ & \quad + \frac{1}{16} (v_1v_2)^3 - \frac{5}{8}
(n_{12}v_1)^2 v_1^2 v_2^2 - \frac{23}{32} (n_{12}v_1) (n_{12}v_2)
v_1^2 v_2^2 + \frac{1}{16} v_1^4 v_2^2 \nonumber \\ & \quad - \frac{1}{32}
v_1^2 (v_1v_2) v_2^2 \bigg) \nonumber \\ & - \frac{3 G^4 m_1^4 m_2}{8
  r_{12}^4} + \frac{G^4 m_1^3 m_2^2}{r_{12}^4} \left( -
\frac{9707}{420} + \frac{22}{3} \ln \left(\frac{r_{12}}{r'_1} \right)
\right) \nonumber \\ & + \frac{G^3 m_1^2 m_2^2}{r_{12}^3} \bigg(
\frac{383}{24} (n_{12}v_1)^2 - \frac{889}{48} (n_{12}v_1) (n_{12}v_2)
\nonumber \\ & \quad - \frac{123}{64} (n_{12}v_1)(n_{12}v_{12}) \pi^2 -
\frac{305}{72} v_1^2 + \frac{41}{64} \pi^2 (v_1v_{12}) +
\frac{439}{144} (v_1v_2) \bigg) \nonumber \\ & + \frac{G^3 m_1^3
  m_2}{r_{12}^3} \bigg( - \frac{8243}{210} (n_{12}v_1)^2 +
\frac{15541}{420} (n_{12}v_1) (n_{12}v_2) + \frac{3}{2} (n_{12}v_2)^2
+ \frac{15611}{1260} v_1^2 \nonumber \\ & \quad - \frac{17501}{1260}
(v_1v_2) + \frac{5}{4} v_2^2 + 22 (n_{12}v_1)(n_{12}v_{12}) \ln \left(
\frac{r_{12}}{r'_1} \right) \nonumber \\ & \quad - \frac{22}{3} (v_1v_{12})
\ln \left( \frac{r_{12}}{r'_1} \right) \bigg) + 1 \leftrightarrow 2\,.
\end{align}
\end{subequations}}\noindent
The logarithms in harmonic coordinates contain the two gauge constants $r'_A$.
Note the appearance of the accelerations at 2PN order.

The 4PN terms have been given in Eqs.~(5.6) of Ref.~\cite{BBBFMa}, but some
quartic ($\propto G^4$) terms there have been later corrected in the Appendix
of Ref.~\cite{BBBFMb} [see Eq.~(A3)]. In fact, for convenience, we use in the
present paper a version of the 4PN Lagrangian that differs from the
one of Refs.~\cite{BBBFMa, BBBFMb} by some unphysical shift (see below). We
thus provide the full 4PN part of the harmonic-coordinates Lagrangian,
following the convention~\eqref{notationQG}:
{\allowdisplaybreaks
\begin{subequations}\label{result4PN}
\begin{align}
L_\text{4PN}^{(0)}&= \frac{7}{256} m_{1} v_1^{10}
 + 1 \leftrightarrow
2\,,\\
%%%%%%%%%%%%%%%%%%%%%%%%%%%%%%%%%%%%%%%%%%%%%%%%%%%%%%%%%%%%%%%%%
L_\text{4PN}^{(1)}&= m_{1} m_{2} \left(\frac{13}{64} (a_2 v_1) (n_{12} v_1)^5
 + \frac{5}{128} (a_2 n_{12}) (n_{12} v_1)^6
 -  \frac{13}{64} (n_{12} v_1)^5 (a_2 v_2)\right.\nonumber\\
& \quad + \frac{11}{64} (a_2 v_1) (n_{12} v_1)^4 (n_{12} v_2)
 + \frac{5}{64} (a_2 n_{12}) (n_{12} v_1)^5 (n_{12} v_2)\nonumber\\
& \quad + \frac{5}{32} (a_2 v_1) (n_{12} v_1)^3 (n_{12} v_2)^2
 -  \frac{5}{32} (a_1 n_{12}) (n_{12} v_1)^3 (n_{12} v_2)^3\nonumber\\
& \quad -  \frac{1}{16} (a_2 v_1) (n_{12} v_1)^3 (v_1 v_2)
 + \frac{11}{64} (a_2 n_{12}) (n_{12} v_1)^4 (v_1 v_2)\nonumber\\
& \quad + \frac{5}{16} (a_2 n_{12}) (n_{12} v_1)^3 (n_{12} v_2) (v_1 v_2)
 + \frac{5}{16} (n_{12} v_1)^2 (a_1 v_2) (n_{12} v_2) (v_1 v_2)\nonumber\\
& \quad -  \frac{3}{16} (a_2 v_1) (n_{12} v_1) (v_1 v_2)^2
 + \frac{1}{16} (a_1 v_1) (n_{12} v_2) (v_1 v_2)^2\nonumber\\
& \quad + \frac{5}{16} (a_1 n_{12}) (n_{12} v_1) (n_{12} v_2) (v_1 v_2)^2
 -  \frac{77}{96} (a_2 v_1) (n_{12} v_1)^3 v_1^{2}
 -  \frac{27}{128} (a_2 n_{12}) (n_{12} v_1)^4 v_1^{2}\nonumber\\
& \quad + \frac{77}{96} (n_{12} v_1)^3 (a_2 v_2) v_1^{2}
 -  \frac{13}{32} (a_2 v_1) (n_{12} v_1)^2 (n_{12} v_2) v_1^{2}
 -  \frac{11}{32} (a_2 n_{12}) (n_{12} v_1)^3 (n_{12} v_2) v_1^{2}\nonumber\\
& \quad -  \frac{7}{32} (a_2 v_1) (n_{12} v_1) (n_{12} v_2)^2 v_1^{2}
 -  \frac{27}{64} (a_2 n_{12}) (n_{12} v_1)^2 (n_{12} v_2)^2 v_1^{2}
 -  \frac{19}{32} (a_1 v_1) (n_{12} v_2)^3 v_1^{2}\nonumber\\
& \quad -  \frac{3}{16} (a_2 v_1) (n_{12} v_1) (v_1 v_2) v_1^{2}
 -  \frac{13}{32} (a_2 n_{12}) (n_{12} v_1)^2 (v_1 v_2) v_1^{2}\nonumber\\
& \quad + \frac{33}{16} (n_{12} v_1) (a_2 v_2) (v_1 v_2) v_1^{2}
 -  \frac{7}{16} (a_2 n_{12}) (n_{12} v_1) (n_{12} v_2) (v_1 v_2) v_1^{2}\nonumber\\
& \quad + \frac{1}{16} (a_1 v_2) (n_{12} v_2) (v_1 v_2) v_1^{2}
 + \frac{123}{64} (a_2 v_1) (n_{12} v_1) v_1^{4}
 + \frac{53}{128} (a_2 n_{12}) (n_{12} v_1)^2 v_1^{4}\nonumber\\
& \quad -  \frac{123}{64} (n_{12} v_1) (a_2 v_2) v_1^{4}
 + \frac{49}{64} (a_2 v_1) (n_{12} v_2) v_1^{4}
 + \frac{31}{64} (a_2 n_{12}) (n_{12} v_1) (n_{12} v_2) v_1^{4}\nonumber\\
& \quad + \frac{49}{64} (a_2 n_{12}) (v_1 v_2) v_1^{4}
 -  \frac{75}{128} (a_2 n_{12}) v_1^{6}
 + \frac{17}{96} (n_{12} v_1)^3 (a_1 v_2) v_2^{2}\nonumber\\
& \quad -  \frac{23}{32} (a_1 v_1) (n_{12} v_1)^2 (n_{12} v_2) v_2^{2}
 + \frac{15}{32} (a_1 n_{12}) (n_{12} v_1)^3 (n_{12} v_2) v_2^{2}\nonumber\\
& \quad + \frac{17}{32} (a_1 n_{12}) (n_{12} v_1)^2 (v_1 v_2) v_2^{2}
 + \frac{33}{32} (a_2 v_1) (n_{12} v_1) v_1^{2} v_2^{2}
 + \frac{93}{32} (a_1 v_1) (n_{12} v_2) v_1^{2} v_2^{2}\nonumber\\
& \quad \left.  -  \frac{23}{32} (a_1 n_{12}) (n_{12} v_1) 
(n_{12} v_2) v_1^{2} v_2^{2}\right)\nonumber\\
 &+ \frac{m_1 m_2}{r_{12}} \left(\frac{35}{128} (n_{12} v_1)^5 (n_{12} v_2)^3
 -  \frac{35}{256} (n_{12} v_1)^4 (n_{12} v_2)^4\right.\nonumber\\
& \quad -  \frac{15}{128} (n_{12} v_1)^4 (n_{12} v_2)^2 (v_1 v_2)
 + \frac{15}{32} (n_{12} v_1)^3 (n_{12} v_2)^3 (v_1 v_2)\nonumber\\
& \quad -  \frac{15}{32} (n_{12} v_1)^3 (n_{12} v_2) (v_1 v_2)^2
 + \frac{5}{32} (n_{12} v_1)^2 (v_1 v_2)^3
 -  \frac{3}{16} (n_{12} v_1) (n_{12} v_2) (v_1 v_2)^3\nonumber\\
& \quad + \frac{1}{32} (v_1 v_2)^4
 -  \frac{5}{32} (n_{12} v_1)^3 (n_{12} v_2)^3 v_1^{2}
 -  \frac{5}{16} (n_{12} v_1) (n_{12} v_2)^3 (v_1 v_2) v_1^{2}\nonumber\\
& \quad + \frac{9}{32} (n_{12} v_1) (n_{12} v_2) (v_1 v_2)^2 v_1^{2}
 -  \frac{15}{32} (n_{12} v_2)^2 (v_1 v_2)^2 v_1^{2}
 + \frac{1}{32} (v_1 v_2)^3 v_1^{2}\nonumber\\
& \quad + \frac{57}{128} (n_{12} v_1) (n_{12} v_2)^3 v_1^{4}
 -  \frac{15}{128} (n_{12} v_2)^4 v_1^{4}
 + \frac{39}{128} (n_{12} v_2)^2 (v_1 v_2) v_1^{4}
 + \frac{3}{4} (v_1 v_2)^2 v_1^{4}\nonumber\\
& \quad -  \frac{11}{32} (n_{12} v_2)^2 v_1^{6}
 -  \frac{5}{4} (v_1 v_2) v_1^{6}
 + \frac{75}{128} v_1^{8}
 -  \frac{75}{128} (n_{12} v_1)^5 (n_{12} v_2) v_2^{2}\nonumber\\
& \quad -  \frac{53}{128} (n_{12} v_1)^4 (v_1 v_2) v_2^{2}
 + \frac{99}{64} (n_{12} v_1)^3 (n_{12} v_2) v_1^{2} v_2^{2}
 -  \frac{21}{64} (n_{12} v_1)^2 (n_{12} v_2)^2 v_1^{2} v_2^{2}\nonumber\\
& \quad + \frac{11}{64} (n_{12} v_1)^2 (v_1 v_2) v_1^{2} v_2^{2}
 + \frac{35}{32} (n_{12} v_1) (n_{12} v_2) (v_1 v_2) v_1^{2} v_2^{2}
 -  \frac{1}{32} (v_1 v_2)^2 v_1^{2} v_2^{2}\nonumber\\
& \quad -  \frac{185}{128} (n_{12} v_1) (n_{12} v_2) v_1^{4} v_2^{2}
 + \frac{23}{64} (n_{12} v_2)^2 v_1^{4} v_2^{2}
 -  \frac{99}{128} (v_1 v_2) v_1^{4} v_2^{2}
 + \frac{5}{8} v_1^{6} v_2^{2}\nonumber\\
&\quad \left. + \frac{3}{256} v_1^{4} v_2^{4}\right)
 + 1 \leftrightarrow
2\,,\\
%%%%%%%%%%%%%%%%%%%%%%%%%%%%%%%%%%%%%%%%%%%%%%%%%%%%%%%%%%%%%%%%%
L_\text{4PN}^{(2)}&= \frac{m_{1}^2 m_{2}}{r_{12}} 
\left(\frac{2099}{288} (a_1 v_1) (n_{12} v_1)^3
 + \frac{3341}{480} (a_2 v_1) (n_{12} v_1)^3
 + \frac{59}{180} (n_{12} v_1)^3 (a_2 v_2)\right.\nonumber\\
& \quad + \frac{2197}{240} (a_1 n_{12}) (n_{12} v_1)^3 (n_{12} v_2)
 + \frac{6661}{720} (a_2 n_{12}) (n_{12} v_1)^3 (n_{12} v_2)\nonumber\\
& \quad + \frac{10223}{480} (n_{12} v_1)^2 (a_1 v_2) (n_{12} v_2)
 + \frac{3059}{96} (a_1 v_1) (n_{12} v_1) (n_{12} v_2)^2\nonumber\\
& \quad -  \frac{3781}{160} (n_{12} v_1) (a_1 v_2) (n_{12} v_2)^2
 + \frac{1337}{240} (a_1 n_{12}) (n_{12} v_1) (n_{12} v_2)^3
 + \frac{4621}{480} (a_1 v_2) (n_{12} v_2)^3\nonumber\\
& \quad + \frac{1133}{960} (a_1 n_{12}) (n_{12} v_2)^4
 + \frac{3613}{48} (a_1 v_1) (n_{12} v_1) (v_1 v_2)
 + \frac{4529}{240} (a_2 v_1) (n_{12} v_1) (v_1 v_2)\nonumber\\
& \quad + \frac{2099}{96} (a_1 n_{12}) (n_{12} v_1)^2 (v_1 v_2)
 -  \frac{23}{60} (a_1 n_{12}) (n_{12} v_1) (n_{12} v_2) (v_1 v_2)\nonumber\\
& \quad + \frac{6499}{240} (a_1 v_2) (n_{12} v_2) (v_1 v_2)
 + \frac{247}{30} (a_1 n_{12}) (n_{12} v_2)^2 (v_1 v_2)
 + \frac{2503}{96} (a_1 n_{12}) (v_1 v_2)^2\nonumber\\
& \quad + \frac{7193}{240} (a_2 v_1) (n_{12} v_1) v_1^{2}
 + \frac{3021}{320} (a_2 n_{12}) (n_{12} v_1)^2 v_1^{2}
 + \frac{4723}{96} (n_{12} v_1) (a_1 v_2) v_1^{2}\nonumber\\
& \quad + \frac{3679}{480} (n_{12} v_1) (a_2 v_2) v_1^{2}
 + \frac{13549}{480} (a_1 v_1) (n_{12} v_2) v_1^{2}
 + \frac{8849}{480} (a_2 v_1) (n_{12} v_2) v_1^{2}\nonumber\\
& \quad + \frac{2063}{96} (a_1 n_{12}) (n_{12} v_1) (n_{12} v_2) v_1^{2}
 + \frac{166}{15} (a_2 n_{12}) (n_{12} v_1) (n_{12} v_2) v_1^{2}\nonumber\\
& \quad + \frac{4621}{320} (a_2 n_{12}) (n_{12} v_2)^2 v_1^{2}
 + \frac{8849}{480} (a_2 n_{12}) (v_1 v_2) v_1^{2}
 + \frac{6943}{384} (a_1 n_{12}) v_1^{4}\nonumber\\
& \quad + \frac{2293}{160} (n_{12} v_1) (a_1 v_2) v_2^{2}
 + \frac{3733}{160} (a_1 v_1) (n_{12} v_2) v_2^{2}
 + \frac{7}{5} (a_1 n_{12}) (n_{12} v_1) (n_{12} v_2) v_2^{2}\nonumber\\
& \quad -  \frac{3733}{160} (a_1 v_2) (n_{12} v_2) v_2^{2}
 -  \frac{139}{20} (a_1 n_{12}) (n_{12} v_2)^2 v_2^{2}
 -  \frac{5593}{240} (a_1 n_{12}) (v_1 v_2) v_2^{2}\nonumber\\
&\quad \left. + \frac{3613}{192} (a_1 n_{12}) v_1^{2} v_2^{2}
 + \frac{2931}{320} (a_1 n_{12}) v_2^{4}\right) \nonumber\\
 %%%%%%%%%%%%%%%%%%%%%%%%%%%%%%%%%%
& + \frac{m_{1}^2 m_{2}}{r_{12}^{2}} \left(- \frac{4027}{800} (n_{12} v_1)^6 
+ \frac{3227}{800} (n_{12} v_1)^5 (n_{12} v_2)
 -  \frac{6301}{240} (n_{12} v_1)^4 (n_{12} v_2)^2 \right.\nonumber\\
& \quad + \frac{6661}{240} (n_{12} v_1)^3 (n_{12} v_2)^3 
 -  \frac{2221}{64} (n_{12} v_1)^4 (v_1 v_2)
 + \frac{25267}{720} (n_{12} v_1)^3 (n_{12} v_2) (v_1 v_2)\nonumber\\
& \quad -  \frac{6661}{480} (n_{12} v_1)^2 (n_{12} v_2)^2 (v_1 v_2)
 -  \frac{23401}{480} (n_{12} v_1)^2 (v_1 v_2)^2\nonumber\\
& \quad + \frac{4529}{240} (n_{12} v_1) (n_{12} v_2) (v_1 v_2)^2
 -  \frac{8369}{480} (v_1 v_2)^3
 + \frac{17393}{2880} (n_{12} v_1)^4 v_1^{2}\nonumber\\
& \quad -  \frac{26237}{720} (n_{12} v_1)^3 (n_{12} v_2) v_1^{2}
 + \frac{4561}{160} (n_{12} v_1)^2 (n_{12} v_2)^2 v_1^{2}
 + \frac{691}{240} (n_{12} v_1) (n_{12} v_2)^3 v_1^{2}\nonumber\\
& \quad + \frac{4621}{240} (n_{12} v_2)^4 v_1^{2}
 -  \frac{14987}{960} (n_{12} v_1)^2 (v_1 v_2) v_1^{2}
 + \frac{2601}{160} (n_{12} v_1) (n_{12} v_2) (v_1 v_2) v_1^{2}\nonumber\\
& \quad + \frac{14649}{320} (n_{12} v_2)^2 (v_1 v_2) v_1^{2}
 + \frac{97}{5} (v_1 v_2)^2 v_1^{2}
 -  \frac{4879}{320} (n_{12} v_1)^2 v_1^{4}\nonumber\\
& \quad + \frac{5399}{192} (n_{12} v_1) (n_{12} v_2) v_1^{4}
 + \frac{83}{15} (n_{12} v_2)^2 v_1^{4}
 + \frac{749}{128} (v_1 v_2) v_1^{4}
 + \frac{20389}{1920} v_1^{6}\nonumber\\
& \quad + \frac{107}{180} (n_{12} v_1)^4 v_2^{2}
 -  \frac{1823}{240} (n_{12} v_1)^3 (n_{12} v_2) v_2^{2}
 -  \frac{1}{2} (n_{12} v_1)^2 (n_{12} v_2)^2 v_2^{2}\nonumber\\
& \quad + \frac{1021}{120} (n_{12} v_1)^2 (v_1 v_2) v_2^{2}
 + \frac{1}{2} (n_{12} v_1) (n_{12} v_2) (v_1 v_2) v_2^{2}
 + \frac{67}{4} (v_1 v_2)^2 v_2^{2}\nonumber\\
& \quad -  \frac{21709}{960} (n_{12} v_1)^2 v_1^{2} v_2^{2}
 -  \frac{1873}{480} (n_{12} v_1) (n_{12} v_2) v_1^{2} v_2^{2}
 -  \frac{4621}{320} (n_{12} v_2)^2 v_1^{2} v_2^{2}\nonumber\\
& \quad -  \frac{42017}{960} (v_1 v_2) v_1^{2} v_2^{2}
 + \frac{11119}{960} v_1^{4} v_2^{2}
 -  \frac{21}{8} (n_{12} v_1)^2 v_2^{4}
 -  \frac{3}{8} (n_{12} v_1) (n_{12} v_2) v_2^{4}\nonumber\\
&\quad \left. + \frac{3}{16} (n_{12} v_2)^2 v_2^{4}
 -  \frac{105}{8} (v_1 v_2) v_2^{4}
 + \frac{105}{16} v_1^{2} v_2^{4}
 + \frac{115}{32} v_2^{6}\right)
 + 1 \leftrightarrow
2\,,\\
%%%%%%%%%%%%%%%%%%%%%%%%%%%%%%%%%%%%%%%%%%%%%%%%%%%%%%%%%%%%%%%%%
L_\text{4PN}^{(3)}&= \frac{m_{1}^2 m_{2}^2}{r_{12}^{2}} 
\left(- \frac{1099}{144} (a_2 v_1) (n_{12} v_1)
 + \frac{41}{64} \pi^2 (a_2 v_1) (n_{12} v_1)
 + \frac{2005}{96} (a_2 n_{12}) (n_{12} v_1)^2 \right. \nonumber\\
& \quad -  \frac{123}{128} \pi^2 (a_2 n_{12}) (n_{12} v_1)^2
 + \frac{225233}{1800} (a_1 v_1) (n_{12} v_2)
 -  \frac{43}{64} \pi^2 (a_1 v_1) (n_{12} v_2)\nonumber\\
& \quad -  \frac{477941}{3600} (a_1 n_{12}) (v_1 v_2)
 + \frac{21}{16} \pi^2 (a_1 n_{12}) (v_1 v_2)
 + \frac{477941}{7200} (a_1 n_{12}) v_1^{2}\nonumber\\
&\quad \left. -  \frac{21}{32} \pi^2 (a_1 n_{12}) v_1^{2}\right) \nonumber\\
%%%%%%%%%%%%%
& + \frac{m_{1}^2 m_{2}^2}{r_{12}^{3}} \left(- \frac{173617}{2880} (n_{12} v_1)^4
 -  \frac{2155}{1024} \pi^2 (n_{12} v_1)^4\right.\nonumber\\
& \quad + \frac{173587}{720} (n_{12} v_1)^3 (n_{12} v_2)
 + \frac{2155}{256} \pi^2 (n_{12} v_1)^3 (n_{12} v_2)
 -  \frac{85871}{480} (n_{12} v_1)^2 (n_{12} v_2)^2\nonumber\\
& \quad -  \frac{6465}{1024} \pi^2 (n_{12} v_1)^2 (n_{12} v_2)^2
 + \frac{5651}{300} (n_{12} v_1)^2 (v_1 v_2)
 -  \frac{939}{256} \pi^2 (n_{12} v_1)^2 (v_1 v_2)\nonumber\\
& \quad -  \frac{6851}{300} (n_{12} v_1) (n_{12} v_2) (v_1 v_2)
 + \frac{939}{256} \pi^2 (n_{12} v_1) (n_{12} v_2) (v_1 v_2)
 + \frac{49139}{720} (v_1 v_2)^2\nonumber\\
& \quad -  \frac{195}{512} \pi^2 (v_1 v_2)^2
 -  \frac{3677}{1200} (n_{12} v_1)^2 v_1^{2}
 + \frac{447}{512} \pi^2 (n_{12} v_1)^2 v_1^{2}
 -  \frac{222679}{1200} (n_{12} v_1) (n_{12} v_2) v_1^{2}\nonumber\\
& \quad -  \frac{189}{256} \pi^2 (n_{12} v_1) (n_{12} v_2) v_1^{2}
 + \frac{153079}{800} (n_{12} v_2)^2 v_1^{2}
 -  \frac{69}{512} \pi^2 (n_{12} v_2)^2 v_1^{2}\nonumber\\
& \quad -  \frac{61733}{900} (v_1 v_2) v_1^{2}
 -  \frac{55}{256} \pi^2 (v_1 v_2) v_1^{2}
 + \frac{10337}{320} v_1^{4}
 + \frac{133}{1024} \pi^2 v_1^{4}
 -  \frac{116123}{3600} v_1^{2} v_2^{2}\nonumber\\
&\quad \left. + \frac{477}{1024} \pi^2 v_1^{2} v_2^{2}\right) \nonumber\\
%%%%%%%%%%%%%%%%%%%%%
& +  \frac{m_{1}^3 m_{2}}{r_{12}^{2}} \left(\frac{44023}{720} (a_1 v_1) (n_{12} v_1)
 + \frac{562}{9} (a_2 v_1) (n_{12} v_1)\right.\nonumber\\
& \quad + 44 \ln\Big(\frac{r_{12}}{r'_{1}}\Big) (n_{12} v_1) (a_1 v_2)
 - 44 \ln\Big(\frac{r_{12}}{r'_{1}}\Big) (n_{12} v_1) (a_2 v_2)
 + \frac{110}{3} \ln\Big(\frac{r_{12}}{r'_{1}}\Big) (a_1 v_1) (n_{12} v_2)\nonumber\\
& \quad + \frac{6397}{75} (a_1 n_{12}) (n_{12} v_1) (n_{12} v_2)
 + \frac{198097}{4200} (a_2 n_{12}) (n_{12} v_1) (n_{12} v_2)\nonumber\\
& \quad + 22 \ln\Big(\frac{r_{12}}{r'_{1}}\Big) (a_2 n_{12}) (n_{12} v_1) (n_{12} v_2)
 + \frac{14377}{280} (a_1 v_2) (n_{12} v_2)\nonumber\\
& \quad -  \frac{110}{3} \ln\Big(\frac{r_{12}}{r'_{1}}\Big) (a_1 v_2) (n_{12} v_2)
 + \frac{44023}{720} (a_1 n_{12}) (v_1 v_2)
 - 44 \ln\Big(\frac{r_{12}}{r'_{1}}\Big) (a_1 n_{12}) (v_1 v_2)\nonumber\\
&\quad \left. + 44 \ln\Big(\frac{r_{12}}{r'_{1}}\Big) (a_1 n_{12}) v_1^{2}
 + \frac{14377}{560} (a_2 n_{12}) v_1^{2}
 + \frac{937}{1440} (a_1 n_{12}) v_2^{2}\right)\nonumber\\
 %%%%%%%%%%%%%%%%%%%%%%%
& + \frac{m_{1}^3 m_{2}}{r_{12}^{3}} \left(- \frac{30313}{360} (n_{12} v_1)^4
 + \frac{64001}{720} (n_{12} v_1)^3 (n_{12} v_2)
 -  \frac{185917}{1680} (n_{12} v_1)^2 (n_{12} v_2)^2\right.\nonumber\\
& \quad - 55 \ln\Big(\frac{r_{12}}{r'_{1}}\Big) (n_{12} v_1)^2 (n_{12} v_2)^2
 + \frac{179617}{1680} (n_{12} v_1) (n_{12} v_2)^3
 + 55 \ln\Big(\frac{r_{12}}{r'_{1}}\Big) (n_{12} v_1) (n_{12} v_2)^3\nonumber\\
& \quad -  \frac{94667}{400} (n_{12} v_1)^2 (v_1 v_2)
 + \frac{338099}{1400} (n_{12} v_1) (n_{12} v_2) (v_1 v_2)\nonumber\\
& \quad + 22 \ln\Big(\frac{r_{12}}{r'_{1}}\Big) (n_{12} v_1) (n_{12} v_2) (v_1 v_2)
 -  \frac{214897}{8400} (n_{12} v_2)^2 (v_1 v_2)\nonumber\\
& \quad - 11 \ln\Big(\frac{r_{12}}{r'_{1}}\Big) (n_{12} v_2)^2 (v_1 v_2)
 -  \frac{737}{18} (v_1 v_2)^2
 + \frac{903589}{16800} (n_{12} v_1)^2 v_1^{2}\nonumber\\
& \quad - 55 \ln\Big(\frac{r_{12}}{r'_{1}}\Big) (n_{12} v_1)^2 v_1^{2}
 -  \frac{96287}{1120} (n_{12} v_1) (n_{12} v_2) v_1^{2}
 + 55 \ln\Big(\frac{r_{12}}{r'_{1}}\Big) (n_{12} v_1) (n_{12} v_2) v_1^{2}\nonumber\\
& \quad + \frac{426731}{4200} (n_{12} v_2)^2 v_1^{2}
 + 11 \ln\Big(\frac{r_{12}}{r'_{1}}\Big) (n_{12} v_2)^2 v_1^{2}
 + \frac{202687}{3360} (v_1 v_2) v_1^{2}\nonumber\\
& \quad -  \frac{55}{3} \ln\Big(\frac{r_{12}}{r'_{1}}\Big) (v_1 v_2) v_1^{2}
 + \frac{22769}{2016} v_1^{4}
 + \frac{55}{3} \ln\Big(\frac{r_{12}}{r'_{1}}\Big) v_1^{4}
 -  \frac{177}{8} (n_{12} v_1)^2 v_2^{2}\nonumber\\
& \quad + 66 \ln\Big(\frac{r_{12}}{r'_{1}}\Big) (n_{12} v_1)^2 v_2^{2}
 -  \frac{120397}{4200} (n_{12} v_1) (n_{12} v_2) v_2^{2}
 - 88 \ln\Big(\frac{r_{12}}{r'_{1}}\Big) (n_{12} v_1) (n_{12} v_2) v_2^{2}\nonumber\\
& \quad + \frac{7}{4} (n_{12} v_2)^2 v_2^{2}
 -  \frac{43}{2} (v_1 v_2) v_2^{2}
 + 22 \ln\Big(\frac{r_{12}}{r'_{1}}\Big) (v_1 v_2) v_2^{2}
 -  \frac{8357}{560} v_1^{2} v_2^{2}
 - 22 \ln\Big(\frac{r_{12}}{r'_{1}}\Big) v_1^{2} v_2^{2}\nonumber\\
&\quad \left. + \frac{91}{16} v_2^{4}\right)
 + 1 \leftrightarrow
2\,,\\
%%%%%%%%%%%%%%%%%%%%%%%%%%%%%%%%%%%%%%%%%%%%%%%%%%%%%%%%%%%%%%%%%
L_\text{4PN}^{(4)}&= \frac{m_{1}^4 m_{2}}{r_{12}^4} 
\left(\frac{282629}{900} (n_{12} v_1)^2
 -  \frac{880}{3} \ln\Big(\frac{r_{12}}{r'_{1}}\Big) (n_{12} v_1)^2
 -  \frac{283979}{900} (n_{12} v_1) (n_{12} v_2)\right.\nonumber\\
& \quad + \frac{880}{3} \ln\Big(\frac{r_{12}}{r'_{1}}\Big) (n_{12} v_1) (n_{12} v_2)
 + \frac{9}{4} (n_{12} v_2)^2
 + \frac{208529}{3600} (v_1 v_2)
 -  \frac{220}{3} \ln\Big(\frac{r_{12}}{r'_{1}}\Big) (v_1 v_2)\nonumber\\
&\quad \left. -  \frac{211229}{3600} v_1^{2}
 + \frac{220}{3} \ln\Big(\frac{r_{12}}{r'_{1}}\Big) v_1^{2}
 + \frac{15}{16} v_2^{2}\right) \nonumber \\
 %%%%%%%%%%%%%%%%%%%%%
& + \frac{m_{1}^3 m_{2}^2}{r_{12}^4} \left(- \frac{1268557}{50400} (n_{12} v_1)^2  
+ \frac{659}{96} \pi^2 (n_{12} v_1)^2
 \right.\nonumber\\
& \quad -  \frac{286}{3} \ln\Big(\frac{r_{12}}{r'_{1}}\Big) (n_{12} v_1)^2
 + \frac{11530469}{25200} (n_{12} v_1) (n_{12} v_2) \nonumber\\
& \quad -  \frac{1715}{48} \pi^2 (n_{12} v_1) (n_{12} v_2)
 + 44 \ln\Big(\frac{r_{12}}{r'_{1}}\Big) (n_{12} v_1) (n_{12} v_2)
 + 64 \ln\Big(\frac{r_{12}}{r'_{2}}\Big) (n_{12} v_1) (n_{12} v_2)\nonumber\\
& \quad -  \frac{2233689}{5600} (n_{12} v_2)^2
 + \frac{2771}{96} \pi^2 (n_{12} v_2)^2
 + \frac{110}{3} \ln\Big(\frac{r_{12}}{r'_{1}}\Big) (n_{12} v_2)^2
 - 64 \ln\Big(\frac{r_{12}}{r'_{2}}\Big) (n_{12} v_2)^2\nonumber\\
& \quad -  \frac{959797}{8400} (v_1 v_2)
 + \frac{103}{16} \pi^2 (v_1 v_2)
 -  \frac{154}{3} \ln\Big(\frac{r_{12}}{r'_{1}}\Big) (v_1 v_2)
 - 16 \ln\Big(\frac{r_{12}}{r'_{2}}\Big) (v_1 v_2)\nonumber\\
& \quad + \frac{858533}{50400} v_1^{2}
 -  \frac{15}{32} \pi^2 v_1^{2}
 + \frac{121}{3} \ln\Big(\frac{r_{12}}{r'_{1}}\Big) v_1^{2}
 + \frac{5482669}{50400} v_2^{2}
 -  \frac{191}{32} \pi^2 v_2^{2}
 + \frac{22}{3} \ln\Big(\frac{r_{12}}{r'_{1}}\Big) v_2^{2}\nonumber\\
&\quad \left. + 16 \ln\Big(\frac{r_{12}}{r'_{2}}\Big) v_2^{2}\right)
 + 1 \leftrightarrow
2\,,\\
%%%%%%%%%%%%%%%%%%%%%%%%%%%%%%%%%%%%%%%%%%%%%%%%%%%%%%%%%%%%%%%%%
L_\text{4PN}^{(5)}&= \frac{3}{8} \frac{m_{1}^5 m_{2}}{r_{12}^5}
 + \frac{m_{1}^3 m_{2}^3}{r_{12}^5} \left(\frac{597771}{5600}
 -  \frac{71}{32} \pi^2
 -  \frac{110}{3} \ln\Big(\frac{r_{12}}{r'_{1}}\Big)\right)
 \nonumber\\
& + \frac{m_{1}^4 m_{2}^2}{r_{12}^5} \left(\frac{1734977}{25200}
 + \frac{105}{32} \pi^2 -  \frac{242}{3} \ln\Big(\frac{r_{12}}{r'_{1}}\Big)
 - 16 \ln\Big(\frac{r_{12}}{r'_{2}}\Big)\right) + 1 \leftrightarrow
2\,.
\end{align}
\end{subequations}}\noindent
Besides all previous instantaneous terms, there is also the
non-local tail term given by Eq.~\eqref{Ltail} which is to be added. The Lagrangian ~\eqref{L3PN}--\eqref{result4PN} is manifestly invariant under global Lorentz-Poincar\'e transformations.

Again, in the present paper, we have adopted (somewhat arbitrarily\footnote{We
  have tried to minimize the number of operations (\textit{e.g.}, the length of the shift) with
  respect to our initial, ``brute'' calculation in Ref.~\cite{BBBFMa}.}) a Lagrangian
that is slightly changed with respect to the Lagrangian published in
Refs.~\cite{BBBFMa, BBBFMb}. Of course, the dynamics is equivalent since
the Lagrangian~\eqref{L3PN}--\eqref{result4PN} differs from the one
in~\cite{BBBFMa, BBBFMb} by terms that come from some unphysical shifts
$\bm{\eta}_A$ at the 4PN order (plus a total time derivative). These shifts
are made of $G^3$ and $G^4$ contributions only. They are explicitly given by
(see Ref.~\cite{BBBFMa} for our conventions regarding shifts)
\begin{subequations}\label{shifteta}
\begin{align}
\bm{\eta}^{(3)}_{1\,\text{4PN}}={}
&\frac{\bm{v}_{12}}{r_{12}^2} \Bigl(\frac{769}{24} m_{1}^2 m_{2} (n_{12} v_{12})
 + \frac{561}{35} m_{1} m_{2}^2 (n_{12} v_{12})\Bigr)
 + \frac{\bm{n}_{12}}{r_{12}^2} \biggl[m_{1} m_{2}^2 \Bigl(\frac{21719}{1400} (n_{12} v_{12})^2
 -  \frac{2096}{175} v_{12}^{2}\Bigr)\nonumber\\
& + m_{1}^2 m_{2} \Bigl(- \frac{2119}{50} (n_{12} v_{12})^2
 + \frac{58769}{2100} v_{12}^{2}\Bigr)\biggr]\,,
\\
\bm{\eta}^{(4)}_{1\,\text{4PN}}={}
&\Bigl(\frac{8861}{2100} m_{1}^3 m_{2}
 + \frac{613}{350} m_{1}^2 m_{2}^2
 -  \frac{5183}{2100} m_{1} m_{2}^3\Bigr) \frac{\bm{n}_{12}}{r_{12}^3}
\,,
\end{align}\end{subequations}
together with $1 \leftrightarrow 2$ for the other particle.

\section{The integral of the center of mass (CM)} 
\label{sec:Gi} 

With the general frame harmonic-coordinates 4PN Lagrangian (see
App.~\ref{sec:Lgen}), we have computed the ten invariants associated with the
invariance under the Lorentz-Poincar\'e group. The results being very long, we
only display the integral of the CM, which is necessary to define the frame of
the CM used throughout this paper. The CM position $\bm{G}$ satisfies
\begin{align}\label{dGdt}
\frac{\ud \bm{G}}{\ud t} = \bm{P}\,,
\end{align}
where $\bm{P}$ is the conserved linear momentum, $\ud \bm{P}/\ud t = 0$.
We have thus $\bm{G}=\bm{P} t + \bm{Z}$ for the conserved dynamics, where $\bm{Z}$ denotes the CM integral. The
complete results are
\begin{subequations}\label{G3PN}
\begin{align}
\bm{G}_\text{N} &= m_1 \bm{y}_1 + 1 \leftrightarrow 2\,,\\
%%%%%%%%%%%%%%%%%%%%%%%%%%%%%%%%%%%%%%%%%%%%%%%%%%%%%%%%%%%%%%%%%
\bm{G}_\text{1PN} &= \bm{y}_1\bigg(-\frac{G m_1 m_2}{2 r_{12}} + \frac{ m_1
  v_1^2}{2}\bigg) + 1 \leftrightarrow 2\,,\\
%%%%%%%%%%%%%%%%%%%%%%%%%%%%%%%%%%%%%%%%%%%%%%%%%%%%%%%%%%%%%%%%%
\bm{G}_\text{2PN} &= \bm{v}_1 G
m_1 m_2 \bigg(-\frac{7}{4} (n_{12}v_1) - 
\frac{7}{4} (n_{12}v_2)\bigg)    +  \bm{y}_1
\bigg(-\frac{5 G^2 m_1^2 m_2}{4 r_{12}^2} + \frac{7 G^2 m_1
m_2^2}{4r_{12}^2} \nonumber  \\ & \qquad +  
\frac{3 m_1 v_1^4}{8}  + \frac{G m_1
m_2}{r_{12}} 
\bigg(-\frac{1}{8} (n_{12}v_1)^2 - \frac{1}{4} (n_{12}v_1) (n_{12}v_2) +
\frac{1}{8} (n_{12}v_2)^2 \nonumber \\ & \qquad + \frac{19}{8} v_1^2  -
\frac{7}{4} 
(v_1v_2) - \frac{7}{8} v_2^2\bigg)\bigg) + 1
\leftrightarrow 2\,,\\
%%%%%%%%%%%%%%%%%%%%%%%%%%%%%%%%%%%%%%%%%%%%%%%%%%%%%%%%%%%%%%%%%
\bm{G}_\text{3PN} &= \bm{v}_1 \bigg( \frac{235G^2 m_1^2 m_2}{24r_{12}} \bigg(
(n_{12}v_1) - (n_{12}v_2)\bigg) - \frac{235G^2 m_1 m_2^2}{24r_{12}}
 \bigg( (n_{12}v_1) - 
(n_{12}v_2)\bigg) \nonumber \\ & \qquad + G m_1 m_2 \bigg(\frac{5}{12}
(n_{12}v_1)^3 + \frac{3}{8} 
(n_{12}v_1)^2 (n_{12}v_2) + \frac{3}{8} (n_{12}v_1) (n_{12}v_2)^2 
\nonumber \\ & \qquad +
\frac{5}{12} (n_{12}v_2)^3 
- \frac{15}{8} (n_{12}v_1) v_1^2 - (n_{12}v_2)
v_1^2 + \frac{1}{4} (n_{12}v_1) (v_1v_2) \nonumber \\ & \qquad 
+ \frac{1}{4} (n_{12}v_2) (v_1v_2) -
(n_{12}v_1) v_2^2 - \frac{15}{8} (n_{12}v_2) v_2^2\bigg)
\bigg) \nonumber \\ & +  
\bm{y}_1 \bigg( \frac{5m_1 v_1^6}{16} + 
\frac{G m_1 m_2}{r_{12}}
\bigg(\frac{1}{16} (n_{12}v_1)^4 + \frac{1}{8} (n_{12}v_1)^3 (n_{12}v_2) +
\frac{3}{16} (n_{12}v_1)^2 (n_{12}v_2)^2 \nonumber \\ & \qquad  
+ \frac{1}{4}
(n_{12}v_1) (n_{12}v_2)^3 - \frac{1}{16} 
(n_{12}v_2)^4 - \frac{5}{16} (n_{12}v_1)^2 v_1^2 - \frac{1}{2} (n_{12}v_1)
(n_{12}v_2) v_1^2 \nonumber \\ & \qquad - 
\frac{11}{8} (n_{12}v_2)^2 v_1^2 + \frac{53}{16} v_1^4 + \frac{3}{8}
(n_{12}v_1)^2 (v_1v_2) + \frac{3}{4} (n_{12}v_1) (n_{12}v_2) (v_1v_2) +
\frac{5}{4} (n_{12}v_2)^2 (v_1v_2) \nonumber \\ & \qquad - 5 v_1^2 (v_1v_2) +
\frac{17}{8} 
(v_1v_2)^2 - \frac{1}{4} (n_{12}v_1)^2 v_2^2 - \frac{5}{8} (n_{12}v_1)
(n_{12}v_2) v_2^2 + \frac{5}{16} (n_{12}v_2)^2 v_2^2 \nonumber \\ & \qquad +
\frac{31}{16} v_1^2 
v_2^2 - \frac{15}{8} (v_1v_2) v_2^2 - \frac{11}{16} v_2^4\bigg) + \frac{G^2
m_1^2 m_2}{r_{12}^2} \bigg(\frac{79}{12} (n_{12}v_1)^2 - \frac{17}{3}
(n_{12}v_1) (n_{12}v_2) \nonumber \\ & \qquad + 
\frac{17}{6} (n_{12}v_2)^2 - \frac{175}{24} v_1^2 + \frac{40}{3} (v_1v_2) -
\frac{20}{3} v_2^2\bigg) + \frac{G^2 m_1 m_2^2}{r_{12}^2} 
\bigg(-\frac{7}{3} (n_{12}v_1)^2
\nonumber \\ & \qquad +
\frac{29}{12} (n_{12}v_1) (n_{12}v_2) + \frac{2}{3} (n_{12}v_2)^2 +
\frac{101}{12} v_1^2 - 
\frac{40}{3} (v_1v_2) + \frac{139}{24} v_2^2 \bigg) \nonumber 
\\ & \qquad -\frac{19 G^3 m_1^2 m_2^2}{8 r_{12}^3}+ \frac{G^3 m_1^3
m_2}{r_{12}^3} \bigg(\frac{13721}{1260}  - \frac{22}{3} \ln
\left(\frac{r_{12}}{r'_1} \right)\bigg)  \nonumber \\ & \qquad 
+ \frac{G^3 m_1 m_2^3}{r_{12}^3}
\bigg(-\frac{14351}{1260} + 
\frac{22}{3} \ln \left(\frac{r_{12}}{r'_2} \right)\bigg)\bigg) + 1
\leftrightarrow 2\,.
\end{align}
\end{subequations}
together with
\begin{subequations}\label{G4PN}
\begin{align}
\bm{G}^{(0)}_\text{4PN} &= \frac{35}{128} m_{1} \bm{y}_1 v_1^{8} 
+ 1 \leftrightarrow 2\,,\\
%%%%%%%%%%%%%%%%%%%%%%%%%%%%%%%%%%%%%%%%%%%%%%%%%%%%%%%%%%%%%%%%%
\bm{G}^{(1)}_\text{4PN} &= m_{1} m_{2} \bm{v}_1 \left\{- \frac{13}{64} (n_{12} v_1)^5
 -  \frac{11}{64} (n_{12} v_1)^4 (n_{12} v_2)
 -  \frac{5}{32} (n_{12} v_1)^3 (n_{12} v_2)^2 \right. \nonumber\\
& \qquad -  \frac{5}{32} (n_{12} v_1)^2 (n_{12} v_2)^3
 -  \frac{11}{64} (n_{12} v_1) (n_{12} v_2)^4
 -  \frac{13}{64} (n_{12} v_2)^5
 + \frac{1}{16} (n_{12} v_1)^3 (v_1 v_2)\nonumber\\
& \qquad + \frac{5}{16} (n_{12} v_1)^2 (n_{12} v_2) (v_1 v_2)
 + \frac{5}{16} (n_{12} v_1) (n_{12} v_2)^2 (v_1 v_2)
 + \frac{1}{16} (n_{12} v_2)^3 (v_1 v_2)\nonumber\\
& \qquad + \frac{3}{16} (n_{12} v_1) (v_1 v_2)^2
 + \frac{3}{16} (n_{12} v_2) (v_1 v_2)^2
 + \frac{77}{96} (n_{12} v_1)^3 v_1^{2}
 + \frac{13}{32} (n_{12} v_1)^2 (n_{12} v_2) v_1^{2}\nonumber\\
& \qquad + \frac{7}{32} (n_{12} v_1) (n_{12} v_2)^2 v_1^{2}
 + \frac{17}{96} (n_{12} v_2)^3 v_1^{2}
 + \frac{3}{16} (n_{12} v_1) (v_1 v_2) v_1^{2}
 + \frac{1}{16} (n_{12} v_2) (v_1 v_2) v_1^{2}\nonumber\\
& \qquad -  \frac{123}{64} (n_{12} v_1) v_1^{4}
 -  \frac{49}{64} (n_{12} v_2) v_1^{4}
 + \frac{17}{96} (n_{12} v_1)^3 v_2^{2}
 + \frac{7}{32} (n_{12} v_1)^2 (n_{12} v_2) v_2^{2}\nonumber\\
& \qquad + \frac{13}{32} (n_{12} v_1) (n_{12} v_2)^2 v_2^{2}
 + \frac{77}{96} (n_{12} v_2)^3 v_2^{2}
 + \frac{1}{16} (n_{12} v_1) (v_1 v_2) v_2^{2}
 + \frac{3}{16} (n_{12} v_2) (v_1 v_2) v_2^{2}\nonumber\\
&\left. \qquad -  \frac{33}{32} (n_{12} v_1) v_1^{2} v_2^{2}
 -  \frac{33}{32} (n_{12} v_2) v_1^{2} v_2^{2}
 -  \frac{49}{64} (n_{12} v_1) v_2^{4}
 -  \frac{123}{64} (n_{12} v_2) v_2^{4}\right\}\nonumber\\
& + \frac{m_{1} m_{2} \bm{y}_1}{r_{12}} \left\{- \frac{5}{128} (n_{12} v_1)^6
 -  \frac{5}{64} (n_{12} v_1)^5 (n_{12} v_2)
 -  \frac{15}{128} (n_{12} v_1)^4 (n_{12} v_2)^2 \right.\nonumber\\
& \qquad -  \frac{5}{32} (n_{12} v_1)^3 (n_{12} v_2)^3
 -  \frac{25}{128} (n_{12} v_1)^2 (n_{12} v_2)^4
 -  \frac{15}{64} (n_{12} v_1) (n_{12} v_2)^5
 + \frac{5}{128} (n_{12} v_2)^6\nonumber\\
& \qquad -  \frac{11}{64} (n_{12} v_1)^4 (v_1 v_2)
 -  \frac{5}{16} (n_{12} v_1)^3 (n_{12} v_2) (v_1 v_2)
 -  \frac{15}{32} (n_{12} v_1)^2 (n_{12} v_2)^2 (v_1 v_2)\nonumber\\
& \qquad -  \frac{11}{16} (n_{12} v_1) (n_{12} v_2)^3 (v_1 v_2)
 -  \frac{65}{64} (n_{12} v_2)^4 (v_1 v_2)
 + \frac{5}{32} (n_{12} v_1)^2 (v_1 v_2)^2\nonumber\\
& \qquad + \frac{5}{16} (n_{12} v_1) (n_{12} v_2) (v_1 v_2)^2
 -  \frac{29}{32} (n_{12} v_2)^2 (v_1 v_2)^2
 + \frac{1}{16} (v_1 v_2)^3
 + \frac{27}{128} (n_{12} v_1)^4 v_1^{2}\nonumber\\
& \qquad + \frac{11}{32} (n_{12} v_1)^3 (n_{12} v_2) v_1^{2}
 + \frac{27}{64} (n_{12} v_1)^2 (n_{12} v_2)^2 v_1^{2}
 + \frac{15}{32} (n_{12} v_1) (n_{12} v_2)^3 v_1^{2}\nonumber\\
& \qquad + \frac{137}{128} (n_{12} v_2)^4 v_1^{2}
 + \frac{13}{32} (n_{12} v_1)^2 (v_1 v_2) v_1^{2}
 + \frac{7}{16} (n_{12} v_1) (n_{12} v_2) (v_1 v_2) v_1^{2}\nonumber\\
& \qquad + \frac{81}{32} (n_{12} v_2)^2 (v_1 v_2) v_1^{2}
 + \frac{97}{32} (v_1 v_2)^2 v_1^{2}
 -  \frac{53}{128} (n_{12} v_1)^2 v_1^{4}
 -  \frac{31}{64} (n_{12} v_1) (n_{12} v_2) v_1^{4}\nonumber\\
& \qquad -  \frac{225}{128} (n_{12} v_2)^2 v_1^{4}
 -  \frac{433}{64} (v_1 v_2) v_1^{4}
 + \frac{515}{128} v_1^{6}
 + \frac{15}{128} (n_{12} v_1)^4 v_2^{2}
 + \frac{9}{32} (n_{12} v_1)^3 (n_{12} v_2) v_2^{2}\nonumber\\
& \qquad + \frac{33}{64} (n_{12} v_1)^2 (n_{12} v_2)^2 v_2^{2}
 + \frac{27}{32} (n_{12} v_1) (n_{12} v_2)^3 v_2^{2}
 -  \frac{27}{128} (n_{12} v_2)^4 v_2^{2}\nonumber\\
& \qquad + \frac{7}{32} (n_{12} v_1)^2 (v_1 v_2) v_2^{2}
 + \frac{13}{16} (n_{12} v_1) (n_{12} v_2) (v_1 v_2) v_2^{2}
 + \frac{77}{32} (n_{12} v_2)^2 (v_1 v_2) v_2^{2}\nonumber\\
& \qquad + \frac{67}{32} (v_1 v_2)^2 v_2^{2}
 -  \frac{23}{64} (n_{12} v_1)^2 v_1^{2} v_2^{2}
 -  \frac{23}{32} (n_{12} v_1) (n_{12} v_2) v_1^{2} v_2^{2}
 -  \frac{157}{64} (n_{12} v_2)^2 v_1^{2} v_2^{2}\nonumber\\
& \qquad -  \frac{161}{32} (v_1 v_2) v_1^{2} v_2^{2}
 + \frac{381}{128} v_1^{4} v_2^{2}
 -  \frac{31}{128} (n_{12} v_1)^2 v_2^{4}
 -  \frac{53}{64} (n_{12} v_1) (n_{12} v_2) v_2^{4}\nonumber\\
& \left.\qquad + \frac{53}{128} (n_{12} v_2)^2 v_2^{4}
 -  \frac{123}{64} (v_1 v_2) v_2^{4}
 + \frac{251}{128} v_1^{2} v_2^{4}
 -  \frac{75}{128} v_2^{6}\right\} + 1 \leftrightarrow 2\,,\\
%%%%%%%%%%%%%%%%%%%%%%%%%%%%%%%%%%%%%%%%%%%%%%%%%%%%%%%%%%%%%%%%%
\bm{G}^{(2)}_\text{4PN} &= \bm{v}_1 \left\{\frac{m_{1}^2 m_{2}}{r_{12}} 
\left[- \frac{3341}{480} (n_{12} v_1)^3
 + \frac{10223}{480} (n_{12} v_1)^2 (n_{12} v_2)
 -  \frac{3781}{160} (n_{12} v_1) (n_{12} v_2)^2 \right. \right.\nonumber\\
& \qquad + \frac{4621}{480} (n_{12} v_2)^3
 -  \frac{4529}{240} (n_{12} v_1) (v_1 v_2)
 + \frac{6499}{240} (n_{12} v_2) (v_1 v_2)
 + \frac{9229}{480} (n_{12} v_1) v_1^{2}\nonumber\\
& \left. \qquad -  \frac{8849}{480} (n_{12} v_2) v_1^{2}
 + \frac{2293}{160} (n_{12} v_1) v_2^{2}
 -  \frac{3733}{160} (n_{12} v_2) v_2^{2}\right]
 + \frac{m_{1} m_{2}^2}{r_{12}} \left[\frac{4621}{480} (n_{12} v_1)^3 \right.\nonumber\\
& \qquad -  \frac{3781}{160} (n_{12} v_1)^2 (n_{12} v_2)
 + \frac{10223}{480} (n_{12} v_1) (n_{12} v_2)^2
 -  \frac{3341}{480} (n_{12} v_2)^3\nonumber\\
& \qquad + \frac{6499}{240} (n_{12} v_1) (v_1 v_2)
 -  \frac{4529}{240} (n_{12} v_2) (v_1 v_2)
 -  \frac{3733}{160} (n_{12} v_1) v_1^{2}
 + \frac{2293}{160} (n_{12} v_2) v_1^{2}\nonumber\\
&\left. \left. \qquad -  \frac{8849}{480} (n_{12} v_1) v_2^{2}
 + \frac{9229}{480} (n_{12} v_2) v_2^{2}\right]\right\}
 \nonumber\\ &+ \bm{y}_1 \left\{ \frac{m_{1}^2 m_{2}}{r_{12}^2} 
\left[- \frac{1693}{960} (n_{12} v_1)^4 + \frac{53}{240} (n_{12} v_1)^3 (n_{12} v_2)
 \right.\right.\nonumber\\
& \qquad -  \frac{4079}{480} (n_{12} v_1)^2 (n_{12} v_2)^2
 + \frac{1133}{240} (n_{12} v_1) (n_{12} v_2)^3 \nonumber\\
& \qquad -  \frac{1133}{960} (n_{12} v_2)^4
 -  \frac{509}{60} (n_{12} v_1)^2 (v_1 v_2)
 -  \frac{127}{60} (n_{12} v_1) (n_{12} v_2) (v_1 v_2)\nonumber\\
& \qquad -  \frac{179}{60} (n_{12} v_2)^2 (v_1 v_2)
 + \frac{109}{80} (v_1 v_2)^2
 + \frac{247}{60} (n_{12} v_1)^2 v_1^{2}
 + \frac{451}{60} (n_{12} v_1) (n_{12} v_2) v_1^{2}\nonumber\\
& \qquad + \frac{17}{60} (n_{12} v_2)^2 v_1^{2}
 + \frac{713}{240} (v_1 v_2) v_1^{2}
 -  \frac{3803}{960} v_1^{4}
 + \frac{1709}{120} (n_{12} v_1)^2 v_2^{2}\nonumber\\
& \left. \qquad -  \frac{139}{10} (n_{12} v_1) (n_{12} v_2) v_2^{2}
 + \frac{139}{20} (n_{12} v_2)^2 v_2^{2}
 + \frac{3433}{240} (v_1 v_2) v_2^{2}
 -  \frac{2873}{480} v_1^{2} v_2^{2}
 -  \frac{2931}{320} v_2^{4}\right]\nonumber\\
& \qquad + \frac{m_{1} m_{2}^2}{r_{12}^2} \left[\frac{1133}{960} (n_{12} v_1)^4
 + \frac{1337}{240} (n_{12} v_1)^3 (n_{12} v_2)
 -  \frac{7141}{480} (n_{12} v_1)^2 (n_{12} v_2)^2 \right. \nonumber\\
& \qquad + \frac{2197}{240} (n_{12} v_1) (n_{12} v_2)^3
 -  \frac{4187}{960} (n_{12} v_2)^4
 + \frac{247}{30} (n_{12} v_1)^2 (v_1 v_2)\nonumber\\
& \qquad + \frac{37}{60} (n_{12} v_1) (n_{12} v_2) (v_1 v_2)
 + \frac{59}{60} (n_{12} v_2)^2 (v_1 v_2)
 + \frac{2091}{80} (v_1 v_2)^2
 -  \frac{31}{5} (n_{12} v_1)^2 v_1^{2}\nonumber\\
& \qquad + \frac{2}{5} (n_{12} v_1) (n_{12} v_2) v_1^{2}
 -  \frac{239}{120} (n_{12} v_2)^2 v_1^{2}
 -  \frac{10153}{240} (v_1 v_2) v_1^{2}
 + \frac{5631}{320} v_1^{4}
 -  \frac{83}{15} (n_{12} v_1)^2 v_2^{2}\nonumber\\
& \left. \left. \qquad + \frac{313}{120} (n_{12} v_1) (n_{12} v_2) v_2^{2}
 + \frac{241}{120} (n_{12} v_2)^2 v_2^{2}
 -  \frac{7193}{240} (v_1 v_2) v_2^{2}
 + \frac{9473}{480} v_1^{2} v_2^{2}
 + \frac{9083}{960} v_2^{4}\right]\right\} \nonumber\\& + 1 \leftrightarrow 2\,,\\
%%%%%%%%%%%%%%%%%%%%%%%%%%%%%%%%%%%%%%%%%%%%%%%%%%%%%%%%%%%%%%%%%
\bm{G}^{(3)}_\text{4PN} &= \bm{v}_1 \left\{\frac{m_{1}^2 m_{2}^2}{r_{12}^2} 
\left[\frac{1099}{144} (n_{12} v_1)
 -  \frac{41}{64} \pi^2 (n_{12} v_1)
 + \frac{1099}{144} (n_{12} v_2)
 -  \frac{41}{64} \pi^2 (n_{12} v_2)\right] \right.\nonumber\\
& \qquad + \frac{m_{1}^3 m_{2}}{r_{12}^2} \left[- \frac{562}{9} (n_{12} v_1)
 + 44 \ln\Big(\frac{r_{12}}{r'_{1}}\Big) (n_{12} v_1)
 + \frac{14377}{280} (n_{12} v_2)
 -  \frac{110}{3} \ln\Big(\frac{r_{12}}{r'_{1}}\Big) (n_{12} v_2)\right]\nonumber\\
& \left. \qquad + \frac{m_{1} m_{2}^3}{r_{12}^2} \left[\frac{14377}{280} (n_{12} v_1)
 -  \frac{110}{3} \ln\Big(\frac{r_{12}}{r'_{2}}\Big) (n_{12} v_1)
 -  \frac{562}{9} (n_{12} v_2)
 + 44 \ln\Big(\frac{r_{12}}{r'_{2}}\Big) (n_{12} v_2)\right]\right\}\nonumber\\
& + \bm{y}_1 \left\{\frac{m_{1}^2 m_{2}^2}{r_{12}^3} 
\left[ \frac{2059}{96} (n_{12} v_1)^2
 -  \frac{123}{128} \pi^2 (n_{12} v_1)^2
 -  \frac{3115}{48} (n_{12} v_1) (n_{12} v_2) \right.  \right. \nonumber\\
& \qquad + \frac{123}{64} \pi^2 (n_{12} v_1) (n_{12} v_2)
 + \frac{2317}{96} (n_{12} v_2)^2
 -  \frac{123}{128} \pi^2 (n_{12} v_2)^2
 + \frac{4429}{144} (v_1 v_2)\nonumber\\
& \left. \qquad -  \frac{41}{64} \pi^2 (v_1 v_2)
 -  \frac{1071}{32} v_1^{2}
 + \frac{123}{128} \pi^2 v_1^{2}
 + \frac{439}{288} v_2^{2}
 -  \frac{41}{128} \pi^2 v_2^{2}\right]\nonumber\\
& \qquad + \frac{m_{1}^3 m_{2}}{r_{12}^3} \left[- \frac{9921}{2800} (n_{12} v_1)^2
 + 22 \ln\Big(\frac{r_{12}}{r'_{1}}\Big) (n_{12} v_1)^2
 -  \frac{198097}{4200} (n_{12} v_1) (n_{12} v_2) \right. \nonumber\\
& \qquad - 22 \ln\Big(\frac{r_{12}}{r'_{1}}\Big) (n_{12} v_1) (n_{12} v_2)
 + \frac{198097}{8400} (n_{12} v_2)^2
 + 11 \ln\Big(\frac{r_{12}}{r'_{1}}\Big) (n_{12} v_2)^2
 -  \frac{9875}{1008} (v_1 v_2)\nonumber\\
& \left. \qquad + \frac{154}{3} \ln\Big(\frac{r_{12}}{r'_{1}}\Big) (v_1 v_2)
 + \frac{160193}{10080} v_1^{2}
 - 33 \ln\Big(\frac{r_{12}}{r'_{1}}\Big) v_1^{2}
 -  \frac{937}{1440} v_2^{2}
 - 22 \ln\Big(\frac{r_{12}}{r'_{1}}\Big) v_2^{2}\right]\nonumber\\
& \qquad + \frac{m_{1} m_{2}^3}{r_{12}^3} \left[- \frac{185497}{8400} (n_{12} v_1)^2
 - 11 \ln\Big(\frac{r_{12}}{r'_{2}}\Big) (n_{12} v_1)^2
 + \frac{6397}{75} (n_{12} v_1) (n_{12} v_2) \right. \nonumber\\
& \qquad -  \frac{59501}{1200} (n_{12} v_2)^2
 -  \frac{937}{720} (v_1 v_2)
 - 44 \ln\Big(\frac{r_{12}}{r'_{2}}\Big) (v_1 v_2)
 + \frac{2737}{1440} v_1^{2}
 + 22 \ln\Big(\frac{r_{12}}{r'_{2}}\Big) v_1^{2}\nonumber\\
& \left. \left. \qquad -  \frac{12689}{2016} v_2^{2}
 + \frac{77}{3} \ln\Big(\frac{r_{12}}{r'_{2}}\Big) v_2^{2}\right]\right\} 
+ 1 \leftrightarrow 2\,,\\
%%%%%%%%%%%%%%%%%%%%%%%%%%%%%%%%%%%%%%%%%%%%%%%%%%%%%%%%%%%%%%%%%
\bm{G}^{(4)}_\text{4PN} &=  \bm{y}_1 \left\{\frac{m_{1}^4 m_{2}}{r_{12}^4} 
\left[- \frac{213929}{3600}
 + \frac{220}{3} \ln\Big(\frac{r_{12}}{r'_{1}}\Big)\right]
 + \frac{m_{1} m_{2}^4}{r_{12}^4} \left[\frac{215279}{3600}
 -  \frac{220}{3} \ln\Big(\frac{r_{12}}{r'_{2}}\Big)\right] \right. \nonumber\\
& \qquad + \frac{m_{1}^2 m_{2}^3}{r_{12}^4} \left[\frac{1301639}{12600}
 -  \frac{11}{2} \pi^2
 + 16 \ln\Big(\frac{r_{12}}{r'_{1}}\Big)
 -  \frac{110}{3} \ln\Big(\frac{r_{12}}{r'_{2}}\Big)\right]
 + \frac{m_{1}^3 m_{2}^2}{r_{12}^4} \left[- \frac{144347}{1800} \right. \nonumber\\
& \left. \left. \qquad + \frac{11}{2} \pi^2
 + \frac{88}{3} \ln\Big(\frac{r_{12}}{r'_{1}}\Big)
 - 16 \ln\Big(\frac{r_{12}}{r'_{2}}\Big)\right]\right\} + 1 \leftrightarrow 2\,.
\end{align}
\end{subequations}

The frame of the CM is defined by $\bm{G}=0$, which we solve iteratively with
standard order reduction of accelerations, using the CM equations of motion. This
gives the individual positions of the particles $\bm{y}_A$ in the CM frame as
\begin{subequations}\label{y1y2}
\begin{align}
\bm{y}_1 &= \Big[X_2+\nu (X_1-X_2) P\Big] \bm{x} +\nu
(X_1-X_2)Q\,\bm{v} \,,\\ \bm{y}_2 &= \Big[-X_1+\nu (X_1-X_2)
  P\Big] \bm{x} +\nu (X_1-X_2) Q\,\bm{v} \,,
\end{align}
\end{subequations}
where $\bm{x}=\bm{y}_1-\bm{y}_2$ and $\bm{v}=\ud\bm{x}/\ud t$ are the relative
separation and velocity (see Sec.~\ref{sec:not} for the other notations). The
coefficients $P$ and $Q$ admit the following detailed 4PN
expansions:\footnote{Concerning the dissipative contributions, we write only
  those that appear at the 2.5PN order, which turn out to be only of the type
  $Q_\text{2.5PN}$ (thus, we have $P_\text{2.5PN}=0$). We also expect 3.5PN contributions $P_\text{3.5PN}$ and
  $Q_\text{3.5PN}$, but we do not need these to control the 3.5PN terms in the CM energy and angular momentum [Eqs.~\eqref{Ediss}--\eqref{Gdiss}], and they have not been computed yet.}
\begin{subequations}\label{P}
\begin{align}
P_\text{1PN} &= \frac{v^2}{2} -\frac{G m}{2\,r}\,,\\
%%%%%%%%%%%%%%%%%%%%%%%%%%%%%%%%%%%%%%%%%%%%%%%%%%%%%%%%%%%%%%%%%
P_\text{2PN} &= \frac{3\,v^4}{8} - \frac{3\,\nu\,v^4}{2}
\nonumber\\ & \quad+ \frac{G m}{r}\,\left( -\frac{\dot{r}^2}{8} +
\frac{3\,\dot{r}^2\,\nu}{4} + \frac{19\,v^2}{8} +
\frac{3\,\nu\,v^2}{2} \right)\nonumber\\ &
\quad+\frac{G^2m^2}{r^2}\left(\frac{7}{4} - \frac{\nu}{2} \right)\,,\\
%%%%%%%%%%%%%%%%%%%%%%%%%%%%%%%%%%%%%%%%%%%%%%%%%%%%%%%%%%%%%%%%%
P_\text{3PN} &= \frac{5\,v^6}{16} -
  \frac{11\,\nu\,v^6}{4} + 6\,\nu^2\,v^6 \nonumber\\ & \quad
  +\frac{G m}{r}\left( \frac{\dot{r}^4}{16} -
  \frac{5\,\dot{r}^4\,\nu}{8} + \frac{21\,\dot{r}^4\,\nu^2}{16} -
  \frac{5\,\dot{r}^2\,v^2}{16} + \frac{21\,\dot{r}^2\,\nu\,v^2}{16}
  \right.\nonumber\\ & \quad\quad\quad -\left.
  \frac{11\,\dot{r}^2\,\nu^2\,v^2}{2} + \frac{53\,v^4}{16} -
  7\,\nu\,v^4 - \frac{15\,\nu^2\,v^4}{2} \right) \nonumber\\ & \quad
  +\frac{G^2m^2}{r^2}\left( -\frac{7\,\dot{r}^2}{3} +
  \frac{73\,\dot{r}^2\,\nu}{8} + 4\,\dot{r}^2\,\nu^2 +
  \frac{101\,v^2}{12} - \frac{33\,\nu\,v^2}{8} + 3\,\nu^2\,v^2 \right)
  \nonumber\\ & \quad + \frac{G^3m^3}{r^3}\left( -\frac{14351}{1260} +
  \frac{\nu}{8} - \frac{\nu^2}{2} + \frac{22}{3}\,\ln
  \Big(\frac{r}{r''_0}\Big) \right)\,,\\ 
%%%%%%%%%%%%%%%%%%%%%%%%%%%%%%%%%%%%%%%%%%%%%%%%%%%%%%%%%%%%%%%%%
P^{(0)}_\text{4PN} &= \left(\frac{35}{128}
 -  \frac{125}{32} \nu
 + \frac{145}{8} \nu^2
 -  \frac{55}{2} \nu^3\right) v^{8}
 \,,\\
%%%%%%%%%%%%%%%%%%%%%%%%%%%%%%%%%%%%%%%%%%%%%%%%%%%%%%%%%%%%%%%%%
P^{(1)}_\text{4PN} &= \frac{m}{r} \left(- \frac{5}{128} \dot{r}^6
 + \frac{35}{64} \nu \dot{r}^6
 -  \frac{125}{64} \nu^2 \dot{r}^6
 + \frac{55}{32} \nu^3 \dot{r}^6
 + \frac{27}{128} \dot{r}^4 v^{2}
 -  \frac{115}{64} \nu \dot{r}^4 v^{2}
 + \frac{517}{64} \nu^2 \dot{r}^4 v^{2}\right.\nonumber\\
& \quad-  \frac{213}{16} \nu^3 \dot{r}^4 v^{2}
 -  \frac{53}{128} \dot{r}^2 v^{4}
 + \frac{3}{2} \nu \dot{r}^2 v^{4}
 -  \frac{95}{8} \nu^2 \dot{r}^2 v^{4}
 + 36 \nu^3 \dot{r}^2 v^{4}
 + \frac{515}{128} v^{6}
\nonumber\\
&\quad \left.  -  \frac{749}{32} \nu v^{6} + \frac{91}{4} \nu^2 v^{6}
 + 42 \nu^3 v^{6}\right)
 \,,\\
%%%%%%%%%%%%%%%%%%%%%%%%%%%%%%%%%%%%%%%%%%%%%%%%%%%%%%%%%%%%%%%%%
P^{(2)}_\text{4PN} &= \frac{m^2}{r^2} \left(\frac{1133}{960} \dot{r}^4
 -  \frac{1007}{48} \nu \dot{r}^4
 + \frac{169}{24} \nu^2 \dot{r}^4
 + 9 \nu^3 \dot{r}^4
 -  \frac{31}{5} \dot{r}^2 v^{2}
 + 26 \nu \dot{r}^2 v^{2}
 -  \frac{541}{8} \nu^2 \dot{r}^2 v^{2}\right.\nonumber\\
&\left.\quad -  \frac{83}{2} \nu^3 \dot{r}^2 v^{2}
 + \frac{5631}{320} v^{4}
 -  \frac{139}{4} \nu v^{4}
 + \frac{71}{4} \nu^2 v^{4}
 -  \frac{45}{2} \nu^3 v^{4}\right)
 \,,\\
%%%%%%%%%%%%%%%%%%%%%%%%%%%%%%%%%%%%%%%%%%%%%%%%%%%%%%%%%%%%%%%%%
P^{(3)}_\text{4PN} &= \frac{m^3}{r^3} \left(- \frac{185497}{8400} \dot{r}^2
 -  \frac{64347}{1120} \nu \dot{r}^2
 -  \frac{123}{128} \pi^2 \nu \dot{r}^2
 + \frac{495}{16} \nu^2 \dot{r}^2
 + \frac{55}{4} \nu^3 \dot{r}^2
 + 11 \nu \ln\Big(\frac{r}{r'_{0}}\Big) \dot{r}^2\right.\nonumber\\
&\quad - 11 \ln\Big(\frac{r}{r''_{0}}\Big) \dot{r}^2
 + 33 \nu \ln\Big(\frac{r}{r''_{0}}\Big) \dot{r}^2
 + \frac{2737}{1440} v^{2}
 -  \frac{87181}{3360} \nu v^{2}
 + \frac{123}{128} \pi^2 \nu v^{2}
 -  \frac{117}{8} \nu^2 v^{2}\nonumber\\
&\quad\left. + 5 \nu^3 v^{2}
 - 11 \nu \ln\Big(\frac{r}{r'_{0}}\Big) v^{2}
 + 22 \ln\Big(\frac{r}{r''_{0}}\Big) v^{2}\right)
 \,,\\
%%%%%%%%%%%%%%%%%%%%%%%%%%%%%%%%%%%%%%%%%%%%%%%%%%%%%%%%%%%%%%%%%
P^{(4)}_\text{4PN} &= \frac{m^4}{r^4} \left(\frac{215279}{3600}
 + \frac{22043}{720} \nu
 -  \frac{11}{2} \pi^2 \nu
 -  \frac{3}{2} \nu^2
 -  \frac{1}{2} \nu^3
 + 16 \ln\Big(\frac{r}{r'_{0}}\Big)
 - 60 \nu \ln\Big(\frac{r}{r'_{0}}\Big)\right.\nonumber\\
&\quad\left. -  \frac{268}{3} \ln\Big(\frac{r}{r''_{0}}\Big)
 + 120 \nu \ln\Big(\frac{r}{r''_{0}}\Big)\right)
 \,,
\end{align}
\end{subequations}
and
\begin{subequations}\label{Q}
\begin{align}
Q_\text{2PN} &= -\frac{7\,G\,m\,\dot{r}}{4}\,,\\
%%%%%%%%%%%%%%%%%%%%%%%%%%%%%%%%%%%%%%%%%%%%%%%%%%%%%%%%%%%%%%%%%
Q_\text{2.5PN} &= \frac{4\,G\,m\,v^2}{5}
  -\frac{8\,G^2m^2}{5\,r} \,,\\ 
%%%%%%%%%%%%%%%%%%%%%%%%%%%%%%%%%%%%%%%%%%%%%%%%%%%%%%%%%%%%%%%%%
Q_\text{3PN} &= G\,m\,\dot{r}\left( \frac{5\,\dot{r}^2}{12} -
\frac{19\,\dot{r}^2\,\nu}{24} - \frac{15\,v^2}{8} +
\frac{21\,\nu\,v^2}{4} \right) \nonumber\\ & \quad +
\frac{G^2m^2\,\dot{r}}{r}\left( -\frac{235}{24}- \frac{21\,\nu}{4}
\right) \,,\\
%%%%%%%%%%%%%%%%%%%%%%%%%%%%%%%%%%%%%%%%%%%%%%%%%%%%%%%%%%%%%%%%%
Q^{(1)}_\text{4PN} &= m \left(- \frac{13}{64} \dot{r}^5
 + \frac{25}{32} \nu \dot{r}^5
 -  \frac{17}{32} \nu^2 \dot{r}^5
 + \frac{77}{96} \dot{r}^3 v^{2}
 -  \frac{187}{48} \nu \dot{r}^3 v^{2}
 + \frac{19}{4} \nu^2 \dot{r}^3 v^{2}
 -  \frac{123}{64} \dot{r} v^{4}\right.\nonumber\\
&\left.\quad + \frac{199}{16} \nu \dot{r} v^{4}
 - 21 \nu^2 \dot{r} v^{4}\right)
 \,,\\
%%%%%%%%%%%%%%%%%%%%%%%%%%%%%%%%%%%%%%%%%%%%%%%%%%%%%%%%%%%%%%%%%
Q^{(2)}_\text{4PN} &= \frac{m^2}{r} \left(\frac{4621}{480} \dot{r}^3
 + \frac{113}{24} \nu \dot{r}^3
 + \frac{7}{12} \nu^2 \dot{r}^3
 -  \frac{3733}{160} \dot{r} v^{2}
 + \frac{95}{4} \nu \dot{r} v^{2}
 + 28 \nu^2 \dot{r} v^{2}\right)
 \,,\\
%%%%%%%%%%%%%%%%%%%%%%%%%%%%%%%%%%%%%%%%%%%%%%%%%%%%%%%%%%%%%%%%%
Q^{(3)}_\text{4PN} &= \frac{m^3}{r^2} \left(\frac{14377}{280}
 + \frac{71509}{5040} \nu
 -  \frac{41}{64} \pi^2 \nu
 -  \frac{49}{4} \nu^2
 + \frac{22}{3} \nu \ln\Big(\frac{r}{r'_{0}}\Big)
 -  \frac{110}{3} \ln\Big(\frac{r}{r''_{0}}\Big)
 -  \frac{44}{3} \nu \ln\Big(\frac{r}{r''_{0}}\Big)\right) \dot{r}
 \,.
\end{align}\end{subequations}
The CM velocities $\bm{v}_A$ are obtained by differentiating Eqs.~\eqref{y1y2}
with order reduction of accelerations. Recall, from Sec.~\ref{sec:tails}, that
there are no tail contributions in these expressions. The
formulas~\eqref{P}--\eqref{Q} contain logarithmic terms depending on the gauge
constants $r'_0$ and $r''_0$ (\textit{i.e.}, not affecting physical results)
defined from the two scales $r'_A$ entering the general-frame Lagrangian by
means of Eqs.~\eqref{r'0r''0}. Notice the factors $X_{1}-X_{2}$ introduced in front of our definitions of $P$ and $Q$ in Eqs.~\eqref{y1y2}, which guarantee the well-defined equal-mass limit $X_{1}=X_{2}$.

%\bibliography{/nethome/blanchet/Articles/ListeRef/ListeRef.bib}
\bibliography{ListeRef_BBFM}

\end{document}